\documentclass[fleqn,usenatbib]{mnras}

\usepackage{newtxtext,newtxmath}

\usepackage[T1]{fontenc}
\usepackage{ae,aecompl}
\usepackage{tabularx}

\usepackage{graphicx}	
\usepackage{amsmath}	
\usepackage{amssymb}	

\newcommand{\Msun}{$M_{\odot}$ }

\newcommand{\msun}{M_\odot}
\newcommand{\msunyr}{M_\odot \, {\rm yr}^{-1}}
\newcommand{\kms}{{\rm km \, s}^{-1}}

\newcolumntype{Y}{>{\centering\arraybackslash}X}
\newcommand{\insitu}{{\it in-situ}~}
\newcommand{\rbirth}{r_{\rm birth}}

\title[ELS {\it Redux}]{Stars made in outflows may populate the stellar halo of the Milky Way \vspace{-0.1cm}}
\author[S. Yu et al.]{\hspace{-.01cm}Sijie Yu$^{1}$\thanks{E-mail: sijiey3@uci.edu},  
James S. Bullock$^{1}$,
Andrew Wetzel$^{2}$, 
Robyn E. Sanderson$^{3,4}$, 
\newauthor Andrew S. Graus$^{5}$, 
			Michael Boylan-Kolchin$^{5}$,
			Anna M. Nierenberg$^{6}$,
			Michael Y. Grudi{\'c}$^{7}$,
\newauthor		Philip F.~Hopkins$^{8}$, 
			Du{\v s}an Kere{\v s}$^{9}$, 
			Claude-Andr{\'e} Faucher-Gigu{\`e}re$^{7}$ 
		\vspace*{5pt} \\
$^{1}$Department of Physics and Astronomy, University of California Irvine, CA 92697, USA \\
$^{2}$Department of Physics, University of California, Davis, CA 95616, USA \\
$^{3}$Department of Physics and Astronomy, University of Pennsylvania, 209 S 33rd St., Philadelphia, PA 19104, USA \\ 
$^{4}$Center for Computational Astrophysics, Flatiron Institute, 162 5th Ave., New York, NY 10010, USA \\
$^{5}$Department of Astronomy, The University of Texas at Austin,		2515 Speedway, Stop C1400, Austin, TX 78712-1205, USA \\
$^{6}$Jet Propulsion Laboratory, California Institute of Technology, 4800 Oak Grove Dr, Pasadena, CA 91109, USA\\
$^{7}$Department of Physics and Astronomy and CIERA, Northwestern University, 2145 Sheridan Road, Evanston, IL 60208, USA \\
$^{8}$TAPIR, Mailcode 350-17, California Institute of Technology, Pasadena, CA 91125, USA \\
$^{9}$Department of Physics,  University of California at San Diego, 9500 Gilman Drive, La Jolla, CA 92093, USA}

\date{Accepted XXX. Received YYY; in original form ZZZ}

\pubyear{2019}

\begin{document}
\label{firstpage}
\pagerange{\pageref{firstpage}--\pageref{lastpage}}
\maketitle

\begin{abstract}
We  study stellar-halo formation using six Milky Way-mass galaxies in FIRE-2 cosmological zoom simulations.  We find that $5-40\%$ of the outer ($50-300$ kpc) stellar halo in each system consists of \insitu stars that were born in outflows from the main galaxy. 
Outflow stars originate from gas accelerated by super-bubble winds, which can be compressed, cool, and form co-moving stars.  The majority of these stars remain bound to the halo and fall back with orbital  properties similar to the rest of the stellar halo at $z=0$.
In the outer halo, outflow stars are more spatially homogeneous, metal rich, and alpha-element-enhanced than the accreted stellar halo.   
At the solar location, up to $\sim 10 \%$ of our kinematically-identified halo stars were born in outflows; the fraction rises to as high as $\sim 40\%$ for the most metal-rich local halo stars ([Fe/H] $> -0.5$).   
We conclude that the Milky Way stellar halo could contain local counterparts to stars that are observed to form in molecular outflows in distant galaxies.  Searches for such a population may provide a new, near-field approach to constraining feedback and outflow physics.  A stellar halo contribution from outflows is a phase-reversal of the classic halo formation scenario of Eggen, Lynden-Bell \& Sandange, who suggested that halo stars formed in rapidly {\em infalling} gas clouds.  Stellar outflows may be observable in direct imaging of external galaxies and could provide a source for metal-rich, extreme velocity stars in the Milky Way.

\end{abstract}




\section{Introduction} \label{s:intro}

\begin{table*}
  \caption{Simulations used in this work ordered by stellar mass. All simulations have baryonic mass resolution $m_b = 7,067\msun$} 
	\centering 
	\label{tab:info}
	\begin{tabularx}{\textwidth}{XYYYYYYYYYYYY}
		\hline
		\hline  
		Simulation & $M_{\mathrm{200m}}$ &$R_{\mathrm{200m}}$& $M_{\star}$ & ${R_\mathrm{90}}$ &$V_\mathrm{max}$ & $M_{\star \mathrm{>20}}$ & $f_{20}$ & $M_{\star \mathrm{>50}}$ & $f_{50}$ & $M_{\star \mathrm{>150}}$ & $f_{150}$\\ 
		Name  &  $[M_{\odot}]$ &[kpc] & $[M_{\odot}]$ & [kpc] & [${\rm km}{\cdot}{\rm s}^{-1}$] & $[M_{\odot}]$ & & $[M_{\odot}]$ & & $[M_{\odot}]$\\
		\hline 
        \texttt{m12m} & 1.3e12 & 342 & 1.1e11 & 11.3 & 184 & 5.9e9 & 0.05 & 2.1e9 & 0.04 & 1.4e8 & 0.09 \\
        \texttt{m12f} & 1.4e12 & 355 & 8.6e10 & 11.0 & 183 & 8.7e9 & 0.05 & 1.9e9 & 0.06 & 3.1e8 & 0.07 \\
		\texttt{m12b} & 1.2e12 & 335 & 8.1e10 & 9.8 & 181 & 5.8e9 & 0.05 & 1.9e9 & 0.07 & 2.0e8 & 0.24 \\
		\texttt{m12i} & 9.8e11 & 314 & 6.4e10 & 9.2 & 161 & 2.8e9 & 0.08 & 9.8e8 & 0.07 & 9.8e7 & 0.17 \\
		\texttt{m12c} & 1.1e12 & 328 & 6.0e10 & 9.7 & 156 & 1.9e9 & 0.12 & 7.3e8 & 0.13 & 1.6e8 & 0.13 \\
		\texttt{m12w} & 9.1e11 & 301 & 5.8e10 & 8.7 & 156 & 2.4e9 & 0.10 & 6.0e8 & 0.14 & 4.8e7 & 0.34 \\
		\hline  
	\end{tabularx}
	\raggedright
	Columns include: (1) $M_{200m}$, $R_{200m}$: Mass and radius of the halo at $z=0$ defined at an overdensity of  $200$ times the mean matter density. (2) $M_{\star}$: Stellar mass within 20 kpc of the center of the halo at $z=0$.	(3) $R_{90}$: radius enclosing 90\% of $M_{\rm star}$.	(4)$V_{\rm max}$: Maximum circular velocity. 	(5) $M_{>20}$/$M_{>50}$/$M_{>150}$: Mass of all the stars outside 20/50/150kpc from the center of the galaxy.
	(6) $f_{20}$/$f_{50}$/$f_{150}$: Mass fraction of stars outside of 20/50/150kpc that were born within 20 kpc from the center of the galaxy with birth radial velocity $>200 \kms$.
\end{table*}

Galactic outflows are common in the Universe and are crucial for understanding the enrichment of the intergalactic medium and galaxy evolution in general \citep{Veilleux2005}. Molecular outflows with the densities required for star formation are common \citep{Aalto2015,Sturm2011,Sakamoto2009} and there is growing evidence that stars sometimes form in outflows.  For example, \citet{Maiolino17} reported spectroscopic evidence of star formation inside of a galactic outflow at a rate exceeding $15 \msunyr$ in a $z \simeq 0.05$ galaxy.  These authors go on to suggest that star formation may be occurring commonly in galactic outflows but that it has been missed due to inadequate diagnostics. Similarly,  \citet{Gallagher2018} have used the integral field spectroscopic data from MaNGA-SDSS4 to show that prominent star formation is occurring inside $30\%$ of the galactic outflows in their sample. They also find that star formation inside outflows accounts for $5 - 30 \% $ of the total star formation in the galaxy when detected, consistent with another independent analysis from \cite{Rodrguez_del_Pino_2019}.

One important feature of these observed star-forming outflows is that they are fast ($\sim100-250 ~\kms$) but not so fast that they are unbound to the halo potential.  While the outflow stars will not remain confined to the {\em central} galaxy, most will travel outward and then return on plunging, stellar-halo like orbits.   If only $\sim 1\%$ of a $M_\star \simeq 5\times 10^{10} \msun$ galaxy's stars are formed in (bound) radial outflows, this could contribute an appreciable fraction of a $\sim 10^9 ~\msun$  stellar halo typical of a Milky Way-mass galaxy today.    In this paper, we show that stellar outflows do occur in FIRE-2 simulations\footnote{See the FIRE project web site at http://fire.northwestern.edu.} of Milky Way-mass galaxies and that they contribute to their stellar halos appreciably at $z=0$, especially at large galacto-centric radius.  Searches for and detections of such stars could provide a new tool for near-field cosmology to constrain feedback physics and to trace the outflow history of the Milky Way.

Galactic stellar halos are broadly understood to be good laboratories for testing ideas about galaxy formation and for revealing the specific formation histories of individual galaxies.
The stellar halo of the Milky Way was first discovered and characterized by stars that exist on plunging elliptical orbits distinct from those of the Galactic disk \citep{Oort1922,Lindblad1925}.  Later, authors such as \citet*[][ELS hereafter]{Eggen62} recognized that the timescales for stars on such orbits to exchange energy and angular momentum are long compared to the age of the universe, and used this as an opportunity to investigate the Galactic past.
  
   \cite{Johnston16} proposed a classification system for stars in the stellar halo with three categories associated with dynamical origin.  First, \insitu stars, which are formed in orbits close to their current orbits.  Second, {\it kicked-out} stars, which are formed in
orbits unlike their current ones.  And finally, {\it accreted} stars, which are formed in smaller galaxies outside the dark matter halo they currently occupy.  The original proposal for the Milky Way's stellar halo formation by ELS falls into the \insitu category.  ELS suggested that halo stars were born on eccentric orbits from radially-infalling gas clouds and that these stars preserve their eccentricity today.  Stellar halo stars that are classified as kicked-out are born within the inner galaxy on orbits that initially confine them to the disk but become heated to eccentric orbits by mergers \citep{Zolotov2009,Purcell10} or by potential fluctuations from explosive feedback events \citep{ElBadry16,El-Badry18}.  Kicked-out stars are usually envisioned to contribute to the inner stellar halo ($\lesssim 50$ kpc).

\begin{figure*}
	\includegraphics[height=.45\textwidth , trim = 5cm 0.0 28cm 0.0]{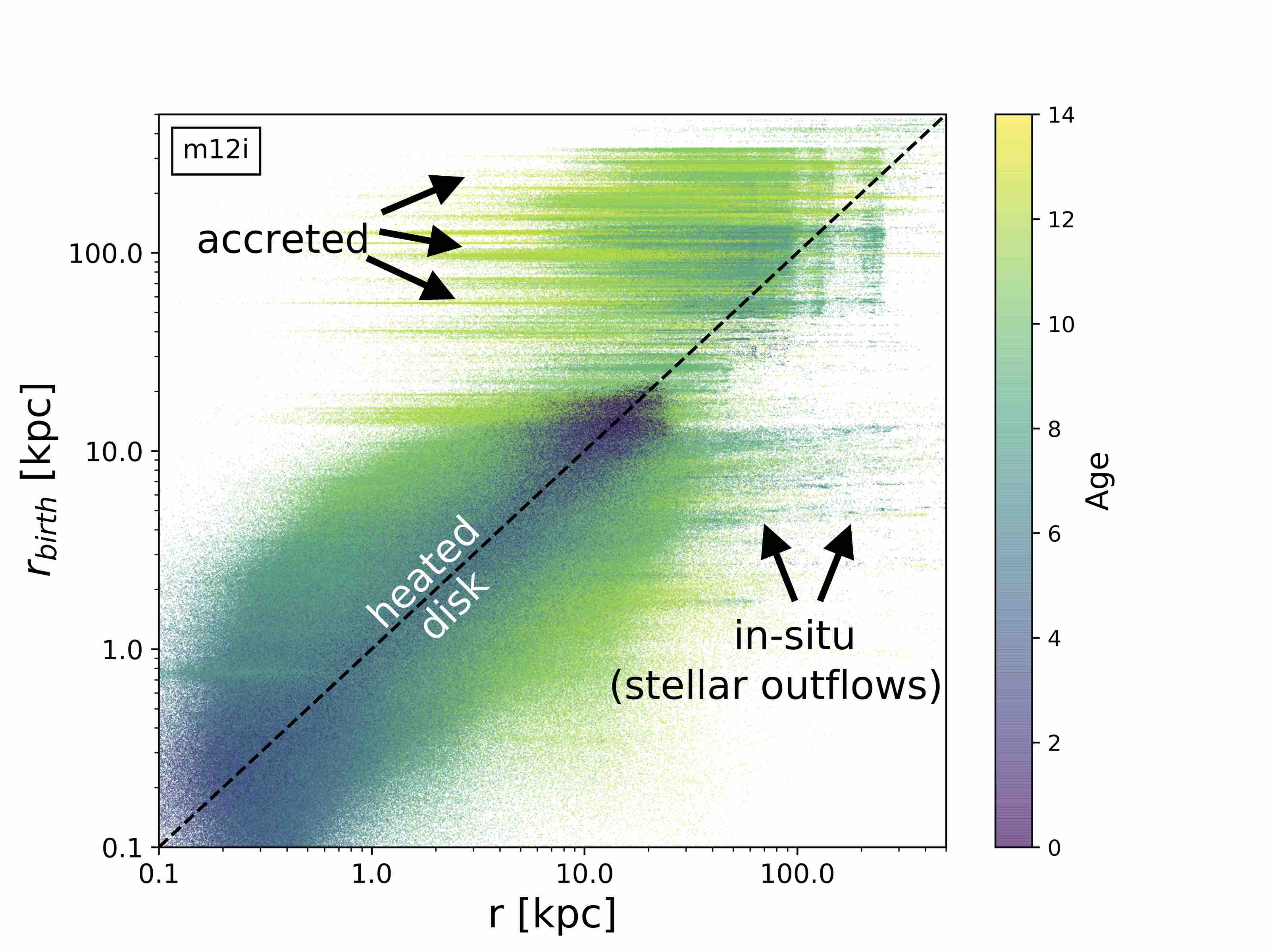}
	\includegraphics[height=0.45\textwidth , trim = 0.0 0.0 20cm 0.0]{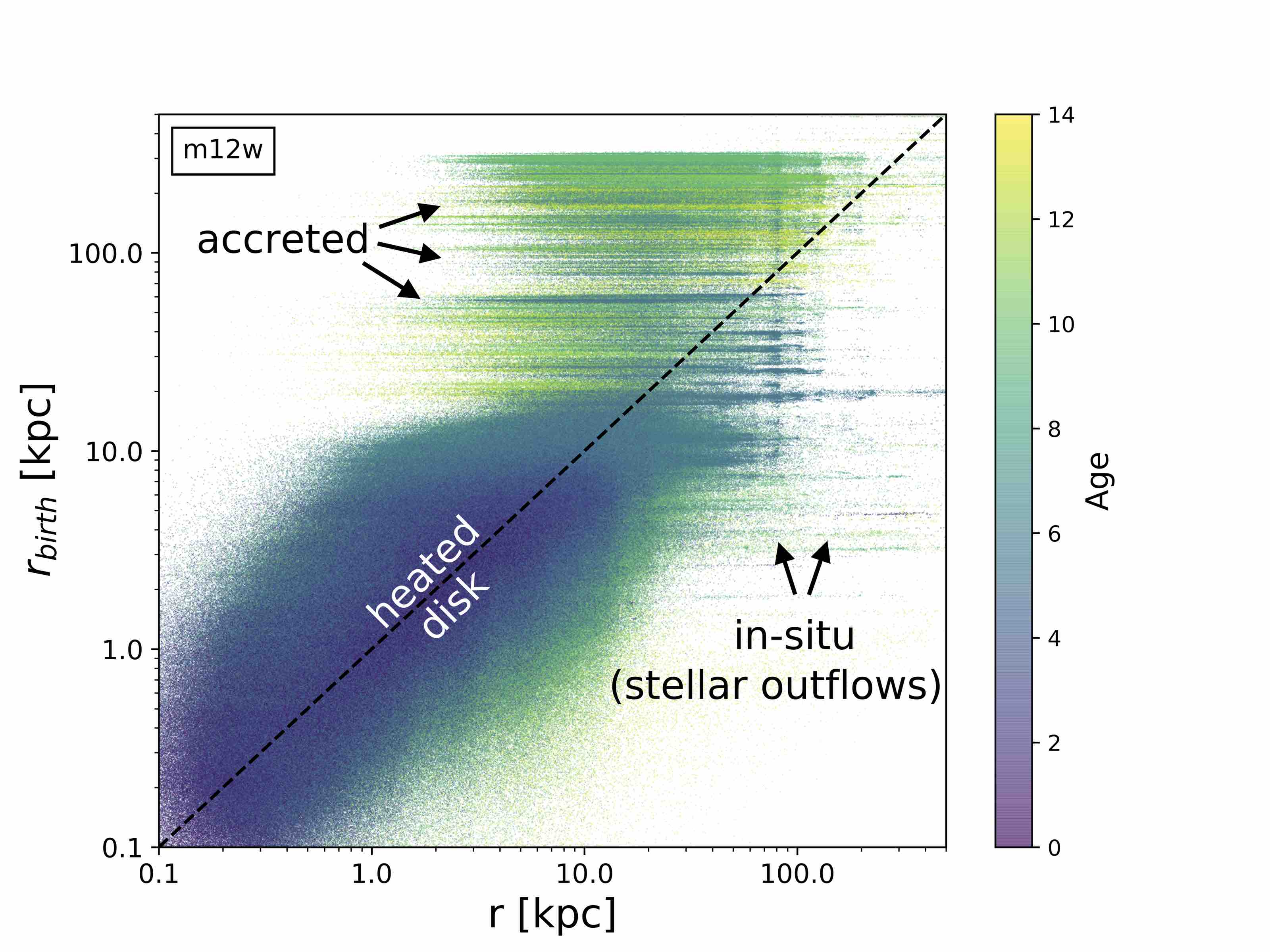}
	\centering
	\caption[Birth radius vs. present radius]{Star particle birth radius vs. current radius color coded by stellar age for \texttt{m12i}(left) and \texttt{m12w} (right). All the star particles that are bound to satellites have been removed from our sample. The dashed black lines in each panel trace $r_{\rm now} = r$.   Three populations are distinguishable here. 1) Accreted stars: these populations dominate at large birth radii ($\gtrsim 30$ kpc) and make thick horizontal bands that extend mostly to the left of the one-to-one line ($r_{\rm birth} > r$).  There is a less dominant population of accreted stars that are flung out after their birth galaxies are destroyed to present radii ($r> r_{\rm birth}$). These stars sometimes pile up in shells of constant $r$, which is the origin of vertical banding (e.g. at $r \simeq 200$ kpc in \texttt{m12i}).  2) Heated/disk stars: these populations mix to create the dense sequence of mostly young stars that is uniformly scattered about the one-to-one line at small radii ($r \lesssim 30$ kpc.  Heated stars are perturbed to both smaller and larger radius.  They sample a continuous population of pre-heated disk stars and leave no horizontal bands.  3) \insitu halo stars: this population is characterized by distinct horizontal bands with characteristic birth radii within the main galaxy ( $r_{\rm birth} \lesssim 10$ kpc) that today have $r > r_{\rm birth}$ as large as $\sim 500$ kpc.  Unlike the accreted stars, the \insitu bands extend to the right of the one-to-one line, which is indicative of an outflow.}
	\label{fig:birth_vs_current}
\end{figure*}

Today there is general agreement that the third category -- accretion -- is responsible for much of the stellar halo.  This idea was first discussed by \cite{Searle&Zinn78} in a scenario that
bears strong anecdotal resemblance to the hierarchical galaxy formation prediction of modern $\Lambda$CDM.   Indeed, one of the strongest pieces of evidence that structure formation is hierarchical on small scales is the success of CDM/accretion-based models of stellar halo formation \citep{Bullock2001,Bullock&Johnston05,DeLucia08,Johnston2008,Cooper10} in predicting the coherent structure observed in the outer stellar halos of the Milky Way and M31 \citep{Bell2008,McConnachie2009}. It is likely that even the central stellar halo of the Milky Way is significantly populated by at least one such accretion event \citep{Helmi_2018,Mackereth_2018,Simon2019,Cunningham_2019,Fattahi_2019,Myeong_2019,Matsuno_2019,necib2019evidence}.

While there is substantial theoretical motivation from a $\Lambda$CDM context to believe that both accreted and kicked-out components populate the stellar halo, little work in the modern framework has discussed the \insitu population advocated by ELS. As discussed above, if star-forming outflows are indeed common, then there is good reason to think that \insitu stars populate the stellar halo after all.  However, unlike the classic ELS conjecture, where halo stars are born on eccentric orbits from {\it inflowing} gas clouds, these \insitu halo stars are created in feedback-driven molecular {\it outflows}.  According to the classification of \citet{Johnston16}, these stars are \insitu because they were created on orbits that are similar to those they inhabit today.

In what follows, we present an analysis of six high-resolution, $\Lambda$CDM cosmological zoom simulations of Milky Way-mass galaxies \citep{Wetzel2016,SGK18,Sanderson2018} from the FIRE-2 collaboration \citep{Hopkins17}. 
In these simulations, clustered supernovae feedback events regularly drive outflows of compressed shells of high-density gas. Occasionally,
the accelerated regions are compressed to densities above our star-formation threshold of $1000 ~{\rm cm}^{-3}$ and become self-gravitating such that they trigger star formation within $\sim 1$ Myr of the acceleration.   Stars formed in these outflows travel ballistically outward and eventually fall back into the halo on eccentric orbits.  
As we show below, this \insitu population can contribute as much as $20-40\%$ of the stellar halo at $r > 250$ kpc and may be distinguished from accreted stars based on their chemical and spatial characteristics.

The outline of this paper is as follows.  Section \ref{s:sims and methods} provides an overview of our simulations and Section \ref{s:def} details our method for identifying outflow stars, while section \ref{s:origin} contains an investigation of their origin.  Section \ref{s:properties} covers properties of outflow stellar halo.  Section \ref{s:discuss} discusses some caveats, and section \ref{s:conclusion} provides our conclusions. 

\section{Simulations and Methods} \label{s:sims and methods}

The simulations that form the basis of our analysis are the hydrodynamic, cosmological zoom-in simulations run with the multi-method gravity plus hydrodynamics code \textsc{Gizmo} \citep{Hopkins15}.   They utilize the FIRE-2 feedback implementation \citep{Hopkins17} and the mesh-free Lagrangian Godunov (MFM) method that provides
adaptive spatial resolution while maintaining conservation of mass, energy, and momentum. The simulations include cooling, as well as heating from an ionizing background and stellar sources, including stellar feedback from OB stars and AGB mass-loss, type Ia and type II supernovae, and photo-heating and radiation pressure, the inputs for which are taken directly from stellar evolution models. Subgrid turbulent metal diffusion is also added, which can produce more realistic metallicity distributions in
dwarf galaxies \citep{Escala18} but does not significantly change other general properties of the galaxies \citep{Hopkins17_2,Su17}.  
Star formation occurs in molecular gas that is locally self-gravitating, sufficiently dense ($ > 1000$ cm$^{-3}$) and Jeans unstable \citep{Krumholz_2011}. The default model requires that the thermal Jeans mass is below the particle mass ($\sim 7000 ~\msun$), which is necessarily satisfied with the virial criterion. We also explore a more conservative model where the thermal Jeans mass is required to be below 1000 $\msun$, the flow must be converging, and the virial criterion is smoothed over time to eliminate spurious cases where the SF criteria is only met in a transient sense. The star formation efficiency of the molecular component of particles that satisfy all star formation criteria is set to $100\%$ per freefall time, ie. $\rm SFR_{\rm particle} = m_{\rm particle} \cdot f_{\rm mol} \, / \, t_{\rm ff}$. Gas particles are converted to stars at this rate probablistically \citep{Katz1996}. Note that this does {\em not} imply that the global efficiency of star formation (even on GMC-scales) is $100\%$. Self-regulated feedback limits star formation to $\sim 1-10\%$ per free-fall time \citep{Hopkins17_2,Orr_2018}.  

We focus on six Milky Way-mass galaxy simulations that are part of the ``Latte Suite'' \citep{Wetzel2016,SGK18,Sanderson2018,Samuel19}. These were all simulated with identical resolutions: initial baryonic particle masses of $m_b = 7,067\msun$,  
gas softening lengths fully adaptive down to $ \simeq 0.5-1$ pc, star softening lengths to $ \simeq 4$ pc, and a dark matter force softening of $\simeq 40$ pc. 
Table~\ref{tab:info} summarizes the galaxy/halo properties.  The galaxies are ordered by decreasing galaxy stellar mass at $z=0$, from $M_{\star} = 1.1\times10^{11} \msun$ (\texttt{m12m}, top) to $M_{\star} = 5.8\times10^{10} \msun$ (\texttt{m12w}, bottom).

\begin{figure*}
	\includegraphics[width=0.49\textwidth , trim = 0.0 0.0 0.0 15.0]{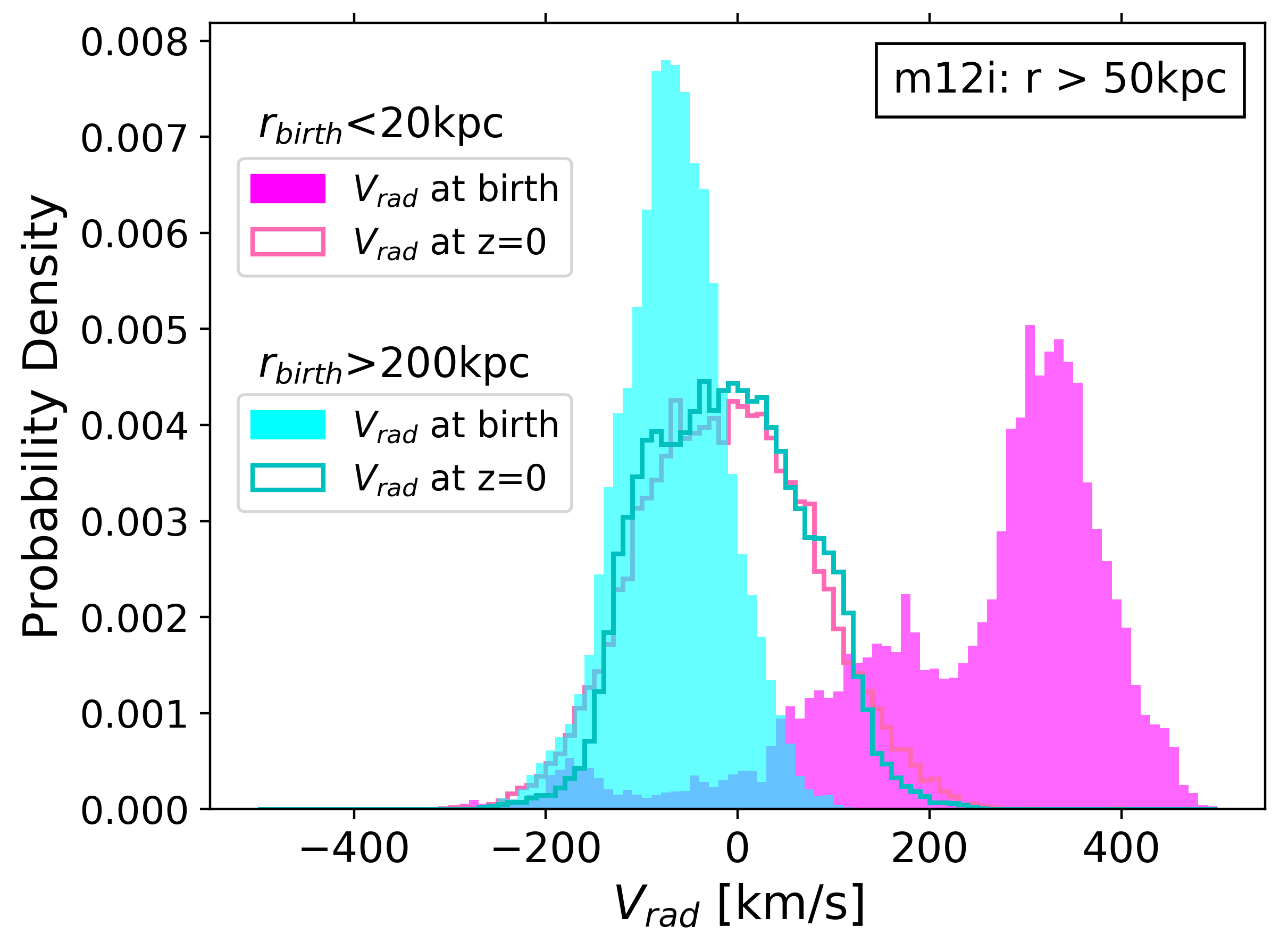}
	\includegraphics[width=0.49\textwidth , trim = 0.0 0.0 0.0 15.0]{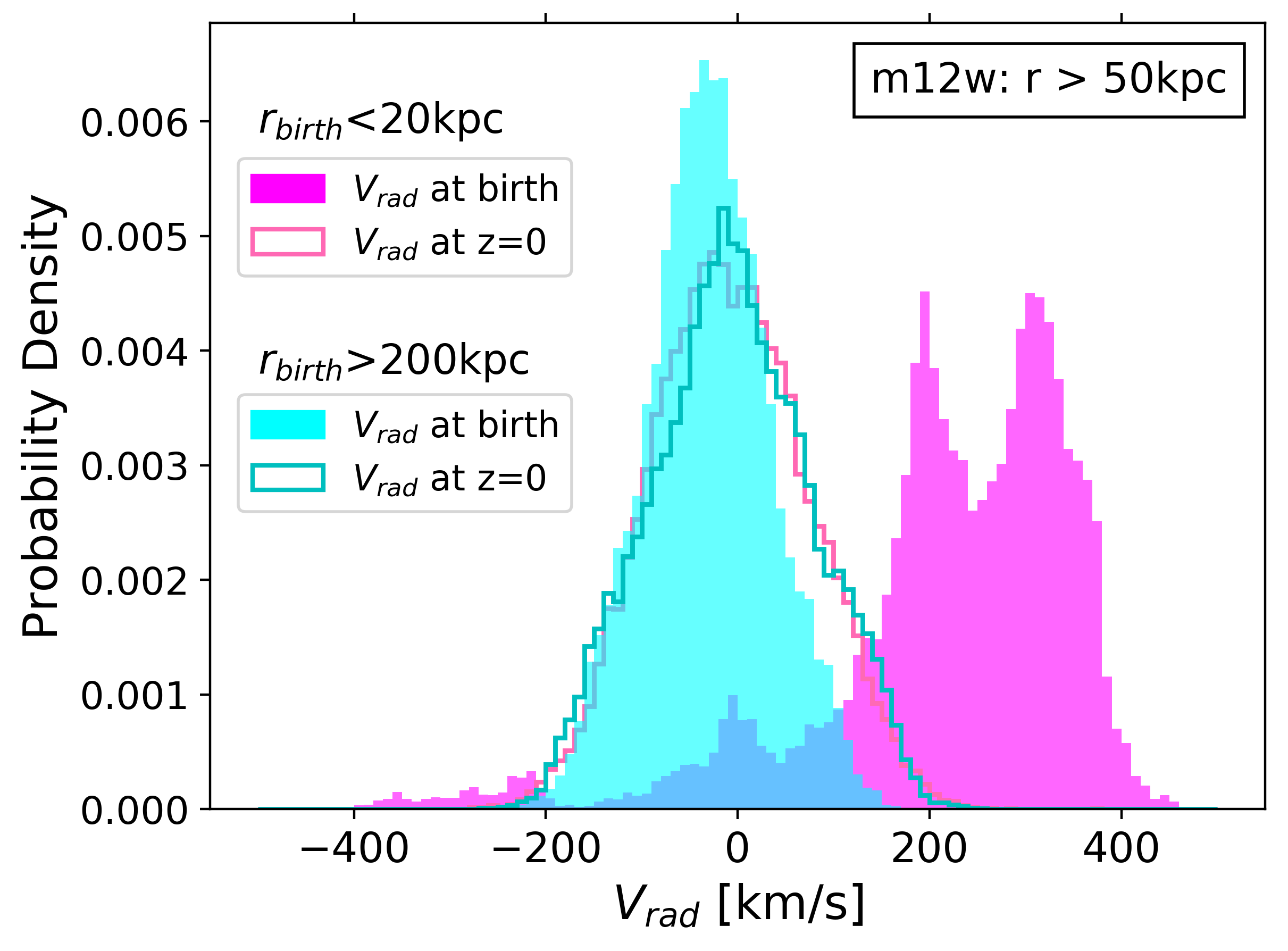}
	\centering
	\caption[Radial velocity at birth]{Radial velocities at birth (shaded) and at the present time (lines) of $z=0$ for outer stellar halo stars (with current radii $r > 50$ kpc) in  \texttt{m12i} (left)  and \texttt{m12w} (right). The shaded magenta histograms show birth velocities of the subset of halo stars that were born within the central galaxy ($r_{\rm birth} < 20$ kpc) while the shaded cyan histograms show birth radial velocities for accreted halo stars ($r_{\rm birth} > 200$ kpc).  We see that outer halo stars that were born within the central galaxy tend to have been created with large positive radial velocities, which means that they were born outflowing.  The accreted halo stars have a bias to have been born with negative radial velocity, consistent with an infalling population.     The radial velocities of these two groups are quite similar at $z=0$ (open histograms).  While the accreted and \insitu populations have different origins, their kinematic properties today do not readily reveal an observable difference.}
	\label{fig:Vbirth}
\end{figure*}

\section{Stellar Birth Velocities and Radii} \label{s:def}

In order to characterize the origin of stellar halo material, we use methods described in \citet{Sanderson2017} to trace all star particles back to their formation time.  The birth radius $r_{\rm birth}$ is defined as the distance of a star particle from the center of the main progenitor host galaxy at the time of their formation.\footnote{Note that \citet{Sanderson2017} were motivated to distinguish accreted stars from stars born within the galaxy (they used $r_{\rm birth} < 30$ kpc).  For this reason they used the term ``\insitu" to include all stars that were formed within or near the galaxy -- this is different than our dynamical definition of \insitu because it includes both stars that are ``kicked-out" by mergers and those that we identify as forming in outflows.}  
We also measure the radial velocity at birth $V_{\rm rad}^{\rm birth}$ in order to roughly characterize the initial trajectory of stars at formation.  We begin by exploring the content of two of our halos (\texttt{m12i} and \texttt{m12w}).

Figure \ref{fig:birth_vs_current} shows the birth radii ($r_{\rm birth}$) and current ($z=0$) radii ($r$) for star particles in the vicinity of two of our simulations \texttt{m12i} (left) and \texttt{m12w} (right). We have removed all stars that are bound to satellite galaxies in the vicinity of the host, so only main-galaxy stars and halo stars are shown here.  The particles are color coded by their ages, as indicated by the color bar, with the youngest stars in blue/purple and the oldest stars in yellow.  The dashed black line in each panel traces $r_{\rm birth} = r$. This representation provides insight into the origin of various populations of the stellar halo.

\begin{figure*}
	\includegraphics[width=0.49\textwidth , trim = 0.0 0.0 30.0 20.0]{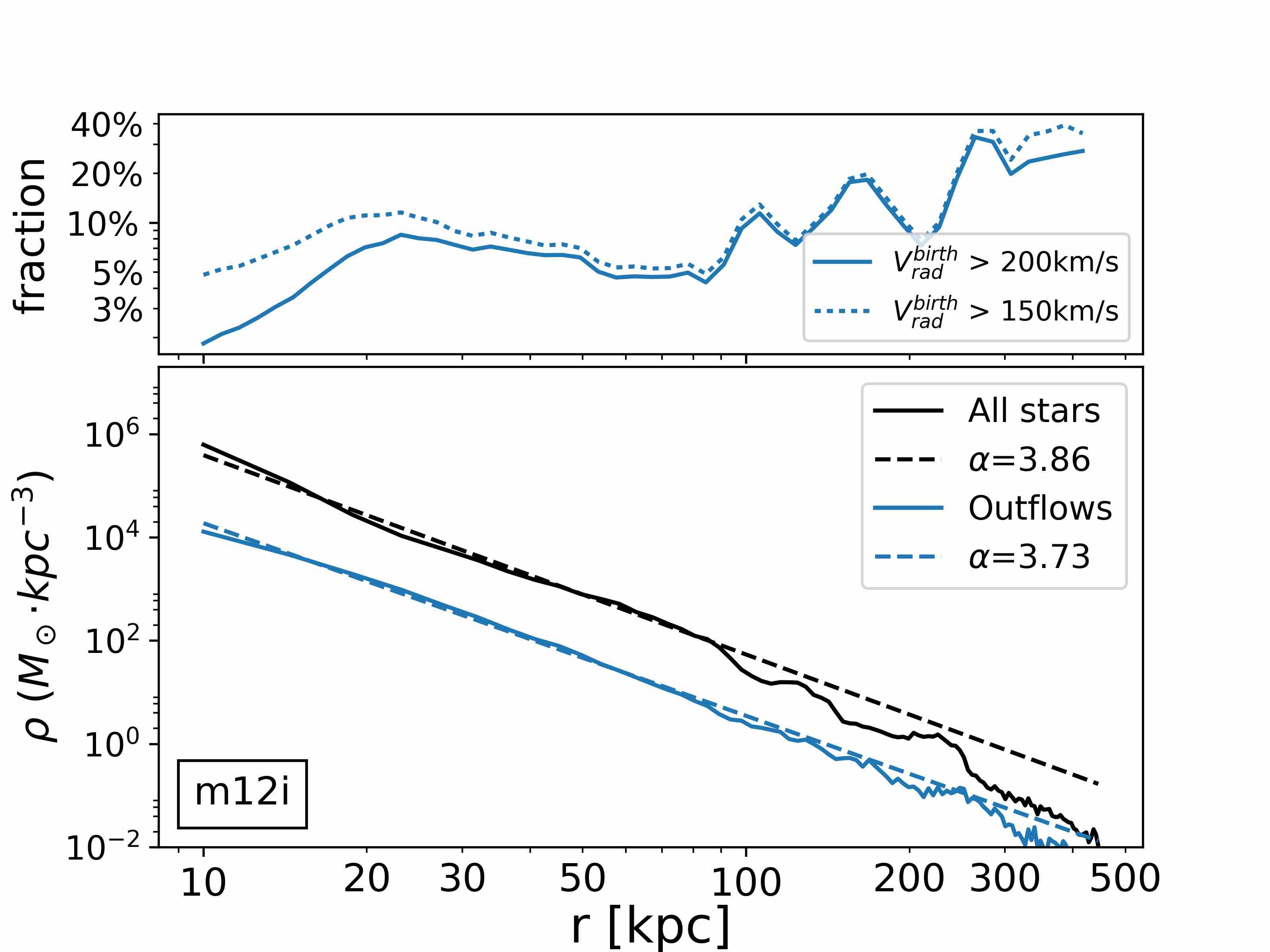}
	\includegraphics[width=0.49\textwidth , trim = 0.0 0.0 30.0 20.0]{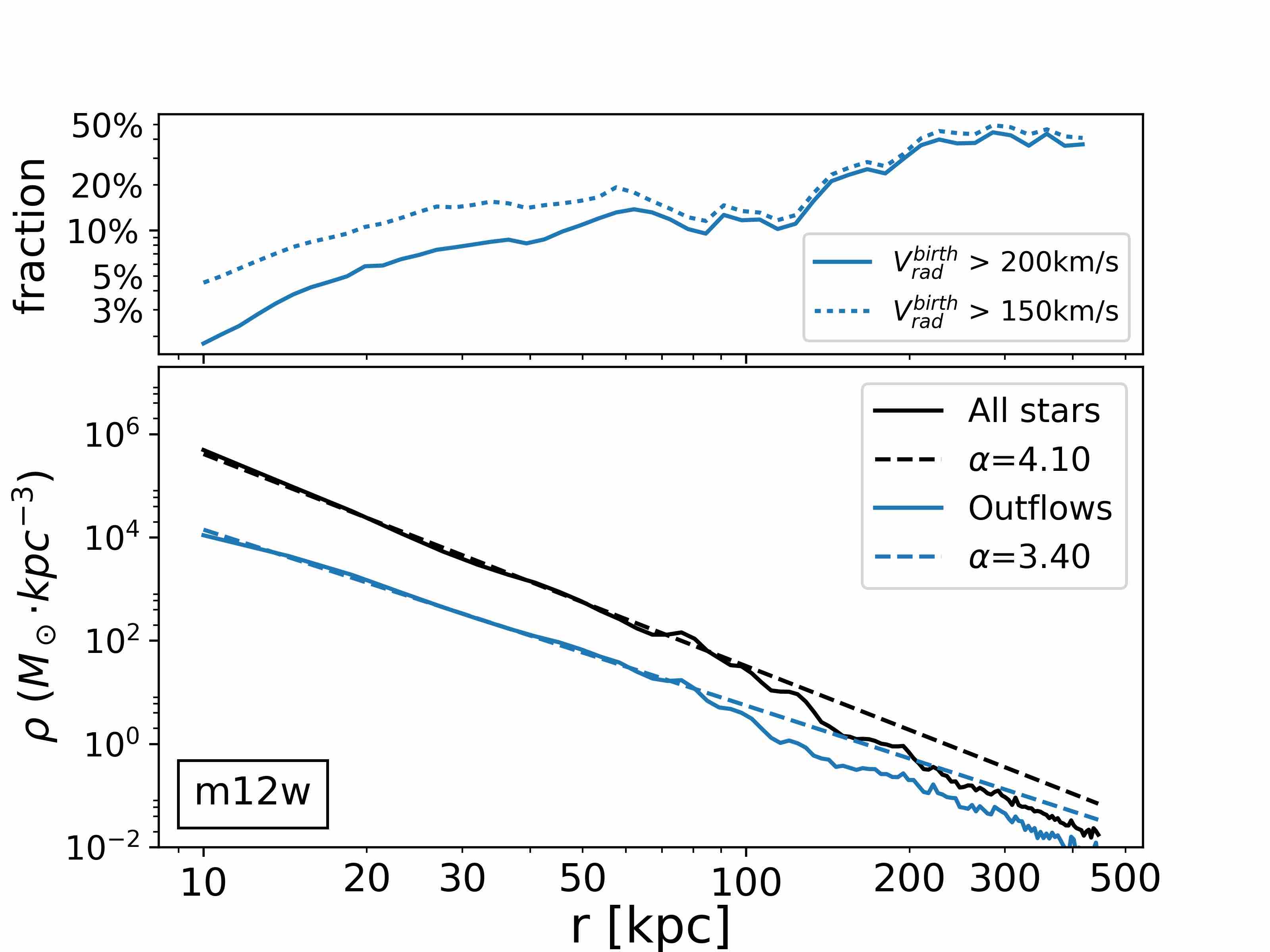}
	\centering
	\caption[Density Profiles]{Stellar density profiles for all stars (black) and for stars born in outflows (colored, defined as $V_{\rm rad}^{\rm birth} > 200 \, \kms$ and $r_{\rm birth} < 20$ kpc) as a function of radius for \texttt{m12i} (left) and \texttt{m12w} (right).  The dashed lines illustrate power-law fits ($\rho \propto r^{- \alpha}$) done for $r = 10 - 85$ kpc in each component and then extrapolated.  The top sections of each panel show the fraction of stars born in outflows, $\rho_{\rm outflow}(r)/\rho_{\rm total}(r)$, as a function of present radius $r$ for two ways of identifying outflow: $V_{\rm rad}^{\rm birth} > 200 \, \kms$ (solid) and  $> 150 \, \kms$ (dotted), both with $r_{\rm birth} < 20$ kpc.   The  fraction of stars born in outflows can become quite substantial at large radius, reaching $\sim 25-50 \%$ of all stars at $r > 250$ kpc, while at $50$ kpc the fraction is fairly modest $\sim 5-10 \%$. }
	\label{fig:m12i_fraction_density_fit_profile}
\end{figure*}

\begin{figure*}
	\includegraphics[width=\textwidth , trim = 0.0 15.0 0.0 0.0]{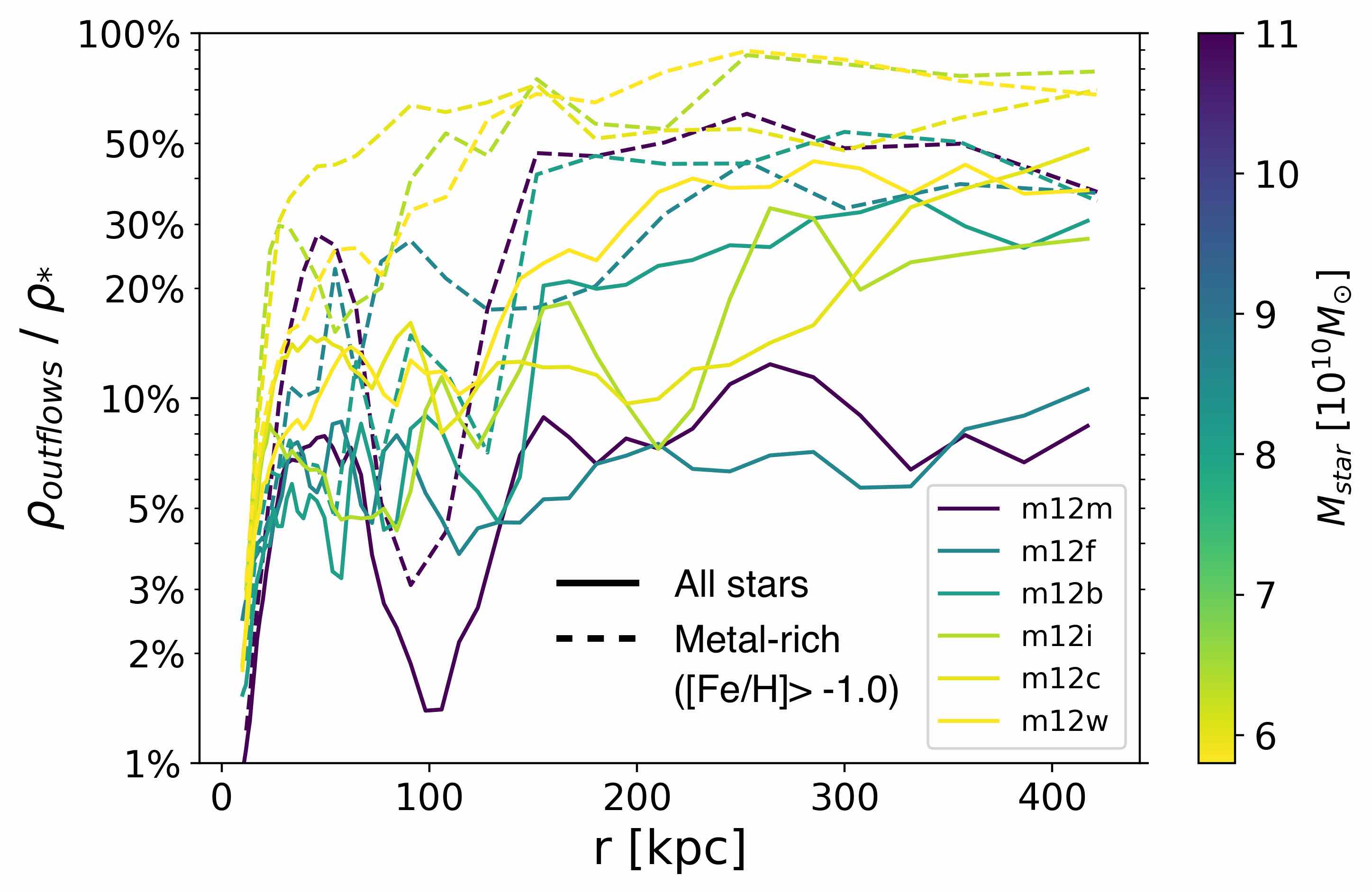}
	\caption[]{The fraction of stars born in outflows as a function of present radius for all six simulations color coded by central galaxy stellar mass. Outflow stars are defined to be those with $V_{\rm rad}^{\rm birth} > 200 \, \kms$ and $r_{\rm birth} < 20$ kpc.  
	The solid lines show the result for all stars while the dashed lines only calculate the ratio the density profiles of the metal-rich ([Fe/H] > -1.0) stars born in outflows to the density profile of all metal-rich. The fraction of metal-rich stars born in outflows can reach $>50 \%$ at $r > 150$ kpc for many systems.}
	\label{fig:all_frac}
\end{figure*}

One clear population visible in each panel of Figure \ref{fig:birth_vs_current} consists of central galaxy (disk) stars at $r \lesssim 20$ kpc, which scatters about the $r = r_{\rm birth}$ line.  Stellar migration and heating processes have driven stars to exist at radii that are either smaller (to the left of the line) or larger (to the right of the line) than their birth radii.

A second population is that of accreted stars (most visible at $r_{\rm birth} \gtrsim 30$ kpc), which are evident as horizontal bands that extend mainly to the left of the one-to-one lines.  The leftward extension means that they were born at a much greater galactocentric distance from the radius where they currently reside.  Stars in this population were formed in  smaller galaxies that were accreted and incorporated into the stellar halo of the main host.   Note that in some cases these horizontal bands have less-prominent extensions to the right of the one-to-one line (with $r>r_{\rm birth}$).  These are stars flung out to large radii after their birth galaxies are destroyed.  In a few places we see vertical bands (e.g. at $r \simeq 100$ kpc and $200$ kpc in \texttt{m12i} and at $r \simeq 80$ kpc in \texttt{m12w}).  We have studied these bands and find that they are ``shells'' left over from galaxy mergers that pile up at orbital apocenter. 

Most relevant to this work is a third population of  halo stars that can be seen as distinct horizontal bands with characteristic birth radii within the main galaxy ( $r_{\rm birth} \lesssim 20$ kpc).   Unlike the accreted stars, these bands (labeled ``\insitu") exist primarily to the right of the one-to-one line ($r > r_{\rm birth}$).  This population is intriguing because it consists of star particles that are now at much greater galactocentric distance relative to their birth places; the banding at near-constant\footnote{Some of these bands show a slight slope towards larger $r_{\rm birth}$ at increasing $z=0$ radius.  As we discuss below, these \insitu stars are born in dense, shell-like outflows.  The stars that form last have emerged from gas that has had more time to be accelerated and this pushes them towards slightly larger birth radii and more extended orbits today.} $r_{\rm birth}$ suggest multiple populations with unique episodes of origin.   Note that there are stars that were born within $20$ kpc that now populate distances out to $ r \sim 500$ kpc.

Figure \ref{fig:Vbirth} demonstrates that most of these \insitu stars that now exist at large radii were indeed born with large positive (outward) radial velocities $\gtrsim 150 \, \kms$, indicative of outflow.  The shaded magenta histograms show $V_{\rm rad}^{\rm birth}$  for outer halo stars with $r > 50$ kpc that were born within the central galaxy region $r_{\rm birth} < 20$ kpc for both \texttt{m12i} (left) and \texttt{m12w} (right). Compare these distributions to the shaded cyan histograms, which show birth radial velocities for stars at the same current radius ($r > 50$ kpc) that were born at large radii $r > 200$ kpc.  These stars tend to be slightly infalling at birth, which is consistent with an accreted population.  Interestingly, these two populations, which had significantly different kinematic properties at formation, today have radial velocity distributions that are very similar (open histograms).   This would be expected for stars that have similar apocenters orbiting in the same potential for several dynamical times. Section \ref{sec:SFR} provides an exploration of the physical origin of these stellar outflows and demonstrates that they tend to occur during bursty star formation episodes, specifically in conjunction with super-bubble winds that accelerate dense molecular gas that goes unstable to star formation after being accelerated outward.

The definition of "outflow" is inherently subjective.  
For the sake of concreteness, in what follows we will identify stars that formed in outflows to be represented by star particles that originated within the central galactic region,  $r_{\rm birth} < 20$ kpc, and that had large, positive velocities at the time they were created: $V_{\rm rad}^{\rm birth} > V_{\rm O}$.  For most of our analysis will use $V_{\rm O} = 200 \, \kms$ as a conservative choice based on the results shown in Figure \ref{fig:Vbirth} and similar analyses done for our other four halos.  As we discuss in Section \ref{sec:local}, more than $95\%$ of stars that exist in the central disk at $z=0$ were born with velocities  $V_{\rm rad}^{\rm birth} < 150 \, \kms$.  Based on this we also explore how our results change if we adopt $V_{\rm O} = 150 \kms$.  Our qualitative conclusions do not change with this choice, though the fraction of halo stars in the inner halo identified as originating in outflows does increase as $V_{\rm O}$ decreases.

We show below that our adopted strategy for identifying stars born in outflows -- stars with small birth radius and large positive birth velocity --  seems to select stars that are born within dense gas that has been accelerated in feedback-driven winds from the main galaxy.  In principle, however, the selection does not preclude other possibilities.  For example, one could imagine stars born within a gas-rich satellite that experiences a starburst episode just after pericenter crossing.  On rare occasions, such stars could meet our selection criteria and then become liberated into the stellar halo after the satellite is tidally disrupted.  We have looked for instances of these events and find them to be rare.  As discussed in subsequent sections, the outflow stars we identify tend to be metal-rich (like the central galaxy), with higher metallicity than the vast majority of satellite galaxies, rendering this possibility statistically unlikely.  
 
With  this definition of outflow stars in hand, we can ask how they populate the stellar halos of our galaxies.  Figure \ref{fig:m12i_fraction_density_fit_profile} shows that they contribute significantly, especially at large radius. The solid black lines in each lower panel present the differential density profiles of all stars in \texttt{m12i} (left) and \texttt{m12w} (right). Stars bound to satellite galaxies have been removed. For reference, the black dashed lines are best-fit power-law slopes ($\rho \propto r^{-\alpha}$) that fit the profile of all the halo stars in each galaxy within $r = 10 - 85$ kpc.  
The solid blue lines show the density profiles of stars born in outflows (using $V_{\rm O} = 200 \kms$), with corresponding power-law fits within $10 - 85$ kpc shown as dashed lines. We see that the outflow profiles tend to be flatter than the total stars, such that the fraction of all stars born in stellar outflows tends to rise with galacto-centric radius. The upper panels show the fraction of stars identified as being born in outflows as the ratio of the two density profiles (solid blue).  We see that outflows contribute a substantial fraction to all the halo stars at large radius in these two halos ($\sim 10-50\%$ beyond $50$ kpc).  The dotted lines show the fraction for halo stars identified with $V_{\rm O} = 150 \kms$.  With this selection, the fraction identified as outflow is slightly higher within $50$ kpc but remains similar at larger radius. 

The solid lines in Figure \ref{fig:all_frac} show the fraction of stars born in outflows for all six halos, defined in the same way as the solid lines in the upper panels of Figure \ref{fig:m12i_fraction_density_fit_profile}.   The systems are colored from lowest to highest in the stellar mass of the central galaxy, starting with \texttt{m12w}, which has $M_{\star} = 5.6 \times 10^{10} \msun$ and proceeding to \texttt{m12m} with $M_\star = 1.1 \times 10^{11} \msun$. Four of the six have $\gtrsim 20 \%$ of the stars in their outer (> 250 kpc) stellar halos composed of stars that were formed in radial outflows from the main galaxy.  The two lowest outflow fractions happen to correspond to the two most massive galaxies \texttt{m12m} and \texttt{m12f}, but given the small sample size is it not possible to say if this is representative of a mass trend or simply halo-to-halo scatter. Both of these galaxies experience late-time mergers that help populate the outer halo with accreted stars.  This may be the reason why the outflow fraction is small -- it is that the accreted mass is larger than average, rather than the outflow mass being smaller than average.  The dip in outflow fraction at $\sim 100$ kpc in \texttt{m12m} is associated with a bump in the total stellar profile at that radius that comes from a remnant of a fairly large, recently-accreted dwarf galaxy that has deposited stars in that area.

\begin{figure*}
\includegraphics[width=0.244\textwidth]{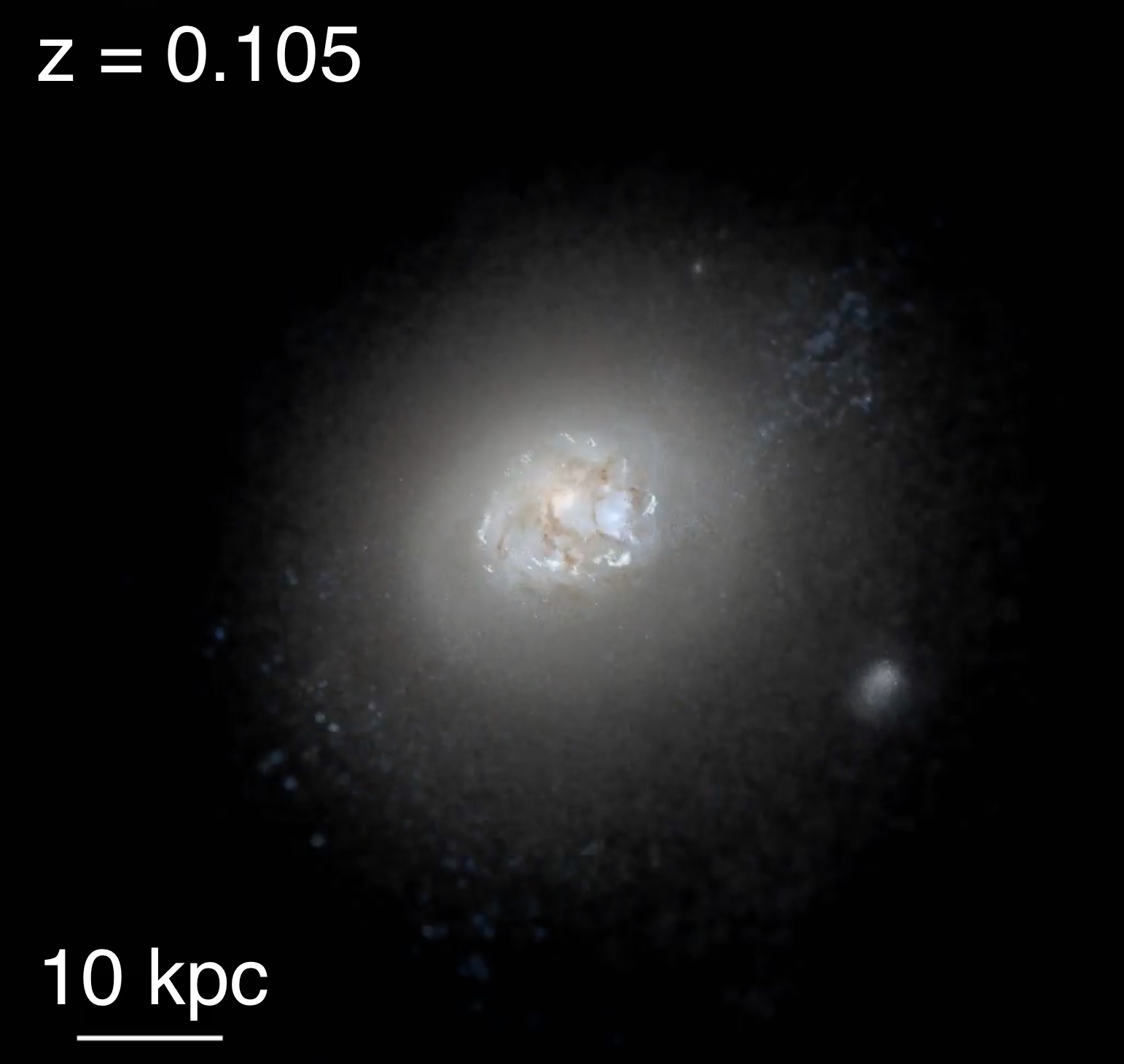}
\includegraphics[width=0.244\textwidth]{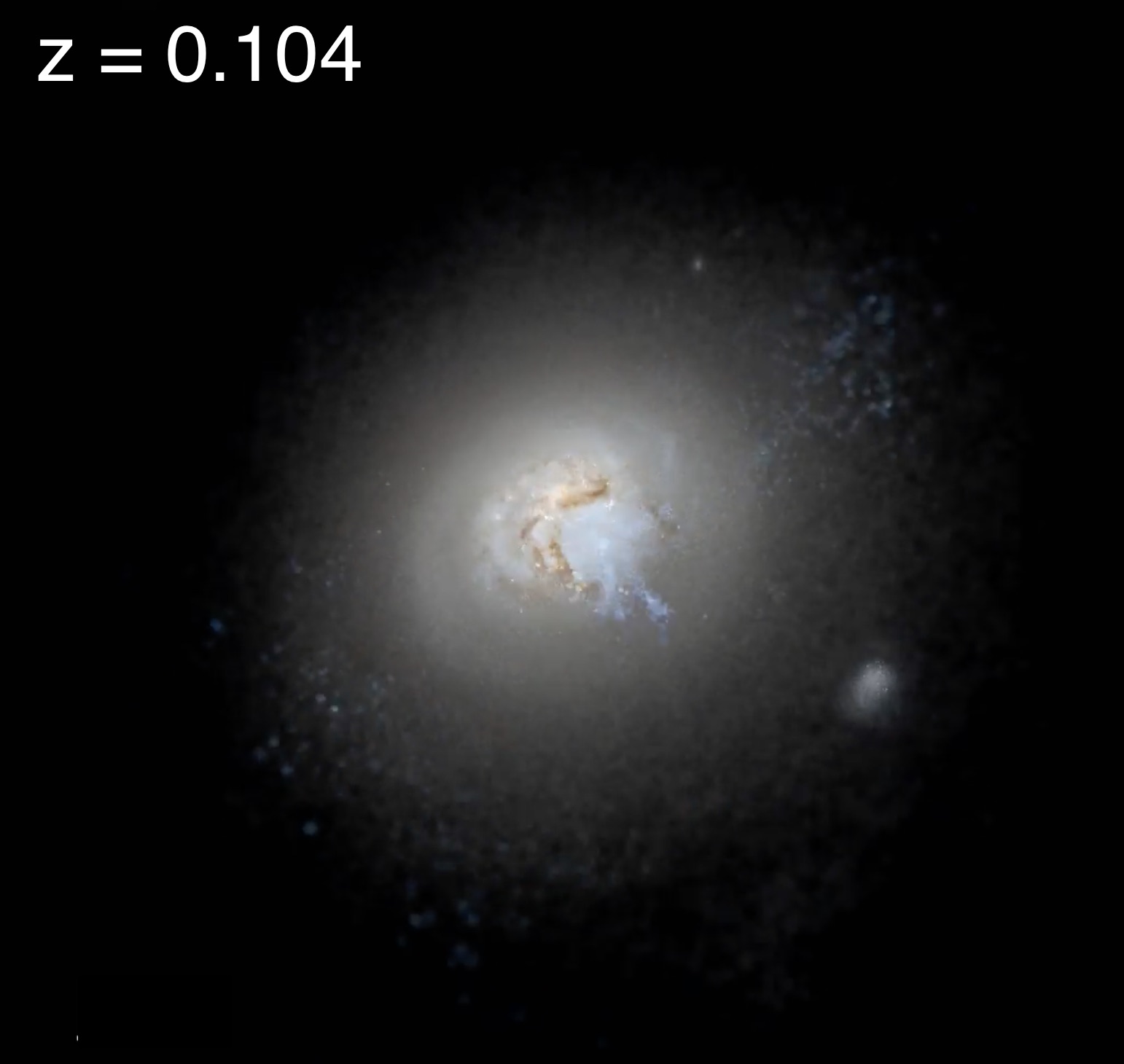}
\includegraphics[width=0.244\textwidth]{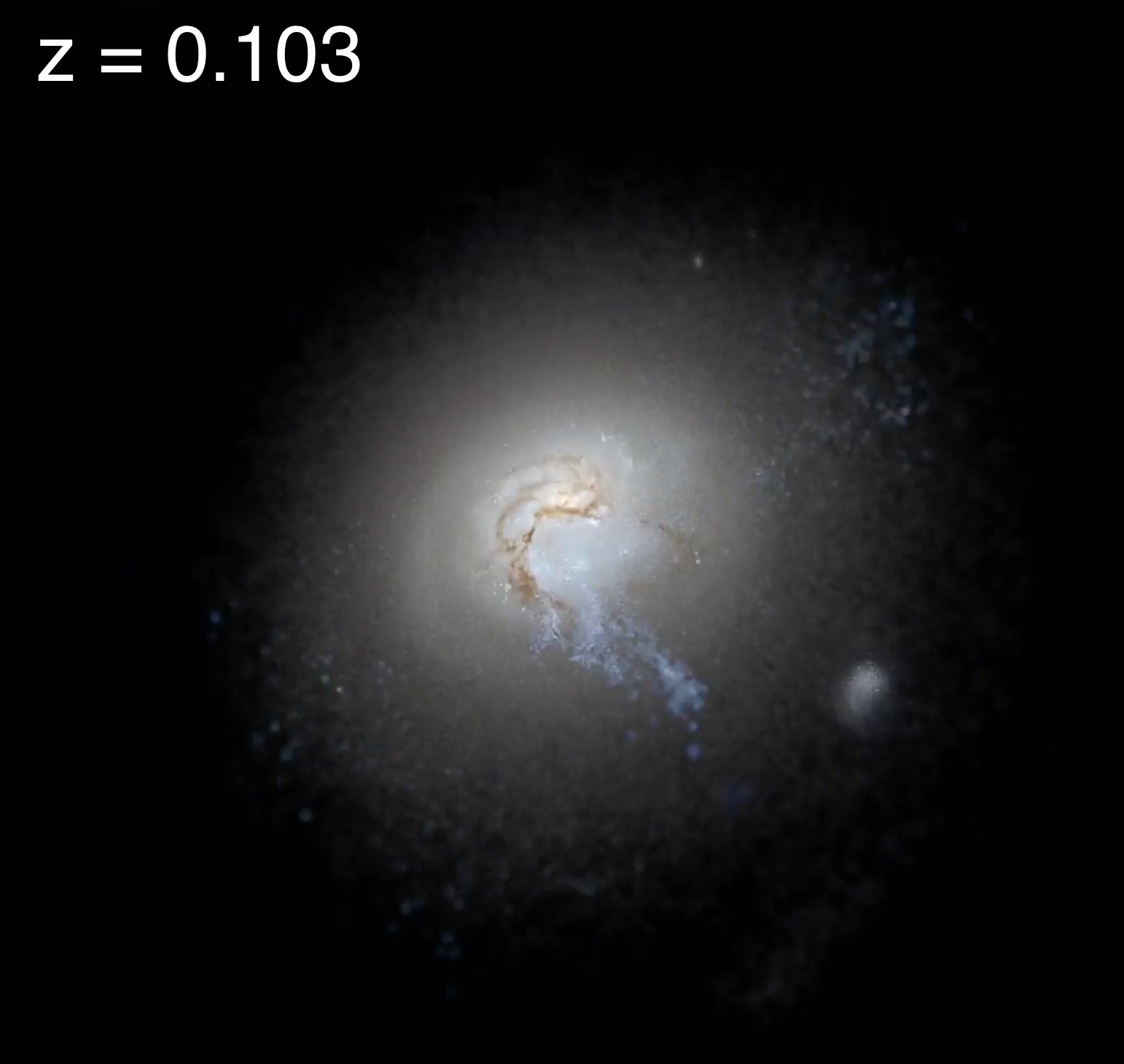}
\includegraphics[width=0.244\textwidth]{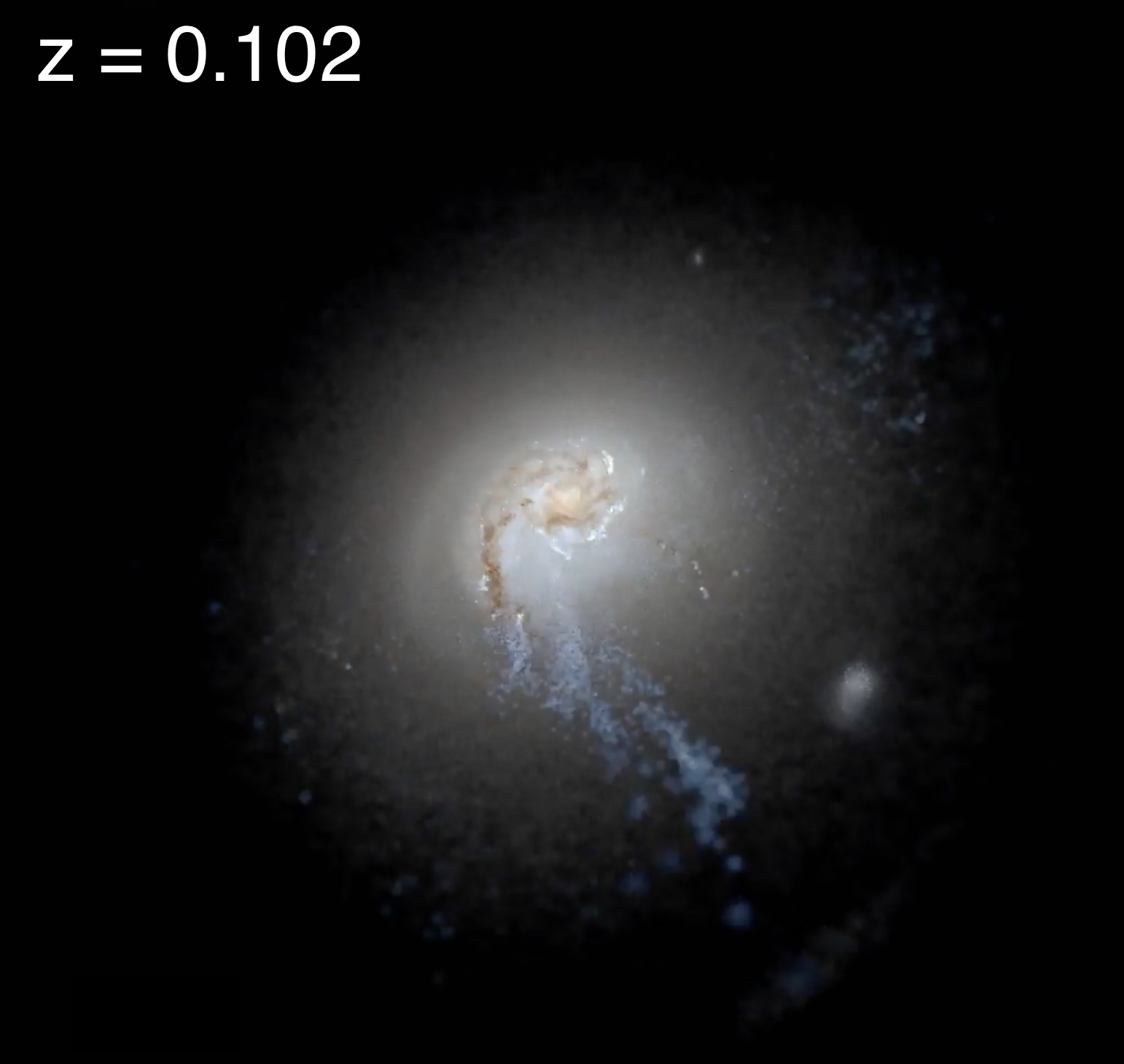}
	\caption[Outflow image]{Star-forming outflows developing in halo \texttt{m12w} over a $\sim 40$ Myr period around redshift $z \simeq 0.1$  (from left). These are mock {\it Hubble Space Telescope} u/g/r composite images created as described in \citet{Hopkins17} and span $80$ kpc on a side. We see several prominent plumes of young blue stars that were born in gas that was originally rotating and then blown radially outward just prior to star formation. The bulk of these stars are bound to the main halo and should eventually fall back on radial orbits indicative of halo stars.}
	\label{fig:m12w_image}
\end{figure*}

\begin{figure*}
\includegraphics[width=\textwidth]{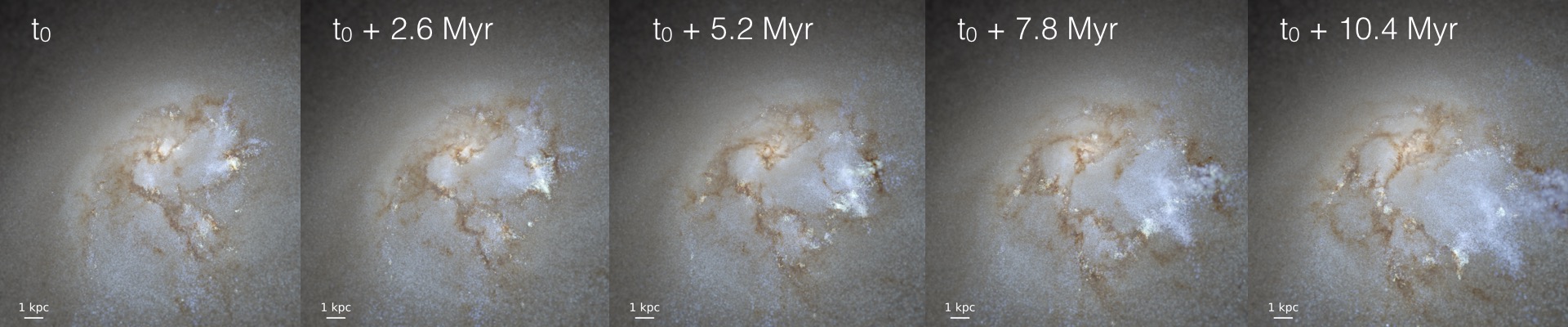}
\includegraphics[width=\textwidth]{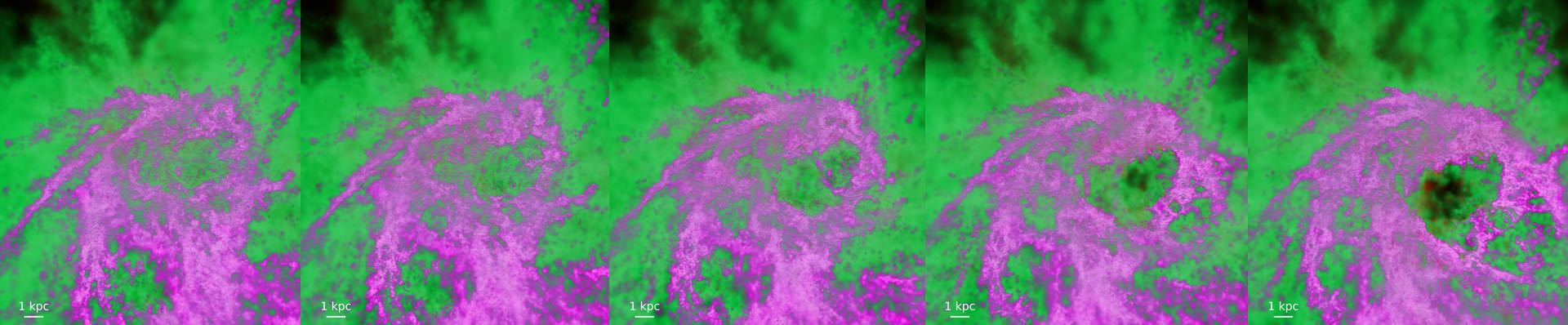}
	\caption[Star/gas outflow image]{An outflow event in stars (upper) and gas (lower) developing in halo \texttt{m12w} over a 10.4 Myr time period.  The left most panel starts at time $t_0$ (1,342.4 Myr before $z=0$) and progresses forward in time in steps of $2.6$ Myr from left to right.  Each image is 20 kpc square.  The upper panels are u/g/r composites as in Figure \ref{fig:m12w_image}. In the lower panel, magenta is cold molecular/atomic gas ($<1000$ K) and green is warm ionized gas ($10^4-10^5$ K). We see that stellar outflows develop along with cold-gas outflows, as stars form from compressed gas that has been accelerated outward at the edges of evacuated super-bubbles.}
	\label{fig:m12w_visual}
\end{figure*}

\begin{figure*}
	\includegraphics[width=\textwidth , trim = 30 0.0 0.0 0.0]{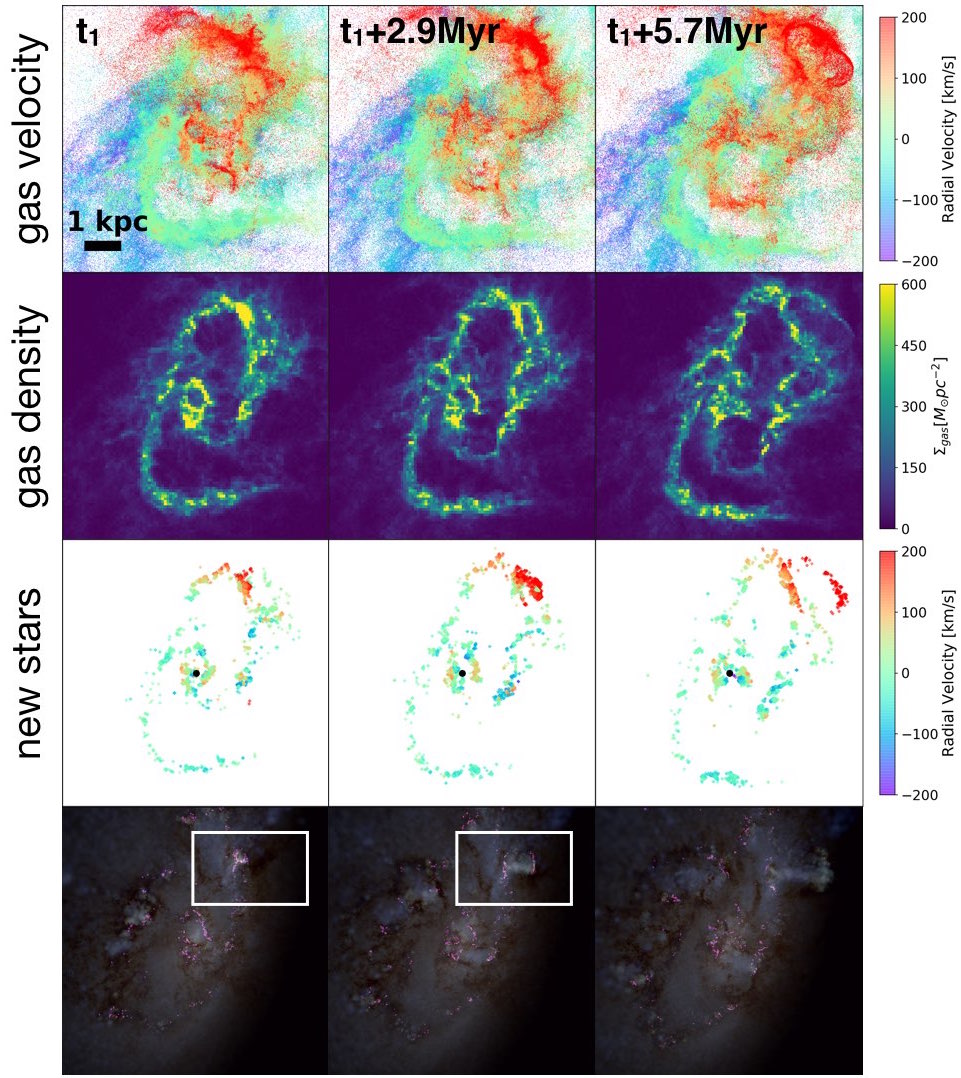}
	\caption[]{Development of a stellar outflow in \texttt{m12w} starting starts at time $t_1$ (1,343.5 Myr before $z=0$) and progressing in $\sim 2.9$ Myr timesteps from left to right.  The top row shows gas particles within a 10 kpc region of the galaxy color coded by radial velocity, with red indicative of fast radial outflow.  The second row shows gas surface density and the third row shows new stars (formed within the previous 2.9 Myr) color coded by radial velocity. The red regions in the third row are young stars that are outflowing, which overlap spatially with high-density gas at the edges of expanding superbubbles. The lowest set of images show color composite similar to those shown in Figure \ref{fig:m12w_image}, now with pink overlaid to trace star formation rate density.  A clear pair of stellar outflow shells is being generated in the white-boxed regions (enlarged in Figure \ref{fig:gas_visual_step}) in the upper left portion of the galaxy. }
	\label{fig:gas_step}
\end{figure*}

The dashed lines in in Figure \ref{fig:all_frac} show the fraction of metal rich ([Fe/H] > -1.0) halo stars born in outflows as a function of radius. We see that among the most metal rich stars in the outer stellar halo, outflows typically dominate.  We discuss the origin of this difference in Sec \ref{sec:outer_chem} and Sec \ref{sec:sum_solar_neighbor}.

 Table~\ref{tab:info} provides more detailed information on the stellar halos of each of our galaxies.  
In the last six columns we list the total stellar mass and fraction of that mass in outflow stars outside of various radial cuts ($r>20$, 50, and 150 kpc) .  
Note that the total mass of outflow stars outside of 20 kpc ranges from $\sim 10^8-10^9$ \Msun and  that this makes up $\sim 5-10\%$ of all stars in the stellar halo beyond 20 kpc.
 
\begin{figure*}
	\includegraphics[width=\textwidth , trim = 80 0.0 0.0 0.0]{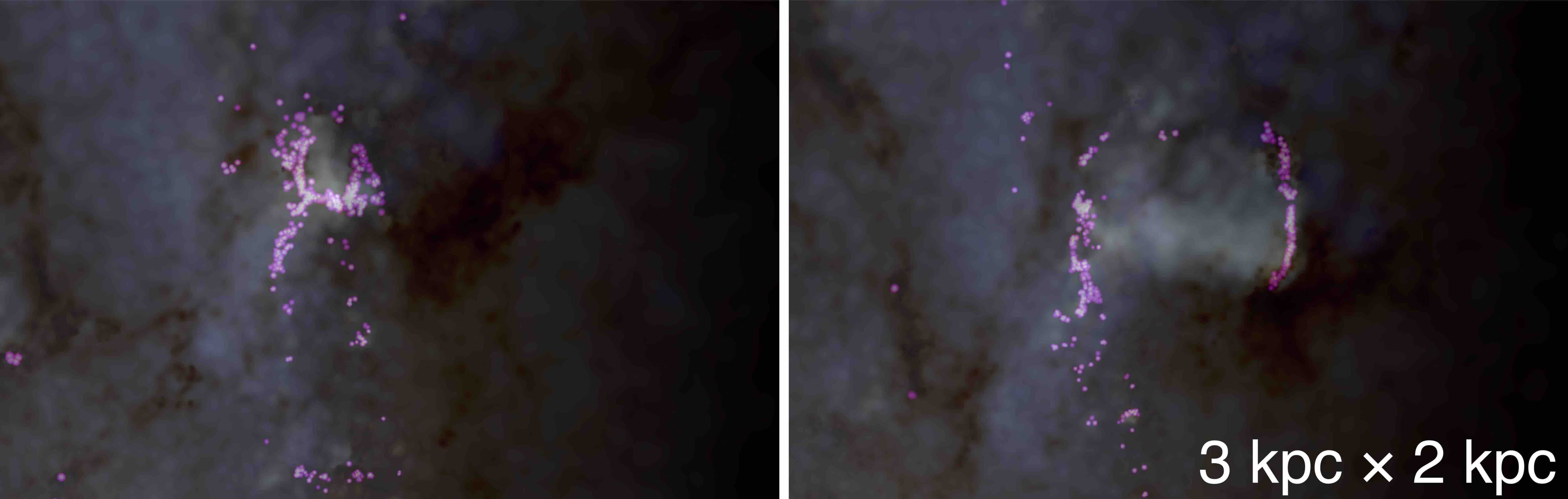}
	\caption[]{Zoomed-in  mock {\it Hubble Space Telescope} u/g/r composite images of the first-two panels in Figure \ref{fig:gas_step}, illustrating an expanding bubble of triggered star formation (pink). Each image spans $\sim 3$ kpc in width and $\sim 2$ kpc in height.}
	\label{fig:gas_visual_step}
\end{figure*}

\section{Origin of stellar outflows} \label{s:origin}

In this section we discuss the identification of outflow stars in our simulations and provide a brief exploration of their origin.

As seen in Figure \ref{fig:birth_vs_current}, the \insitu component of outer ($r > 50$ kpc) halo stars in \texttt{m12i} and \texttt{m12w} show multiple populations with narrow horizontal bands of roughly discrete birth radii. This is consistent with distinct outflow events and would not be expected for stars that were kicked out in merger events or heating.  Figure \ref{fig:Vbirth} shows that these stars are born with large positive radial velocities, which is again consistent with radial outflow and not stars that were kicked out after formation.   By examining the formation histories of our galaxies, we find that  these stars tend to be born in shells of gas compressed and accelerated in discrete outflow events. We provide examples below.  These events are quite obvious in visualizations and movies.~\footnote{For movies and images of Milky Way and Andromeda - like systems in FIRE2 simulations, see: \url{http://www.tapir.caltech.edu/~phopkins/Site/animations/a-gallery-of-milky-way-/} }  

\subsection{Example Stellar Outflow Events}

Figure \ref{fig:m12w_image} presents mock u/g/r composite images of a stellar outflow developing over $40$ Myr in \texttt{m12w}.  Note that this image shows stars and dust only (not gas).  A stream of blue (young) stars extends outward towards the lower right corner of the image as time progresses from left to right.  These stars were born in dense gas that was rotating within the disk and subsequently accelerated in an outflow prior to star formation.   

Figure \ref{fig:m12w_visual} shows a zoomed-in (20 kpc) region around \texttt{m12w} as this prominent outflow develops over the first $10.4$ Myr of its evolution.  The upper panels show a false-color visualization and lower panels present cold (pink) and warm (green) gas at the same epochs.  Clustered supernovae events are driving bubbles of cold-gas outflows in the bottom panels.  Stellar outflows are apparent at the edges of these bubbles where compressed, radially-accelerated gas is converted into stars.

\begin{figure*}
	\centering
	\includegraphics[width=0.49\textwidth , trim = 0.0 0.0 0.0 0.0]{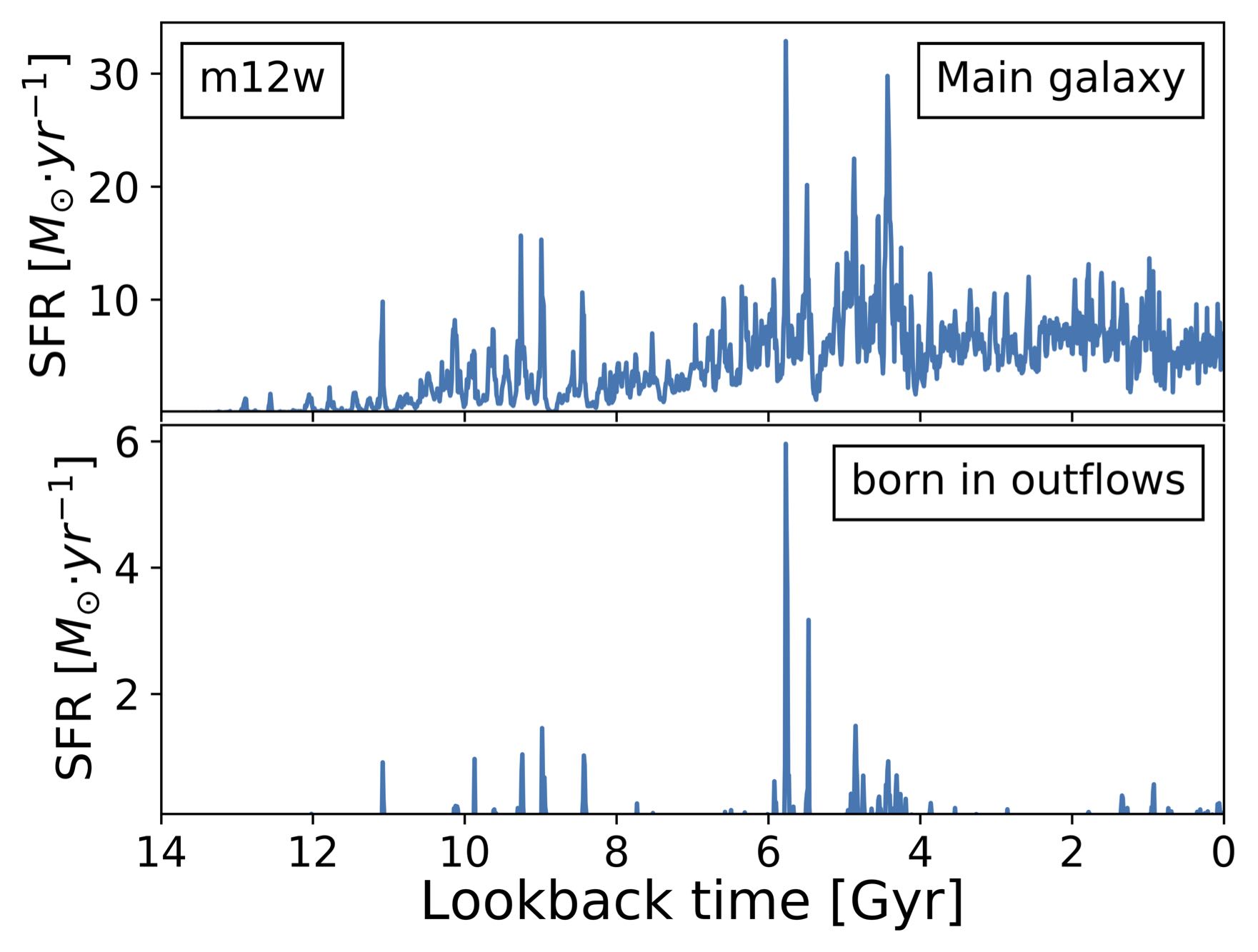}
	\includegraphics[width=0.49\textwidth , trim = 0.0 0.0 0.0 0.0]{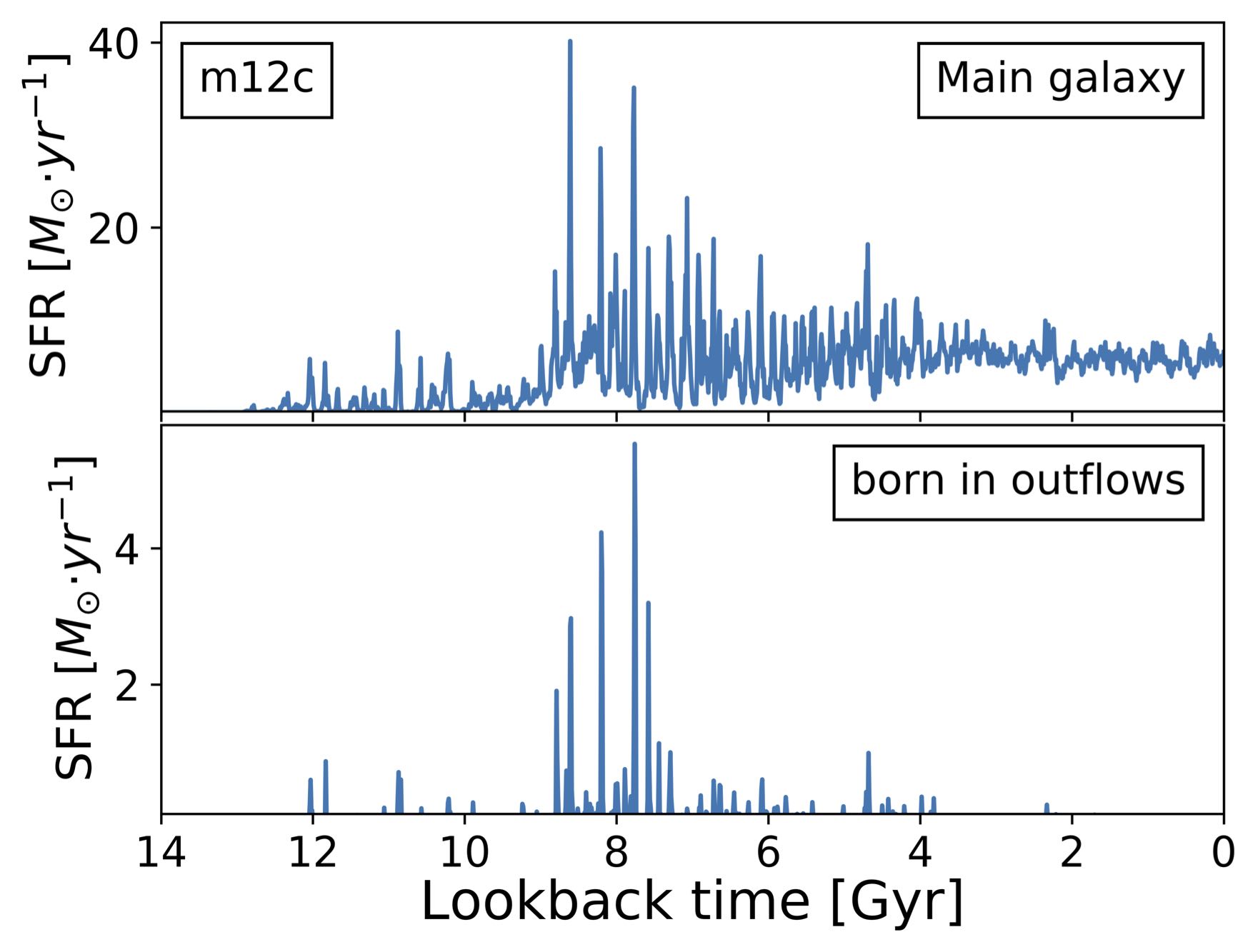}
	\includegraphics[width=0.49\textwidth , trim = 0.0 0.0 0.0 0.0]{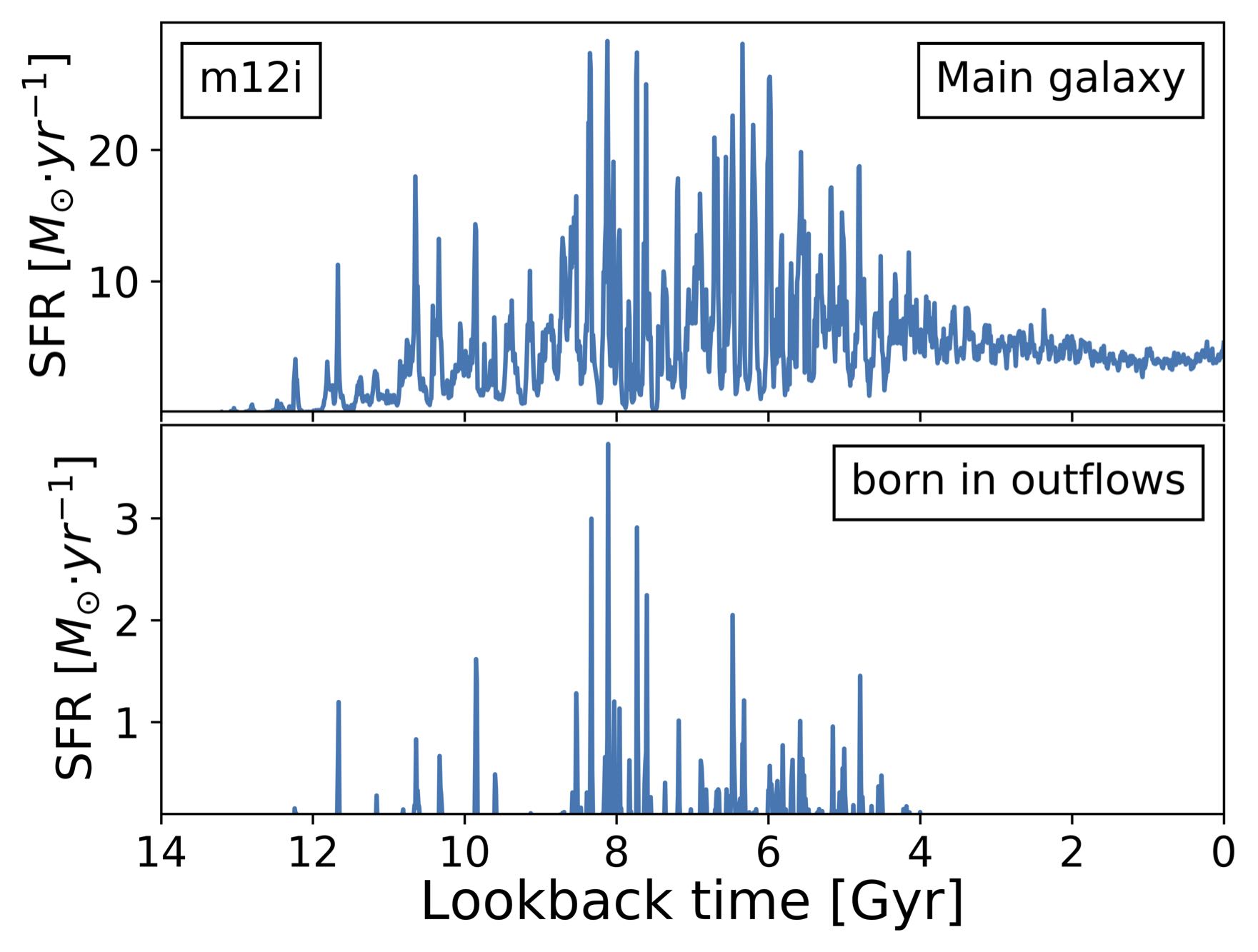}
	\includegraphics[width=0.49\textwidth , trim = 0.0 0.0 0.0 0.0]{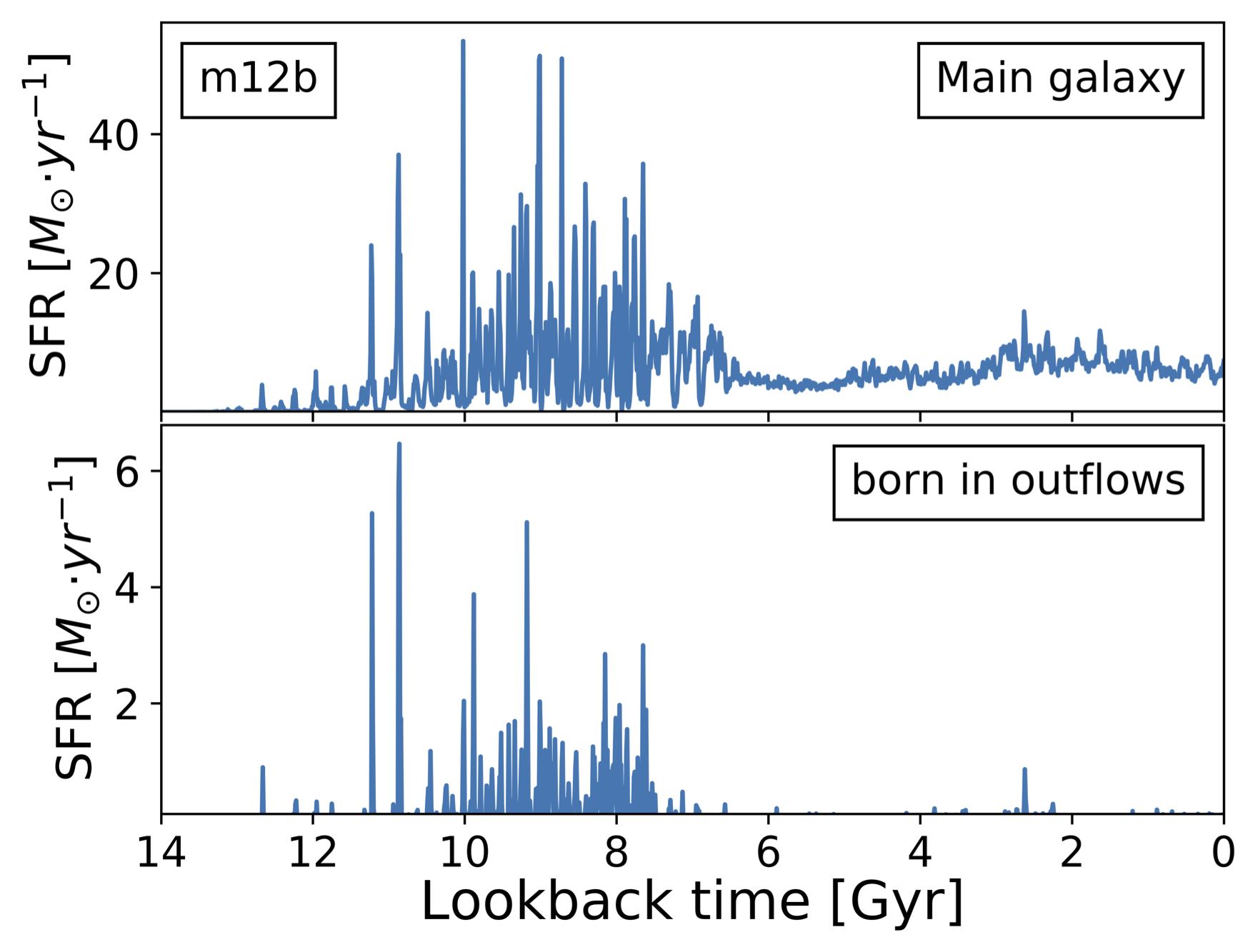}
	\includegraphics[width=0.49\textwidth , trim = 0.0 0.0 0.0 0.0]{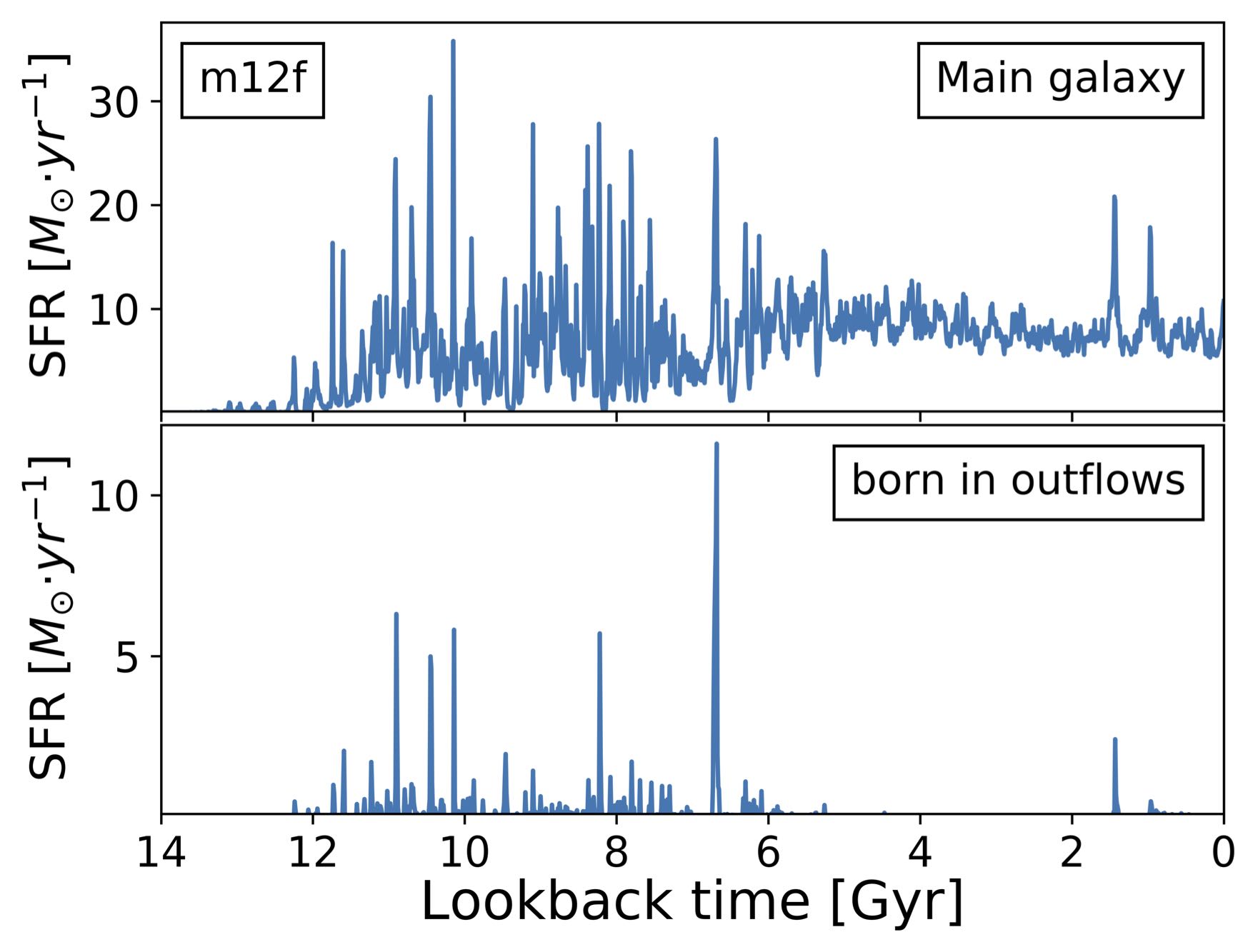}
	\includegraphics[width=0.49\textwidth , trim = 0.0 0.0 0.0 0.0]{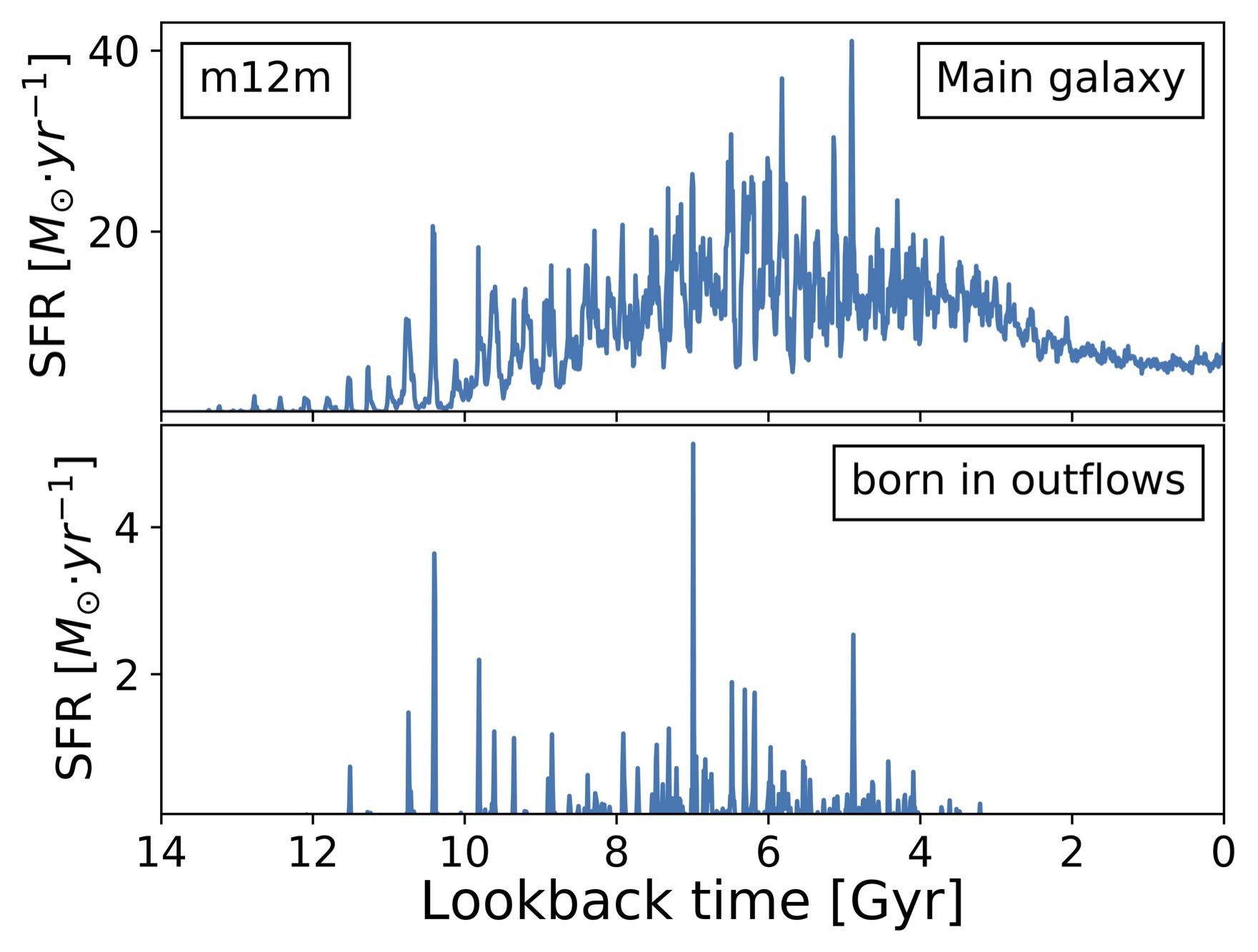}
	\caption[SFH]{Instantaneous star formation rates for the main galaxy (top panels) compared to stellar outflow star formation rate (bottom panels) for all six of our simulated galaxies, as indicated. Star formation rates are averaged over 10 Myr.  Outflows are defined as stars that were formed with $r_{\rm birth} < 20$ kpc and $V_{\rm rad}^{\rm birth} > 200 \, \kms$.   The galaxies are ordered from lowest to highest in total stellar mass from upper left to bottom right. Stellar outflows tend to correlate with bursty star formation and become rare at late time as star formation becomes more constant.   Overall, stars made in outflows account for only $\sim 1 \%$ of all stars formed in the galaxy, but during some starbursts the instantaneous outflow fraction can be as high as $\sim 20-30 \%$. }
	\label{fig:SFH}
\end{figure*}

Figure \ref{fig:gas_step} focuses on the beginning of same outflow event in  {\tt{m12w}}, one that occurs just slightly prior to the one in Figure \ref{fig:m12w_visual}.  The panels are time-ordered from left to right, beginning at a time roughly $1$ Myr earlier than the first panel of Figure \ref{fig:m12w_visual}.  The top row shows all gas particles within a $10$ by $10$ kpc region around the central galaxy, color coded by radial velocity, with red indicative of radial gas outflow according to the color bar.  The second row shows the surface density of that gas color coded according to the bar on the right.  The third row plots {\em new stars} that were formed between the snapshots shown (i.e. those formed in the previous $\sim 3$ Myr) color coded by radial velocity, such that the red points are indicative of young stellar outflows.   The dense gas that spawns these stars accelerates over $\sim 1$ Myr timescales just prior to star formation.  The stars then travel ballistically outward. The bottom row shows the galaxy as a mock u/g/r composite image with the star formation rate density overlaid in pink.     We see star formation (pink) work to both drive outflows and generate new outflowing stars at the edges of an expanding bubble.  Figure \ref{fig:gas_visual_step} zooms in on the first two panels of this image to show a particularly prominent stellar outflow shell developing.  Stars are clearly forming at the edge of a super bubble that is expanding at high radial velocity.

\begin{figure*}
	\includegraphics[width=0.95 \textwidth , trim = 200.0 60.0 100.0 72.0]{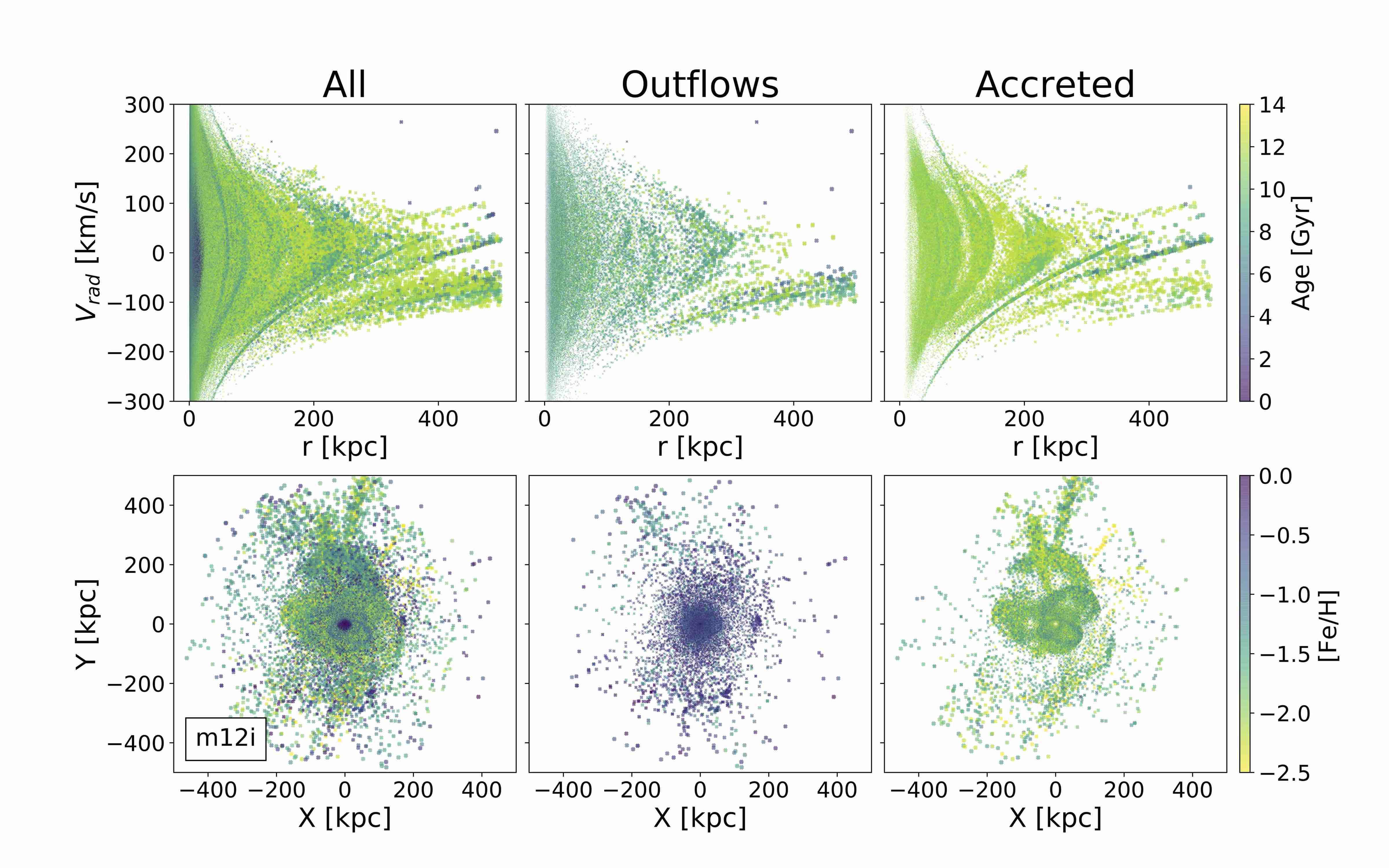}
	\includegraphics[width= 0.95 \textwidth , trim = 200.0 72.0 100.0 12.0]{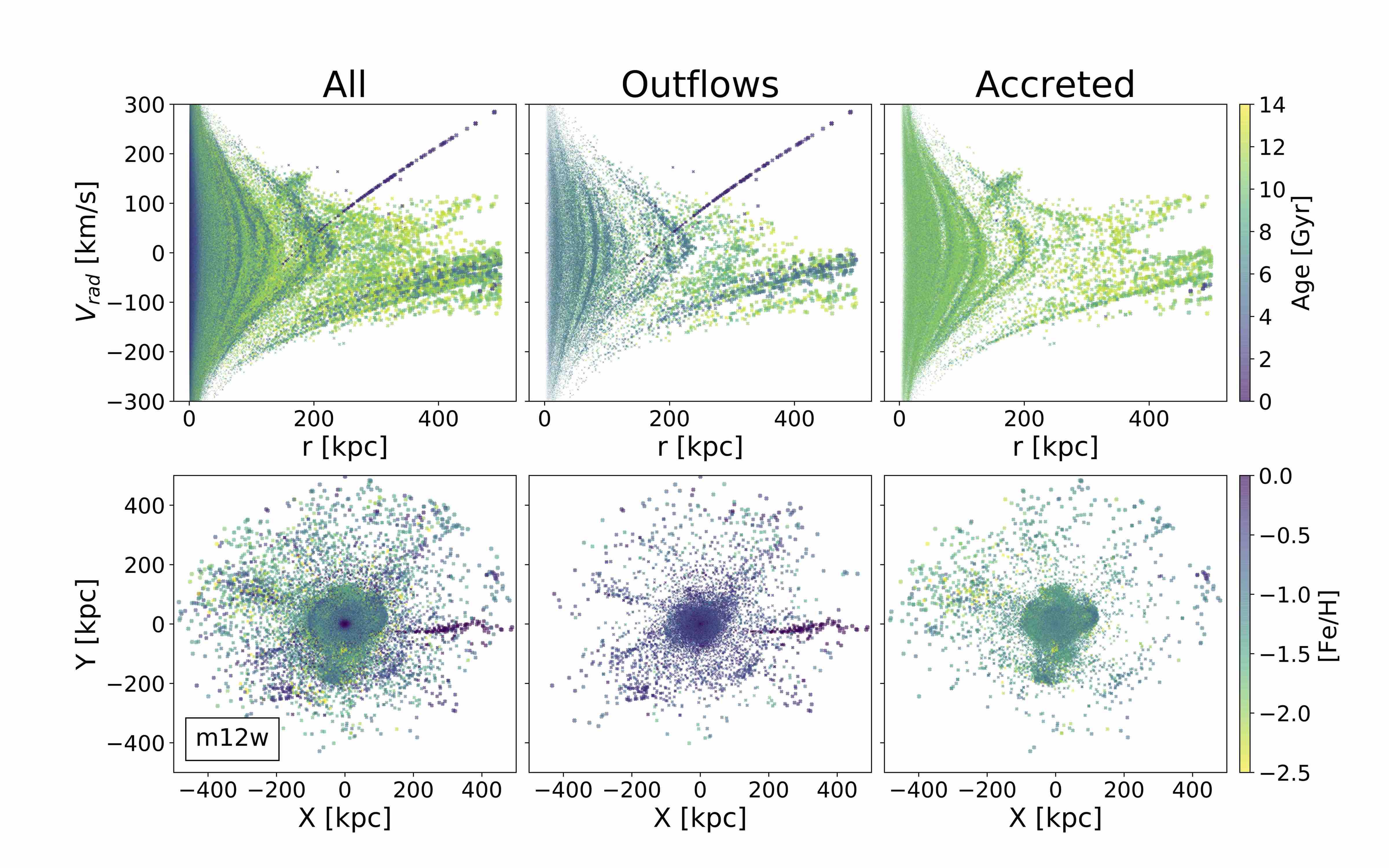}
	\centering
    \caption[]{ Phase space structure of \texttt{m12i} (top set) and \texttt{m12w} (bottom set).  Top rows depict stellar radial velocity vs. radius, color coded by age.   Bottom rows show the spatial distribution of star particles color coded by metallicity. Stars bound to satellite galaxies have been removed.  The left panels includes all star particles. The middle panels includes only outflow stars identified to have $r_{\rm birth} < 20$ kpc and $V_{\rm rad}^{\rm birth} > 200 \, \kms$.  The right panels include only stars with $\rbirth > 200$ kpc (accreted). The \insitu stars tend to be more metal-rich and more smoothly distributed in the stellar halo than accreted stars. A recent, metal-rich stellar outflow is visible in purple in the bottom set of \texttt{m12w}.}
	\label{fig:rvr_spatialiw}
\end{figure*}

\begin{figure*}
    \includegraphics[width=0.45\textwidth, trim = 10.0 5.0 10.0 0.0]{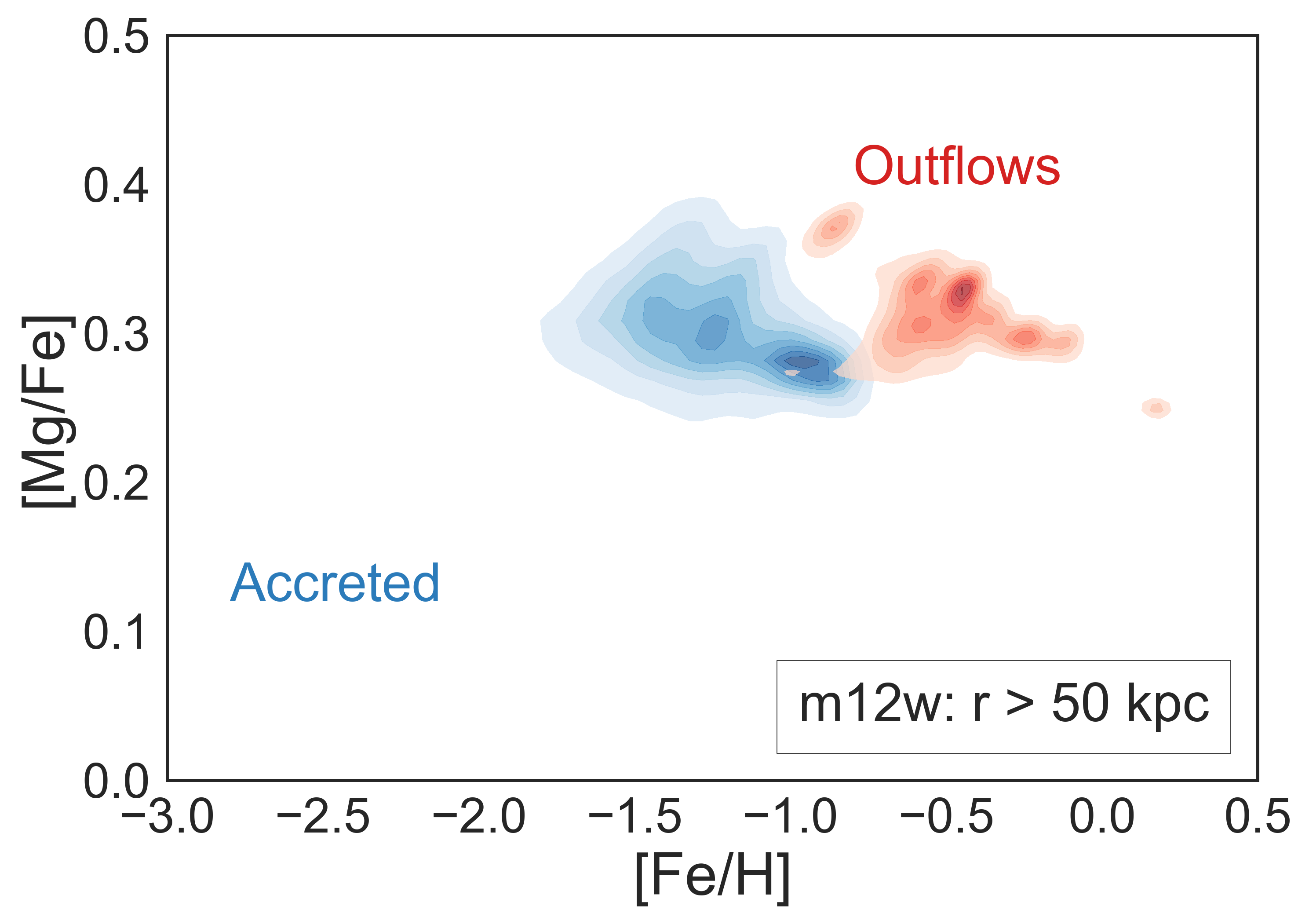}
	\hspace{.2in}
    \includegraphics[width=0.45\textwidth, trim = 10.0 5.0 10.0 0.0]{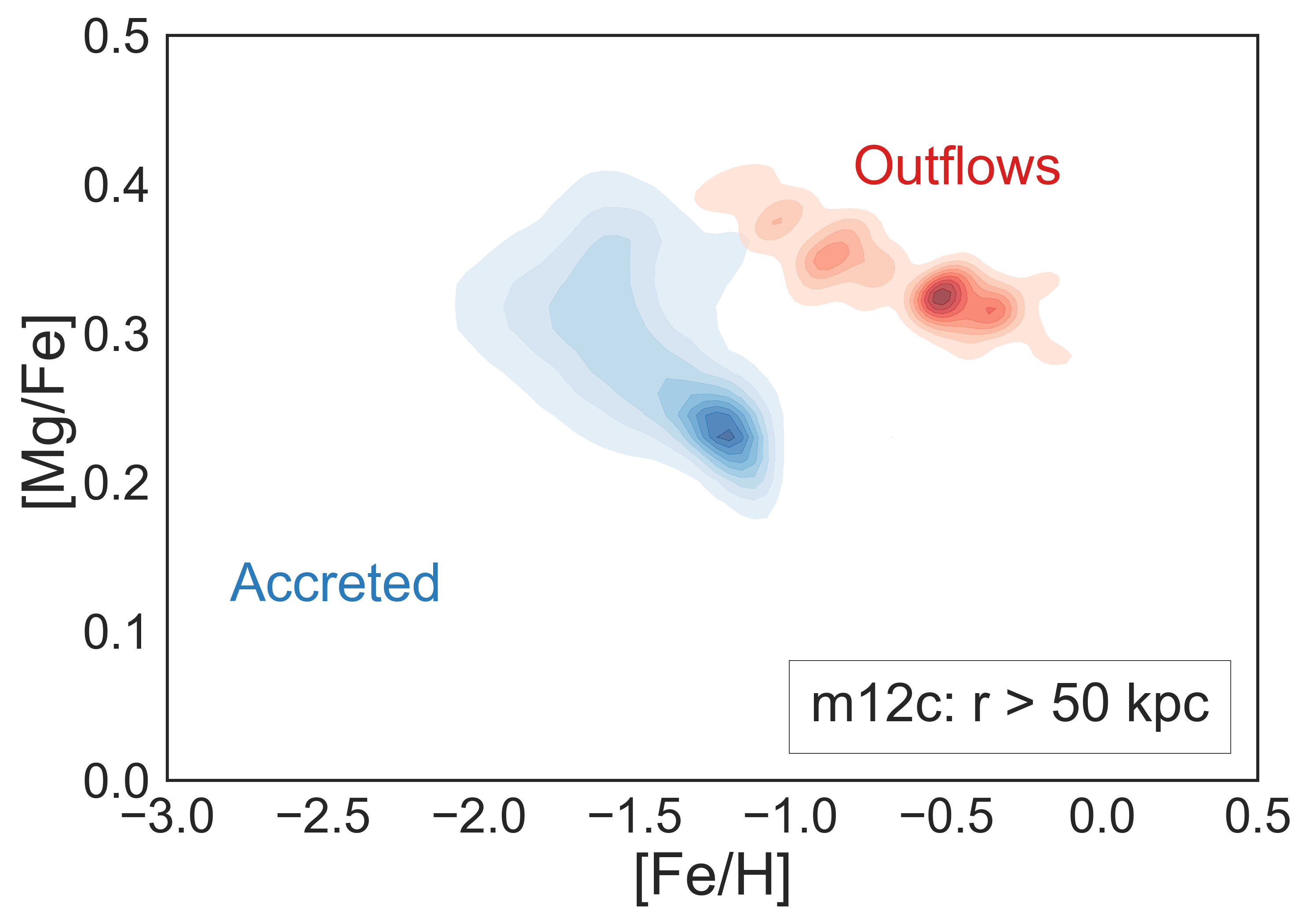}
	\hspace{.2in}
    \includegraphics[width=0.45\textwidth, trim = 10.0 5.0 10.0 0.0]{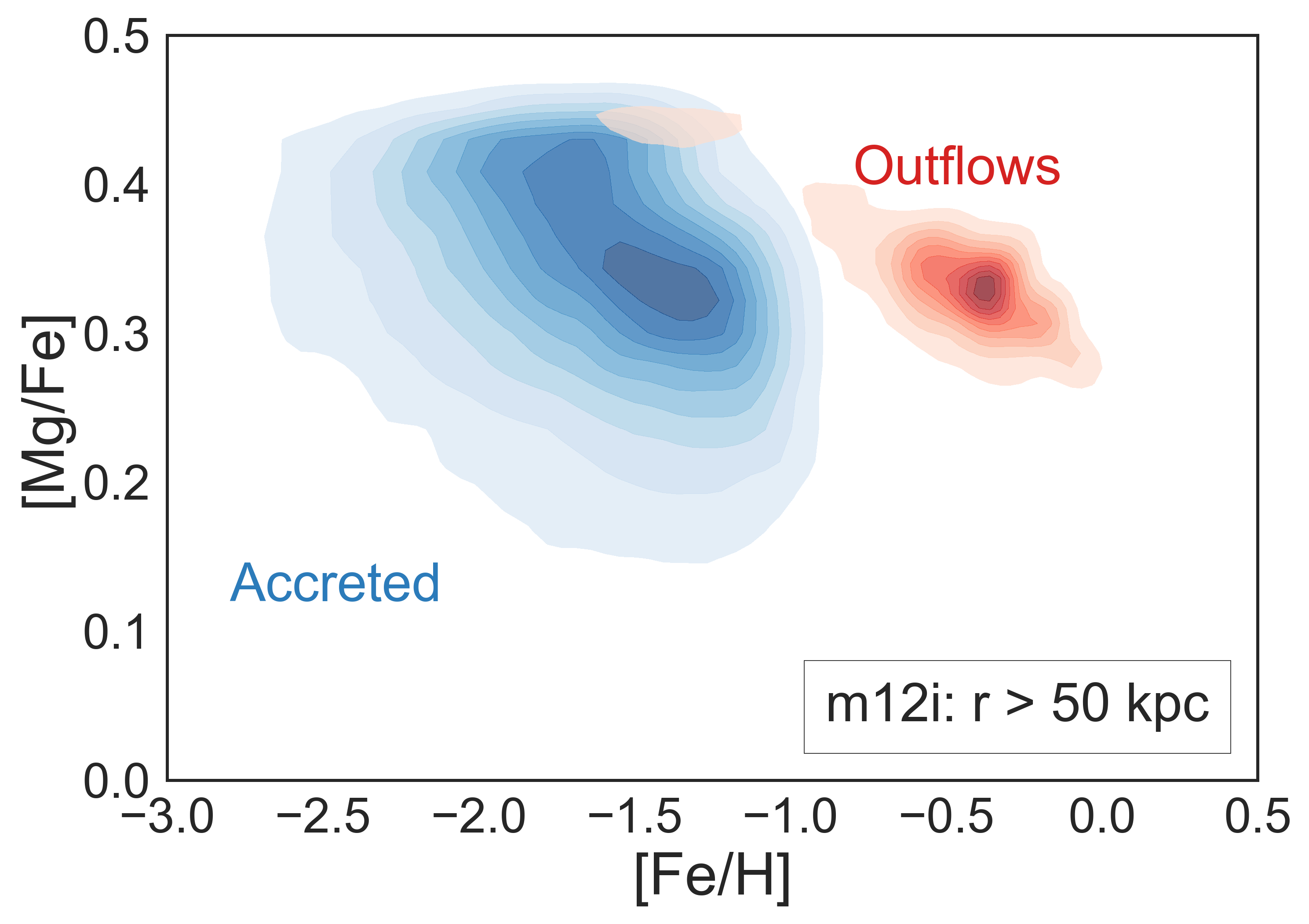}
    \hspace{.2in}
    \includegraphics[width=0.45\textwidth, trim = 10.0 5.0 10.0 0.0]{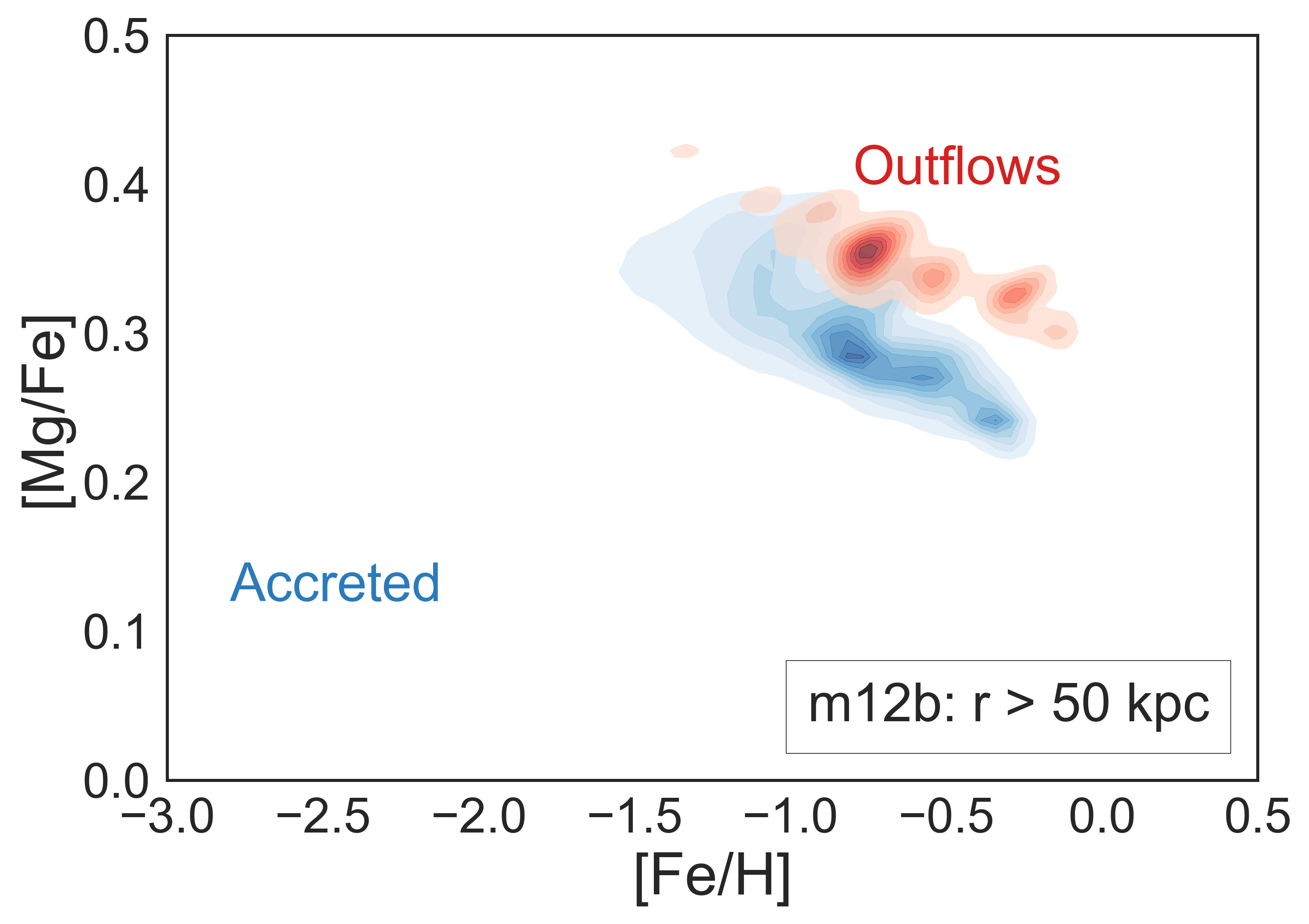}
    \hspace{.2in}
    \includegraphics[width=0.45\textwidth, trim = 10.0 5.0 10.0 0.0]{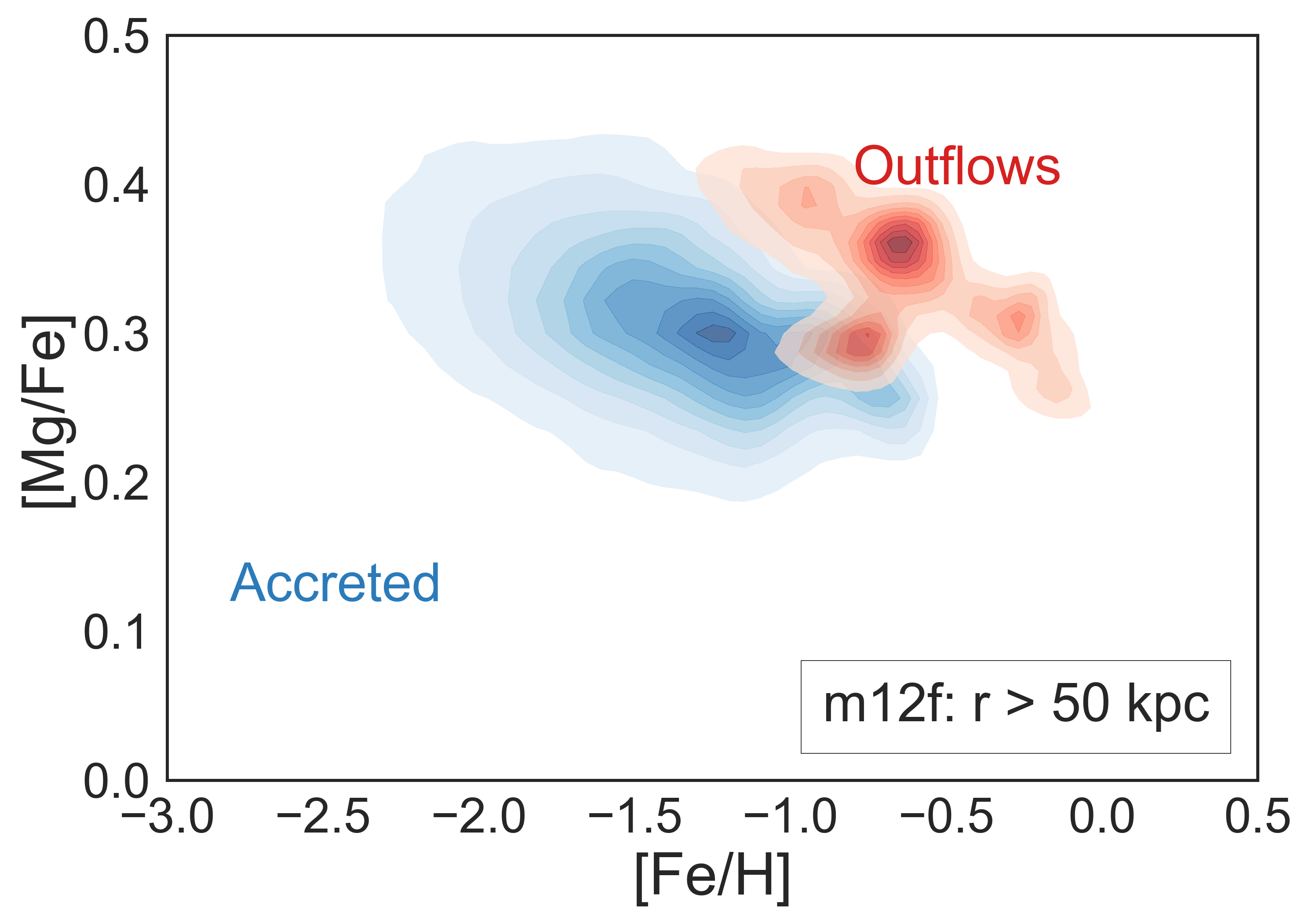}
    \hspace{.2in}
    \includegraphics[width=0.45\textwidth, trim = 10.0 5.0 10.0 0.0]{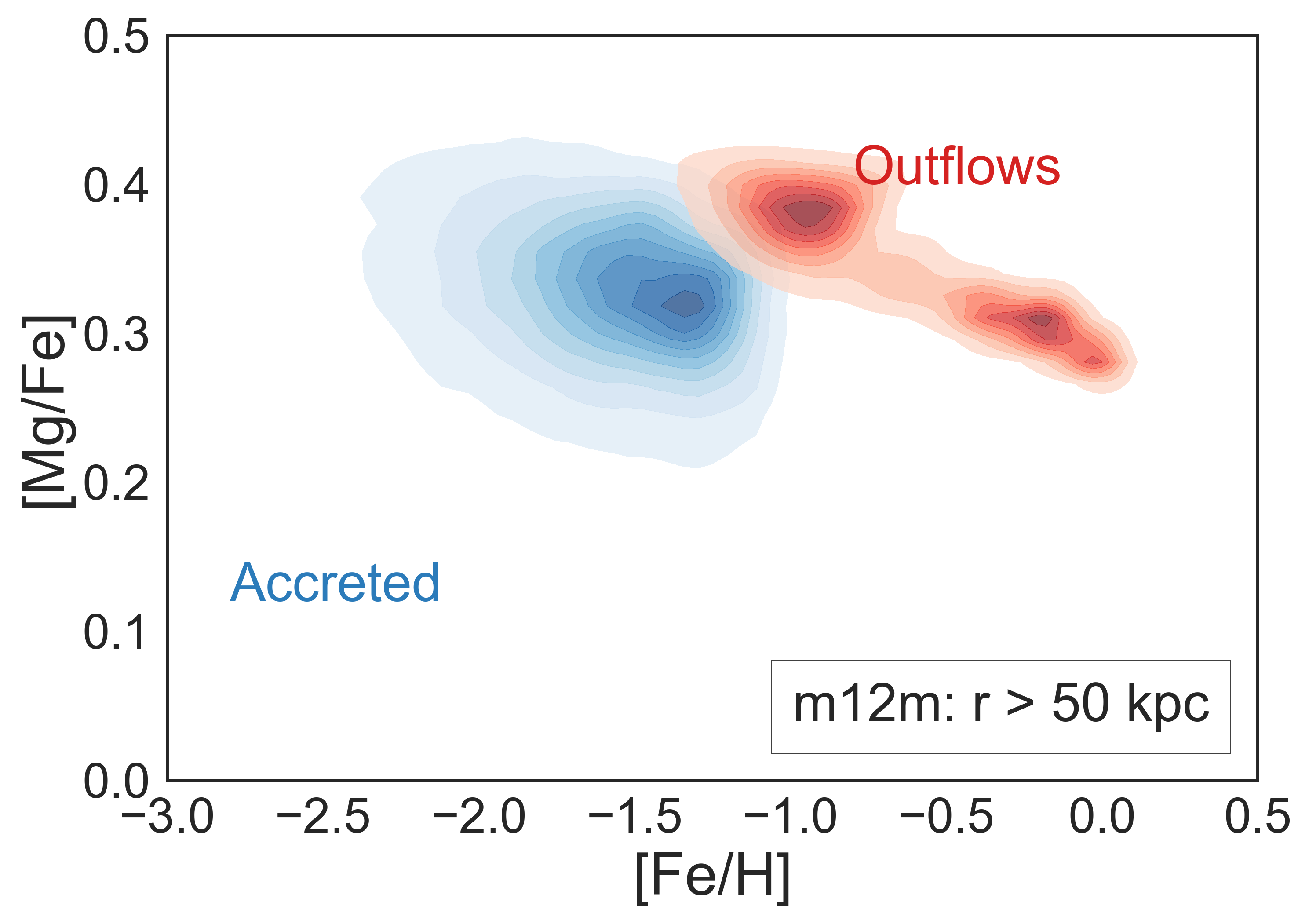}
	\centering
	\caption[Metallicity v.s. R]{Kernel density estimate in [Mg/Fe] versus [Fe/H] for outer halo stars with $r=50-200$ kpc in each system. The galaxies are ordered from lowest to highest in total galaxy stellar mass from upper left to bottom right. The red region shows the outflow population and the blue one shows the distribution for accreted stars born outside of 200 kpc.}
	\label{fig:all_mgfe_feh}
\end{figure*}

\subsection{Stellar Outflows and Total Star Formation} \label{sec:SFR}

Figure \ref{fig:SFH} provides a global view of how these stellar outflows track star formation with time.  The six panels show star formation rates averaged over 10 Myr as a function of lookback time in each galaxy, presented in the order of increasing main-galaxy stellar mass from upper left to lower right. The top panels give the total star formation rate of stars that born in the main galaxy ($r_{\rm birth} < 20$ kpc), while the bottom panels show the formation rate only of stars that we identify as being created in outflows ($r_{\rm birth} < 20$ kpc and $V_{\rm rad}^{\rm birth} > 200 \, \kms$).   Generally, our galaxies have bursty star-formation histories at early times and settle into more steady star formation over the last $\sim 4$ Gyr. Stellar outflows are clearly correlated with bursty star formation and become rare at late times when the main galaxy's star formation rate becomes less bursty. The outflow event we have chosen to show is fairly isolated, which provides us a particularly clean example. The other events that we see, especially those during very bursty star formation episodes appear to be triggered by similar feedback events.
The few stellar outflow events that do occur at late times are correlated with brief periods of elevated star formation in the main galaxy.  Interestingly, while the fraction of all stars that form in outflows is small compared to the total stellar mass of the system ($\sim 1\%$ overall), during starbursts, the instantaneous fraction of stars born in outflows can be as high as $\sim 20-50 \%$.  Interestingly, this result is broadly consistent with the observations \citet{Gallagher2018}, who find that star formation inside outflows accounts for $5 - 30\%$ of the total star formation when detected.  Since the focus of this paper is on the $z=0$ stellar halos of Milky Way-mass galaxies, we defer a more detailed comparison to observation for future work.~\footnote{We note that the $z=0$ global star formation rates of our galaxies ($4-10 \, \msun/$yr) are somewhat higher than estimates for the Milky Way   \citep[1-4 $\msun/$yr][]{Licquia_2015,Zonoozi2019}. However, most of them are relatively steady at late times (like the Milky Way).  These systems lack recent outflow activity, and we would expect that the Milky Way would be similar.}

It is clear from Figure \ref{fig:SFH} that the bulk of stellar outflow activity happened at lookback times greater than $\sim 4-8$ Gyr ago.  Since halo dynamical times are $\sim 2$ Gyr at the virial radius,  there is ample time for most of these outflow stars to have traveled outward and fallen back in to the main halo.   We expect, therefore, that most of these stars will be fairly uniformly distributed in the halo.  However, the small number of late-time  outflow events (that occur within the last $\sim 2$ Gyr) may preserve some spatially coherent structure.  In the following section we investigate these issues in more detail and explore observational differences between outflow stars and other stellar halo material.

\section{Outflow Stars in the Inner and Outer Halo} \label{s:properties}

\subsection{The Outer Stellar Halo: Phase Space Structure}

Figure \ref{fig:rvr_spatialiw} presents orbit diagrams ($V_{\rm rad}$ vs. $r$) and face-on spatial distributions of halo stars in \texttt{m12i} (top set) and \texttt{m12w} (bottom) at $\it{z}$=0.  The color bars map to stellar age for the radial velocity figures (top) and to metallicity for the spatial figures (bottom).  The left columns include all stellar particles, the middle columns include only \insitu stars that were born within the central region of the galaxy ($r_{\rm birth} < 20$ kpc) and with large positive radial velocity ($V_{\rm rad}^{\rm birth} > 200 \, \kms$).  The right panels include only accreted stars that were born beyond 200 kpc of the central galaxy for comparison.   We have removed all stars bound to satellite galaxies in these diagrams. 

There are significant differences between outflow stars and accreted stars in both the orbital diagrams and spatial structures.  While overall the velocity distributions are quite similar at $z=0$ (see the open histograms in Figure \ref{fig:Vbirth}), the phase-space structure is different.  Substructures, visible both spatially and in the orbital diagrams in Figure \ref{fig:rvr_spatialiw}, are dominated by accreted stars born at large radius in distinct dwarf galaxies. These structures offer important constraints on the accretion history. The \insitu outflow population shows many fewer structures and is much more smoothly distributed.  While the orbital properties of accreted and outflow halo stars are {\em not} particularly distinct, the phase-space structure of the two populations is. Although one might have expected outflow stars to follow more radial orbits, we find that by the time most of them have traveled out and fallen back in to the stellar halo at $z=0$, they end up on similar orbits as the rest of the stellar halo.

The main reason why outflow stars are more smoothly distributed is that the outflow events tend to occur long enough ago (Figure \ref{fig:SFH}) that they become well mixed in the halo potential. One counter example to this trend is seen in $\texttt{m12w}$, which had an outflow event at $z\simeq 0.1$ that remains visible as a blue plume at $z=0$.    In the bottom middle panel of Figure \ref{fig:rvr_spatialiw}, we see this as an outflowing stream of metal-rich stars at $Y \simeq 0$ that extends from $X \simeq 0$ to $500$ kpc.  The same stars make up the young streak of particles that have increasing radial velocity with distance in the panel above it.  This behavior is characteristic of an outflow because stars that are liberated from a tidally-destroyed galaxy will tend to trace a characteristic orbital pattern, with radial velocity decreasing in magnitude with distance from the main galaxy (as seen in the upper-right panels). Instead, the stars in this stream are moving more quickly at larger radius, because of the birth velocity gradient\footnote{As we have discussed above, these stars are born in accelerated gas outflows. The stars that form last have emerged from gas that has had more time to be accelerated and this creates a velocity gradient among the stars that have formed.}. They were born within $20$ kpc of the galaxy $t_{\rm age} \simeq 1.34$ Gyr ago and the most distant ones are still on their way out.   We find that $< 10 \%$ of the stars in this young plume are unbound, so that the bulk of these stars are destined to return and inhabit the stellar halo in the future.  In fact, the upper panels for \texttt{m12w} show that stars in the stream with $r \lesssim 150$ kpc have already started to fall back in with negative radial velocity. 

Of the six halos in our simulation sample, $\texttt{m12w}$ and to a lesser extend $\texttt{m12b}$ and \texttt{m12f} have visible outflow streams in phase space at $z=0$.   In both cases, the plumes are less prominent than the big plume in \texttt{m12w}. The vast majority of outflow stars in all six halos have smooth phase-space structure.  

Figure \ref{fig:rvr_spatialiw} also demonstrates that the outflow populations in \texttt{m12i} and \texttt{m12w} are both more metal-rich and moderately younger than the bulk of the stellar halo at large radius.  The age difference is not systematic among our six runs.  Halos \texttt{m12b}, \texttt{m12c}, and \texttt{m12f} all have outflow halo populations that are slightly older than the accreted halo stars.  This is perhaps not unexpected since the outflow star formation rates shown in Figure \ref{fig:SFH} suggests that these three galaxies have more early-time outflow star formation than the others.  However, in all six runs, outflow stars in the outer stellar halo are chemically distinct. We discuss this difference in the next subsection.

\subsection{The Outer Stellar Halo: Chemical Abundances} \label{sec:outer_chem}

Figure \ref{fig:all_mgfe_feh} illustrates [Mg/Fe] versus [Fe/H] distributions for outer halo stars ($r=50 - 200$ kpc) for all six halos as indicated, ordered from upper left to bottom right with increasing central galaxy stellar mass. The different colors are used to represent different populations, with blue showing accreted stars with $r_{\rm birth} > 200$ kpc and red showing outflow stars identified as stars born within $20$ kpc of the main galaxy and with large positive radial velocity ($V_{\rm rad}^{\rm birth} > 200 \, \kms$).  The outflow population tends to be more alpha-enhanced at fixed iron abundance, and typically  more iron-rich than the accreted halo. 

This difference may provide a way to distinguish outflow-produced halo stars from their accreted counterparts. Specifically, outflow stars born within the main galaxy are inherently metal-rich, unlike outer halo stars from accreted, low-mass dwarfs, which tend to be lower metallicity \citep[e.g.][]{Bullock&Johnston05,Johnston2008}.  
These outflow stars also form in bursts, which pushed them towards higher $\alpha$ element abundance at fixed [Fe/H].  Elements like Mg are produced in core-collapse supernovae on timescales that are significantly shorter than SNe Type Ia, which produce the majority of iron \citep[see e.g.,][for discussions]{Robertson2005,Maoz10}.  The accreted dwarfs that form the outer stellar halo tend to have longer periods of star formation, and are less $\alpha$ enhanced at fixed iron abundance than stars formed in bursts.

Note that, while stars from accreted dwarfs in the outer halo are mostly old, they were typically born in low-mass dwarf galaxies that had star formation histories that are extended enough (> 1 Gyr) to have some iron-rich Type-Ia enhancement (and thus Mg/Fe-depletion), as expected from semi-analytic models \citep[e.g.][]{Font08,Johnston2008}.  The outflow stars, on the other hand, were born of gas that was recently enriched in a starburst (via short-lived  massive stars with high alpha yields) but liberated from the main galaxy prior to Type-Ia enrichment. This is consistent with the notion that alpha-enhancement is indicative of short-lived star formation rather than overall stellar age.

\subsection{The Local Stellar Halo} \label{sec:local}

\begin{figure}
\includegraphics[width=0.5\textwidth,trim = 0.0 80.0 0.0 0.0]{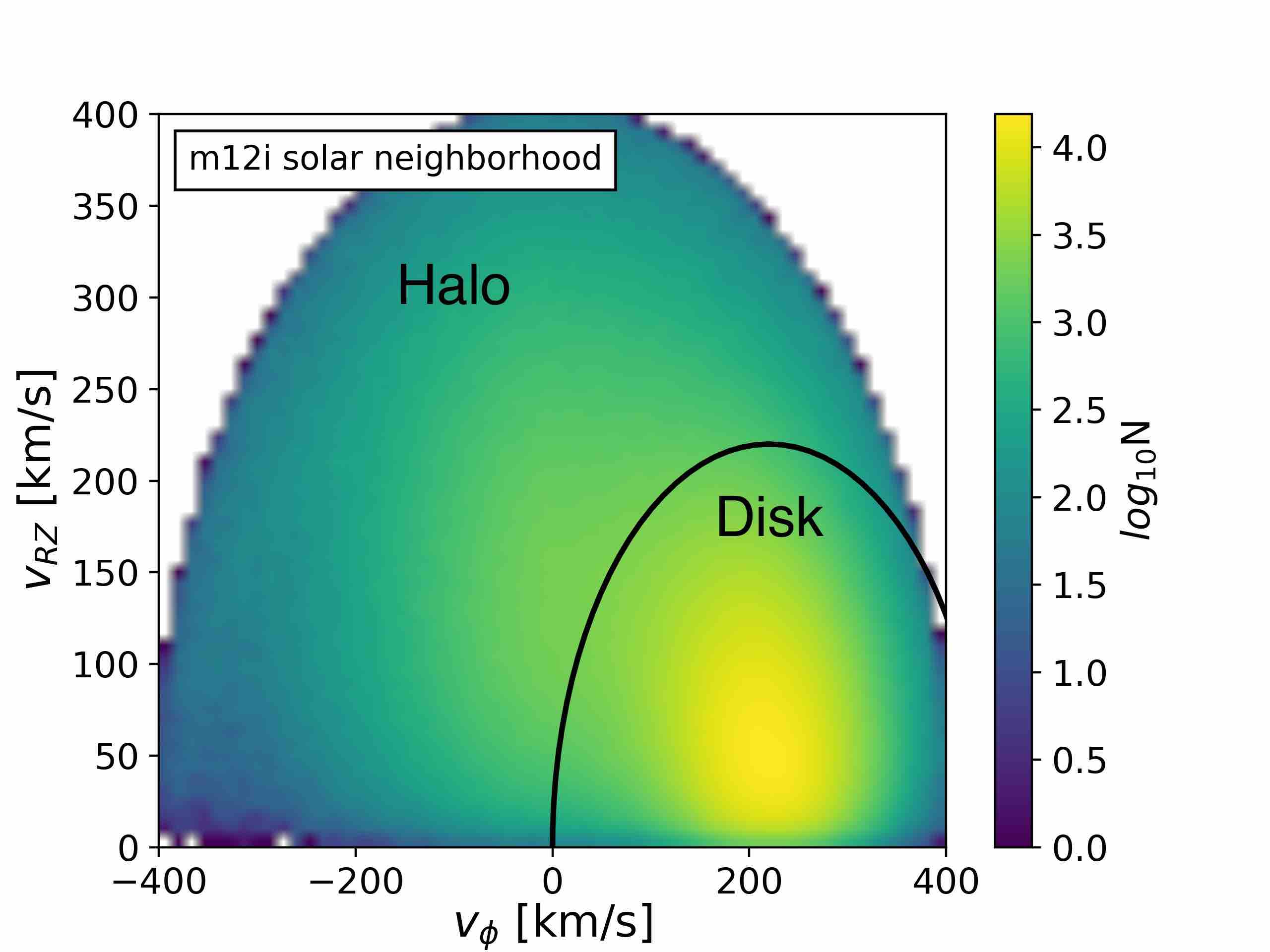}
\caption[Vphi v.s Vrz]{Toomre diagram of stars in an ensemble of solar neighborhoods in \texttt{m12i}, specifically all star particles within $5$ kpc of an $8$ kpc ring about the galactic center. We divide the star particles into disk and halo components, based on their location in this diagram.  Halo stars have $|{\bf V} - {\bf V_{\rm LSR}}|  >\ 220\, \kms$, as indicated by the black line.}  
\label{fig:diskhalo}
\end{figure}

Stars in the local stellar halo offer rich observable diagnostics and this motivates us to explore the degree to which stars formed in outflows may contribute to the local stellar halo and whether their presence might be discerned using observations.

\begin{figure*}
\includegraphics[width=0.462\textwidth,trim = 0.0 0.0 0.0 0.0]{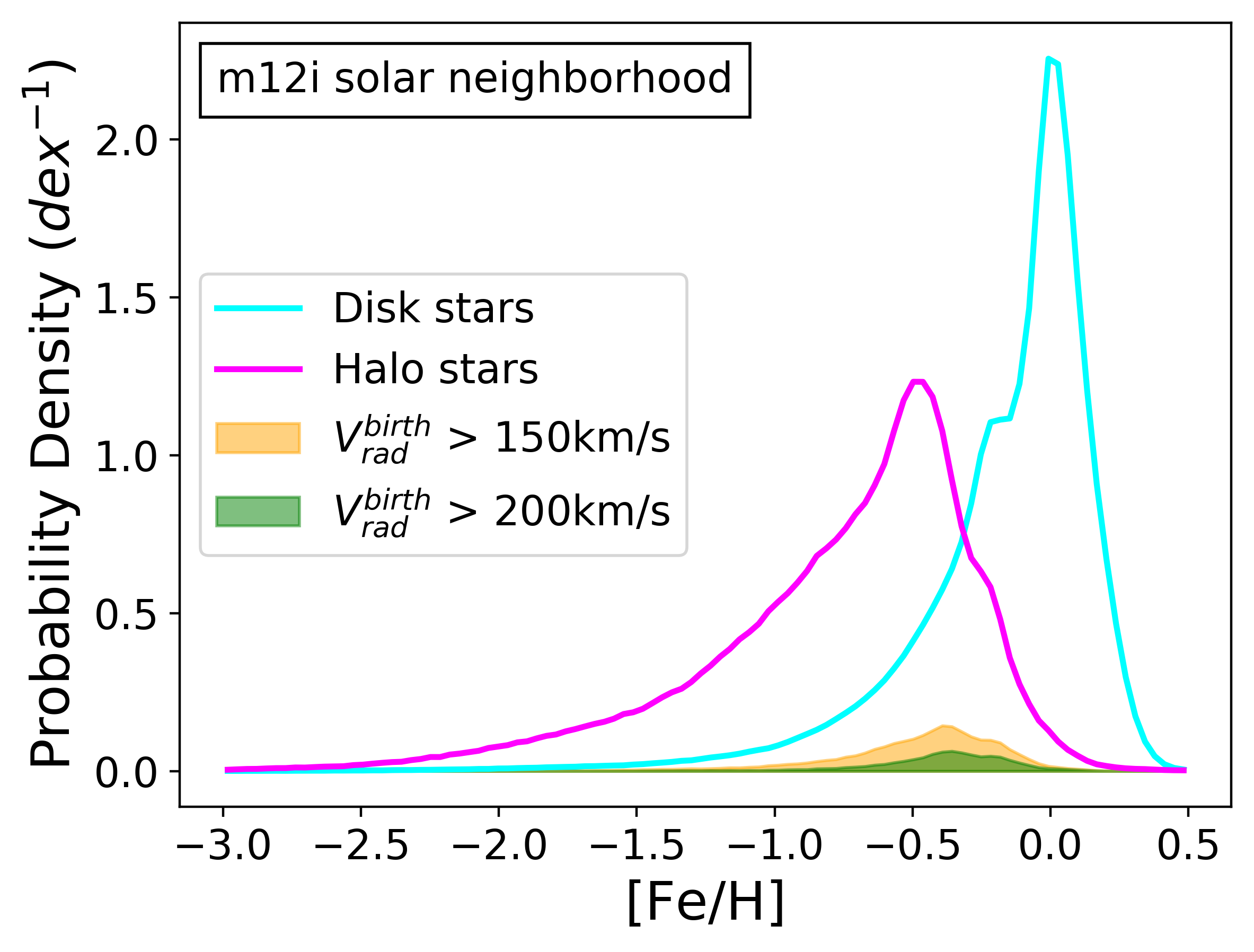}
\includegraphics[width=0.48\textwidth,trim = 0.0 0.0 0.0 0.0]{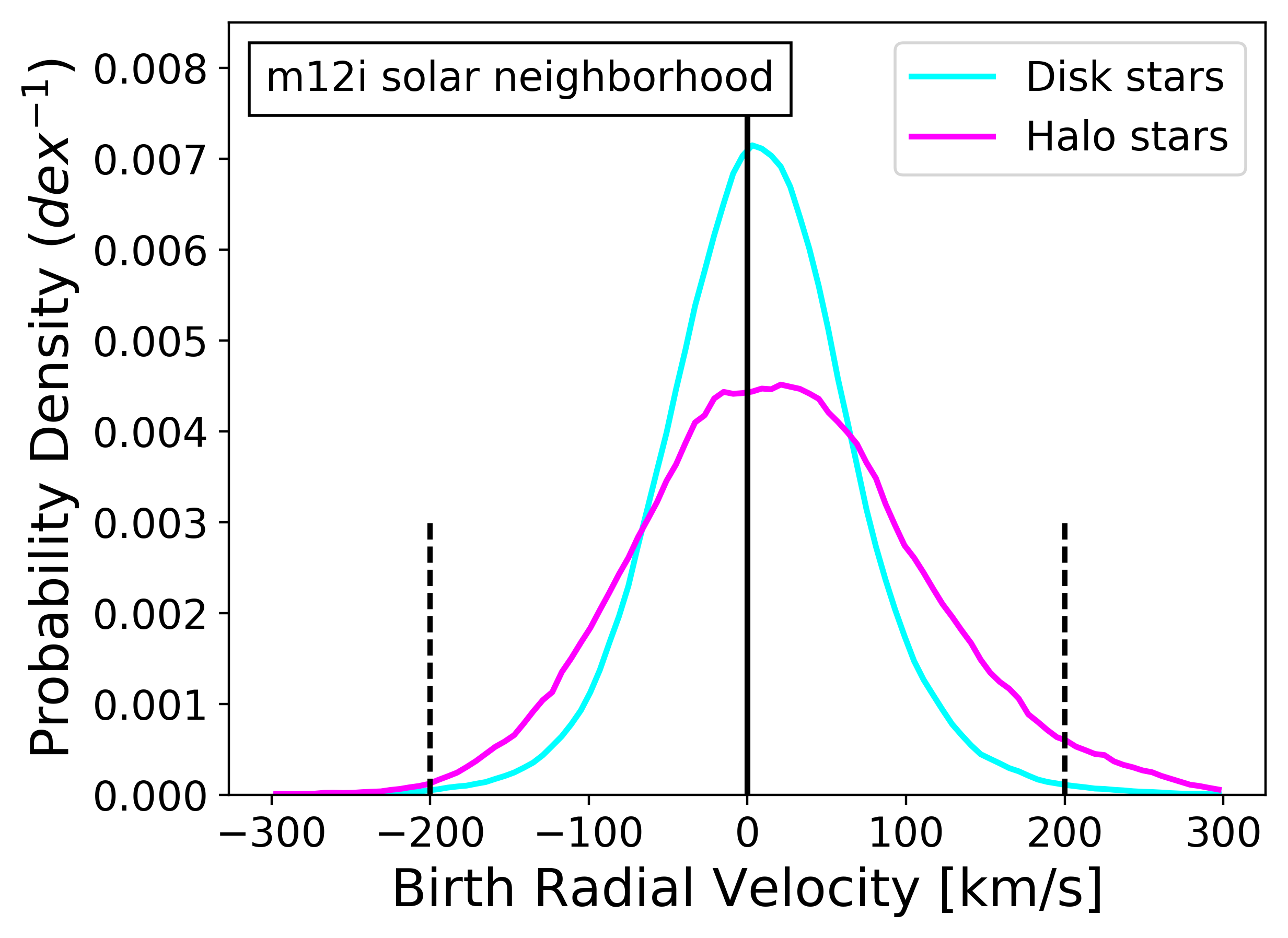}
\caption[]{{\bf Left:} Metallicity distributions for solar neighborhood disk stars (cyan) and halo stars (magenta) in \texttt{m12i}. The disk stars are more metal-rich, as expected.   Sub-components of halo stars that had radial velocities at birth indicative of outflows ($>150 \, \kms$ and $>200 \, \kms$, green and yellow) are shown within the halo star distributions.  We see these outflow-identified stars preferentially inhabit the high-[Fe/H] tail of the local stellar halo. {\bf Right:} Radial velocities at birth for solar neighborhood stars identified as disk (cyan) and halo (magenta).  Disk stars were born with a symmetric distribution of radial velocities, with a $98\%$ width spanning $\pm 150 \, \kms$.  Halo stars show a significant asymmetry towards positive birth velocities, indicative of an outflow contribution. }  
\label{fig:diskhaloFeH}
\end{figure*}

\begin{figure*}
\includegraphics[width=0.465\textwidth,trim = 0.0 0.0 0.0 0.0]{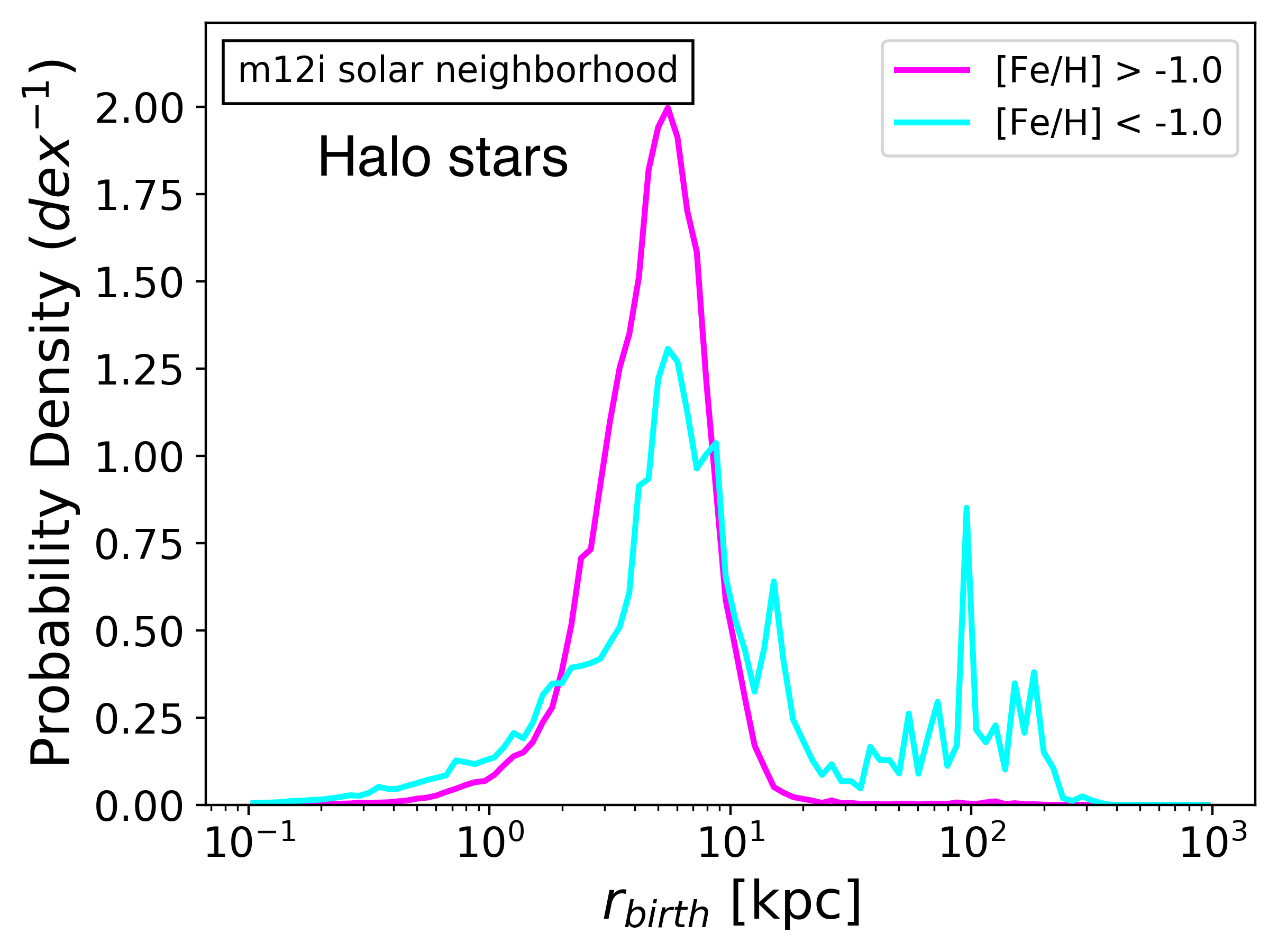}
\includegraphics[width=0.48\textwidth,trim = 0.0 0.0 0.0 0.0]{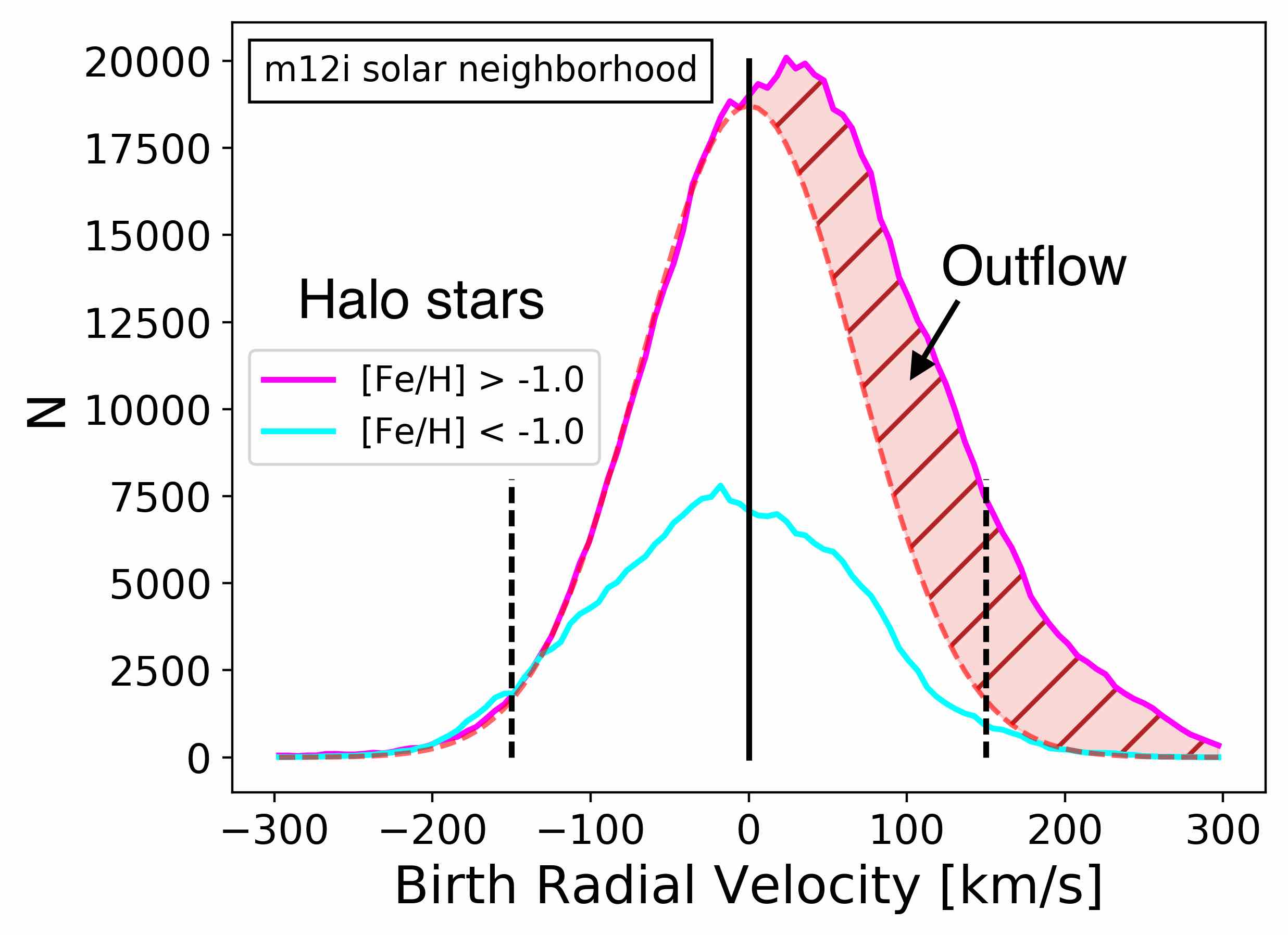}
\caption[]{{\bf Left:} Birth radii distributions of kinematically-identified halo stars in solar neighborhood ensembles divided between metal-rich ([Fe/H] $> -1.0$, magenta) and metal poor ([Fe/H] $< -1.0$, cyan) populations in halo \texttt{m12i}. A significant fraction of local metal-poor halo stars formed at large distances, suggesting that accretion events contributed a major portion of this population. In contrast,  metal-rich local halo stars formed primarily inside the central galaxy. These stars were either born in the central galaxy and subsequently heated or formed in outflows.  {\bf Right:} Radial velocities at birth for the same two populations of metal-rich and metal-poor local halo stars.  Vertical dashed lines at $\pm 150 ~\kms$ are shown for reference. The metal-poor halo stars have a roughly symmetric birth radius distribution, with a slight preference towards infalling birth velocities.  The the metal-rich halo stars, on the other hand, show a strong asymmetry towards positive birth velocities. The dashed line shows a Gaussian fit to the negative part of the metal-rich  $V_{\rm rad}^{\rm birth}$ distribution.  The result is remarkably similar to the birth velocity distribution of disk stars shown by the cyan line in the right panel of  Figure \ref{fig:diskhaloFeH}.  If we interpret this Gaussian part of the distribution as the heated-disk contribution to the metal-rich stellar halo, the remaining shaded region can be interpreted as coming from outflow stars, which is about 27\% of the total metal-rich halo stars locally.  We adopt this approach as one method for estimating the outflow fraction for local halo stars.}  
\label{fig:haloRbirthVbirth}
\end{figure*}

\begin{figure}
\includegraphics[width=0.48\textwidth,trim = 0.0 20.0 0.0 0.0]{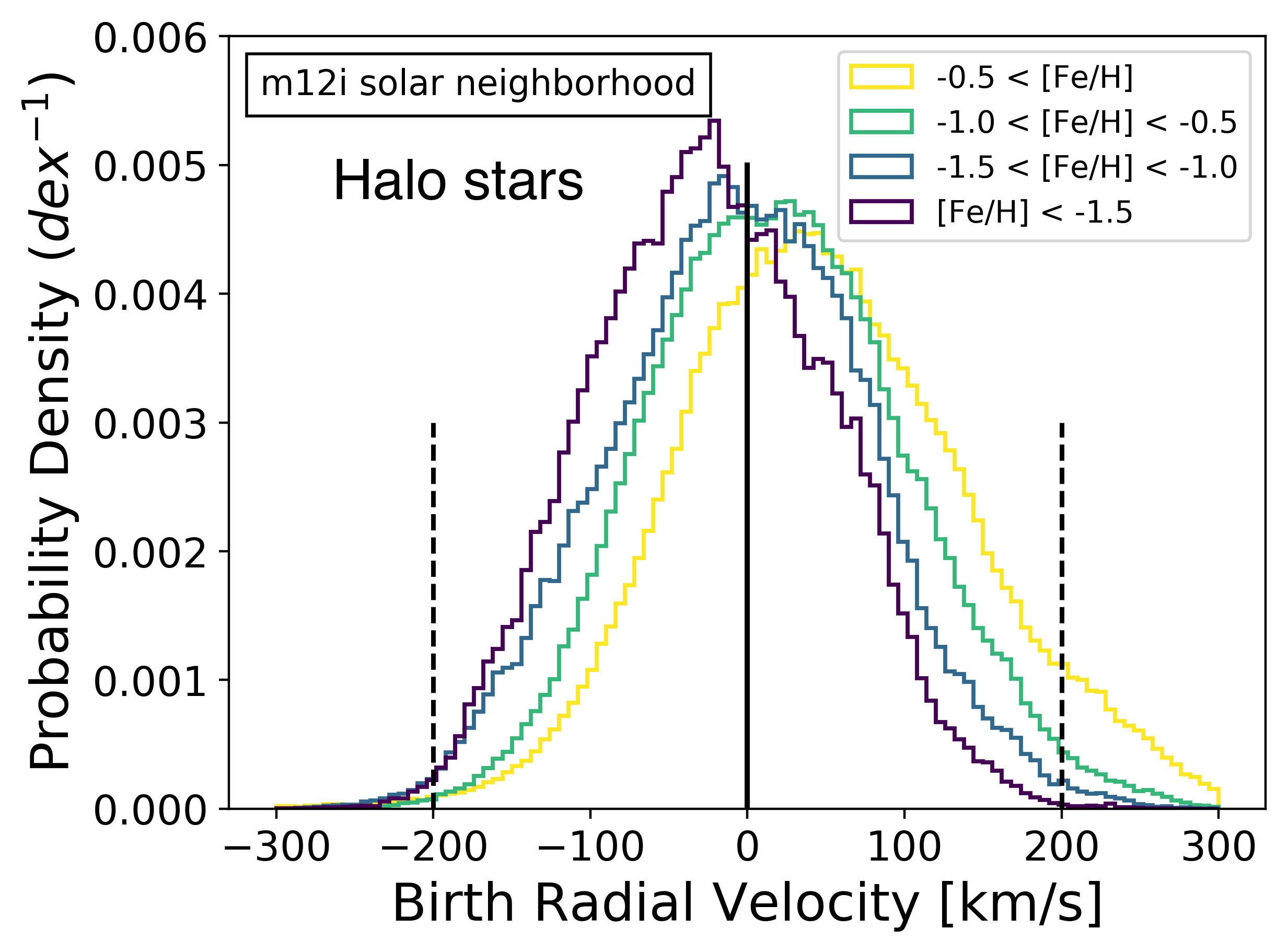}
\caption[]{Radial velocities at birth for solar neighborhood halo stars in different metallicity bins as indicated.  The most metal poor local halo stars have birth velocity distributions that are biased towards negative values, indicative of a large contribution from accreted stars that were born within infalling dwarf galaxies.  As metallicity increases, the birth velocity distributions shift towards positive values, as would be expected if outflows make up an increasing fraction of the local stellar halo at higher metallicities.}  
\label{fig:haloVbirth}
\end{figure}

\subsubsection{Example Case: The Inner Halo of the Latte Galaxy}

Before summarizing results for all six of our halos, we use the \texttt{m12i}, the original {\em Latte} primary halo \citep{Wetzel2016}, as an example to illustrate our approach. 

In order to define the local stellar halo population we construct an ensemble of simulated environments that mimic stellar observations within 5 kpc spheres around the sun. Specifically, we select all star particles that exist within a torus centered on a ring radius 8 kpc in the disk plane from the center of each galaxy. The small radius of the torus is 5 kpc around the main ring. 

Figure \ref{fig:diskhalo} shows the kinematic properties of star particles found in the solar neighborhood torus in \texttt{m12i}. This is a Toomre diagram with the Galactocentric azimuthal component of the velocity vector, $V_{\phi}$, on the horizontal axis and the perpendicular Toomre component, $\sqrt{(V_R^2+V_Z^2)}$, on the vertical axis. Disk stars dominate a large overdensity at $V_{\phi}  \approx 220\, \kms$, corresponding to the circular velocity of the Local Standard of Rest (LSR). Following the selection method of \cite{NissenSchuster10}  and \cite{Bonaca17},  we identify halo stars with the velocity cut $|{\bf V} - {\bf V_{\rm LSR}}|  >\ 220\, \kms$, where ${\bf V_{\rm LSR}}$ is defined to have $ V_\phi = 220 \, \kms$, $V_R = 0$, $V_Z = 0$.  The black line in Figure \ref{fig:diskhalo} marks the boundary where we divide the two populations, as labeled.

The left panel of Figure \ref{fig:diskhaloFeH} shows the metallicity distribution for the two kinematic components identified in Figure \ref{fig:diskhalo}. The distribution of disk stars (cyan) is more metal-rich than the halo (magenta) and peaks at approximately solar metallicity, [Fe/H] = 0. The halo is comparatively metal-poor and shows a peak at [Fe/H] $\sim -0.5$, which is higher than the typical inner halo of the Milky Way \citep[e.g.][]{Allende_Prieto_2006}. Here we also show the sub-components of halo stars with different birth radial velocities. The shaded yellow histogram shows the metallicity distribution for the halos stars with $V_{\rm rad}^{\rm birth}  > 150\, \kms$ while the green one with a slightly stricter criterion, with $V_{\rm rad}^{\rm birth}  >  200\, \kms$. These two components, which have birth velocities indicative of outflows stars, are located in the metal-rich tail of the distribution of halo stars.  This region overlaps significantly with disk star metallicity distribution, which is consistent with the idea that these stars were initially born within the disk but now have a hotter velocity distribution than present-day disk stars.  

We have examined the chemical abundances of local halo stars as a function of birth velocity and birth radius and find no trend that differentiates outflow stars from heated disk stars in this space (see Figure \ref{fig:localalpha}). They also show no significant difference in $z=0$ kinematics from other halo stars (see Figure \ref{fig:localkinematic}).

The right panel of Figure \ref{fig:diskhaloFeH} presents the distribution of radial birth velocities of local stars divided in the same way between disk and halo.  Disk stars show a symmetric distribution about the $V_{\rm rad}^{\rm birth}  \simeq  0\, \kms$, whereas halo stars show a significant asymmetry towards positive birth radial velocities, which suggests that some fraction of these stars were born in outflows. We have argued earlier that outflow stars contribute a significant amount to the outer stellar halo, is is therefore natural to suspect that a fraction of high-velocity halo stars in the solar neighborhood were also born this way. As the energy required to put stars on plunging orbits in the solar neighborhood is potentially lower than that required to make them orbit to the outer stellar halo, we might even expect to see more of these stars in the area around the Sun.

We now focus on kinematically identified halo stars and divide them into a metal-rich component with [Fe/H] $>  -1.0$, and a metal-poor one with [Fe/H] $\leq  -1.0$, in a similar manner as \cite{Bonaca17}. Approximately $80\%$ of the halo sample in \texttt{m12i} is metal-rich by this definition, which is comparable to the $\sim 50\%$ metal-rich fraction of local halo stars identified by \cite{Bonaca17} for the Milky Way.

The left panel of Figure \ref{fig:haloRbirthVbirth} demonstrate that almost all these metal-rich local halo stars were born within $\sim 15$ kpc of the center of the galaxy, with the median of the distribution at $6$ kpc.  These stars were either born in the central disk/galaxy and subsequently heated or they were born in outflows.  The metal poor halo stars, by contrast, shows a significant contribution from stars born at large distances. We find $\sim40\%$ of the metal-poor stars  formed with $r_{\rm birth} > 20$ kpc, suggesting that many of these stars were formed inside dwarf galaxies that later merged with the host galaxy.  

The right panel of Figure \ref{fig:haloRbirthVbirth} shows the birth radial velocity distributions for each of these two populations.  The local metal poor halo in \texttt{m12i} has a roughly symmetric birth velocity distribution (cyan), with a slight preference for infall velocities at birth.  This is again consistent with a significant contribution from accreted dwarfs galaxies.  The metal-rich population, shown in magenta, is significantly skewed towards positive birth velocity. It is clear that the asymmetry in halo star birth velocities seen in the right panel of Figure \ref{fig:diskhaloFeH} mainly comes from metal-rich halo stars.  The dashed line shows a Gaussian fit to the negative portion of the metal-poor distribution, which is remarkably similar to the birth velocity distribution of disk stars shown by the cyan line in the right panel of  Figure \ref{fig:diskhaloFeH}.  If we interpret this part of the distribution as the heated-disk contribution to the stellar halo, the remaining shaded region can be interpreted as coming from outflow stars.  If we do so, we find that these stars contributes approximately $27\%$ 
of metal-rich halo stars locally.

In Figure \ref{fig:haloVbirth}, we further explore this asymmetry by looking at the differences in the birth radial velocity distribution for halo stars with different metallicities. The most metal-rich stars (yellow) are strongly skewed towards being born with large, positive (outflow) velocities.  The most metal-poor stars (purple) tend to have been born with negative radial velocities, indicative of infall inside of accreting dwarf galaxies.  If we fit a Gaussian to the negative side of the [Fe/H] $> -0.5$ distribution (as we did in the right-panel of Figure \ref{fig:haloRbirthVbirth}) we find that the excess outflow fraction is $\sim 40 \%$ in this case.

\subsubsection{Summary of Outflow Fractions in Solar Neighborhoods} \label{sec:sum_solar_neighbor}

\begin{figure}
\includegraphics[width=0.48\textwidth, trim = 0.0 20.0 0.0 0.0]{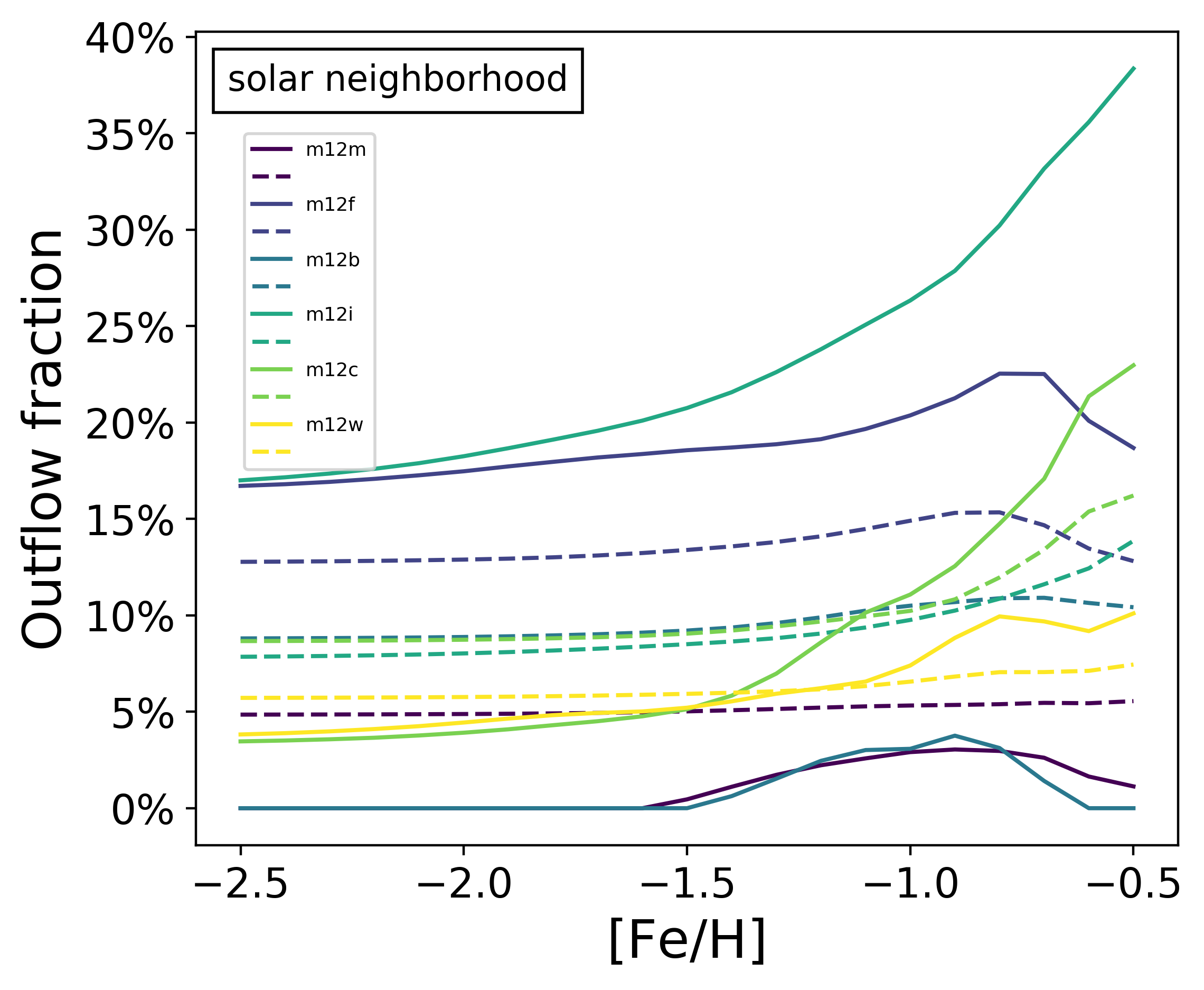}
\caption[]{Fraction of the local stellar halo identified as being formed in outflows as a function of stellar metallicity in each of our simulations.  Fractions include stars more metal-rich than the value of [Fe/H] shown on the horizontal axis. We have identified outflow fractions using two definitions.  First, we associate the outflow fraction with the excess above a symmetric birth velocity distribution as illustrated in the right panel of Figure \ref{fig:haloVbirth}. Solid lines use this definition.  In the second case (dashed) we use the fraction of stars born with radial velocities $> 150 \, \kms$.  
In most cases, the fraction rises towards higher metallicity, with as much as $\sim 40 \%$ of the most metal-rich local stellar halo coming from outflows by this definition.} 
\label{fig:localhalofraction}
\end{figure}

Figure \ref{fig:localhalofraction} provides a summary of local kinematically-identified stellar halo outflow stars for all six of our host halos.  We are specifically quoting the fraction of the local stellar halo identified as being born in stellar outflows, $M_{\rm outflow}(>X)$/$M_{\rm total}(>X)$, plotted as a function of $X=$ [Fe/H] of the stars. We have adopted two methods for identifying stars as being born in outflows.  First, the solid lines show the fraction calculated using the excess positive portion of the birth velocity distribution as illustrated in the right panel of Figure \ref{fig:haloRbirthVbirth}.  The dashed lines adopt a simpler choice, which defines outflow stars to be those born with radial velocities greater than $150\, \kms$.  In most cases the results are similar, with rising outflow fractions towards increasing metallicity.  However, the exact fraction does depends on the accretion history of the main galaxies. One of the exceptions is halo \texttt{m12b}, which has undergone a recent merger.  The resultant birth-velocity distribution of kinematically-identified halo stars has very little skewness and produces a near-zero outflow fraction via the first definition.   

\begin{figure}
\includegraphics[width=0.5\textwidth, trim = 0.0 0.0 10.0 0.0]{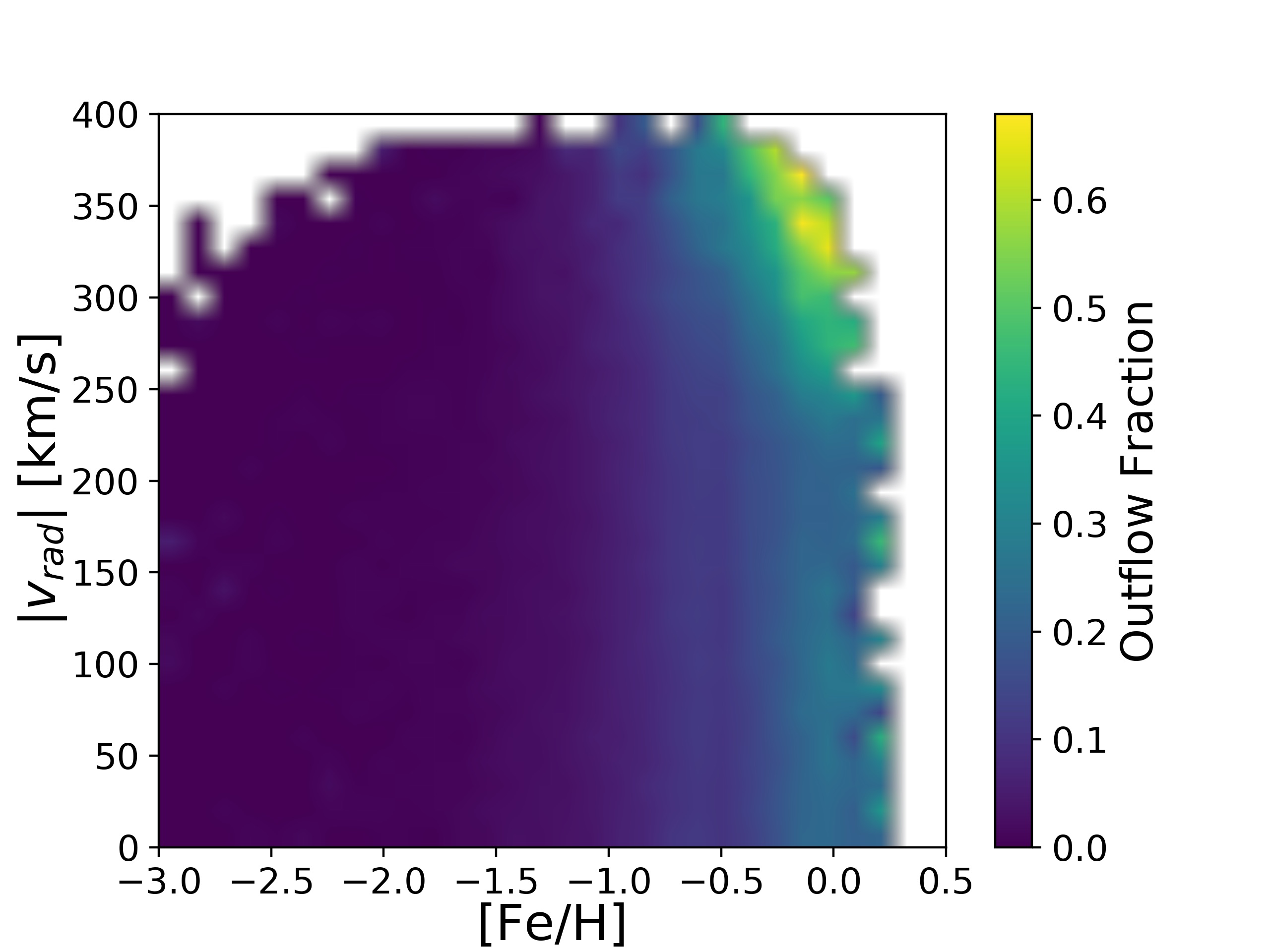}
\caption[]{Fraction of local halo stars that were born in outflows (color bar) as a function of stellar current radial velocity and metallicity. We assume stars with $V_{\rm rad}^{\rm birth} > 150 \, \kms$ were born in outflows.  We see that the majority of stars with both the highest velocities ($|V_{\rm rad}| \gtrsim 300 \, \kms$) {\em and} highest metallicities ([Fe/H] $\gtrsim -0.5$) were created in outflows.}
\label{fig:joint}
\end{figure}

How might we distinguish outflow stars from other stars in the local stellar halo?  While Figure \ref{fig:haloRbirthVbirth} demonstrates that the outflow fraction rises with increasing metallicity, stars born in outflows remain minority populations even at the highest metallicities.  In Appendix \ref{app:2} we show that neither the $z=0$ kinematics of halo stars nor chemical abundances alone provide a strong discriminant, mainly because the overall fraction of outflow stars is small enough that small differences in underlying distributions are washed out. A better determination can be made by exploring the joint distribution of metallicity and radial velocity.  Figure \ref{fig:joint} shows the fraction of stars born in outflows (color bar, using $V_{\rm rad}^{\rm birth} > 150 \, \kms$) as a function of both [Fe/H] and absolute value of current radial velocity.  We see that the outflow fraction dominates for stars that are both fast {\em and} metal-rich. 

\section{Discussion} \label{s:discuss}
  
This paper introduces the possibility that galactic stellar halos may be populated by stars born in dense-gas outflows driven by feedback events. The proposal is motivated both by an analysis of simulations and by spectroscopic evidence for star formation in outflows \citep{Maiolino17,Gallagher2018}.  Specifically, these outflow stars are born with velocities fast enough to take them far from galactic disks but slow enough that they remain bound to galactic halos.  Anecdotally, the outflow scenario bares resemblance to the classic idea of \citet*[][ELS]{Eggen62}, who suggested that halo stars formed in rapidly {\em infalling} gas clouds.  Here we posit a phase-reversal of ELS: some fraction halo stars formed in rapidly outflowing gas. 
  
Before summarizing our quantitative conclusions, it is important to qualify our results given the difficulty of the numerical problem we are considering.  Accelerated cold clouds are subject to hydrodynamical instabilities that could destroy them on timescales shorter than those required for star formation \citep{Bruggen2016,Schneider2017}, though rapid re-cooling as outflows expand \citep{Thompson2016} or radiative shocks with the ambient medium \citep{zubovas14,richings18a} can help explain the prevalence of cold, dense outflow material as observed. As discussed by \citet{Ferrara16}, the degree to which dust grains survive and mix into the cool phase affects rapid molecule formation. 
Simple estimates of the timescales needed to grow dust grains by accretion of metals from the ISM suggest that dust destruction may not be a major problem in the dense gas that ends up forming stars \citep[][]{Richings18b}. 
\citet{Schneider2018} have used the Cholla
Galactic Outflow Simulations to show that the highest velocity cool outflowing gas observed in starburst systems may have a dual origin, where clustered supernova feedback ejects gas out of the disk where it is subsequently accelerated by a hot wind.  Though the episodes they explore correspond to more extreme, high star-formation cases than our systems experience, this picture is qualitatively consistent what we see in our simulations. 

With baryonic particle masses of $\sim 7000 ~\msun$ and adaptive force softening down to $\sim 1$ pc, our simulations do seem to robustly predict the existence of outflows that consist of dense, relatively cold gas. This gas occasionally goes unstable for star formation after acceleration, resulting in stellar outflows.   Star formation in our simulations requires gas to be sufficiently dense and Jeans unstable.  Importantly, we find that stellar outflows are robust in our simulations for re-runs that require more stringent star formation criteria \citep[see Appendix \ref{app:1} and][]{Grudic18}. 
It is worth noting that (although we do not quantitatively analyze these runs here) the qualitative phenomenon of star formation in outflows also appears across our previous FIRE-2 {\em and} FIRE-1 simulations using different numerical resolutions, including or excluding magnetic fields, changing the star formation criteria more broadly, and using an entirely different hydrodynamic method (SPH, in FIRE-1).
Given this, we believe that our conclusions are physically plausible (and consistent with observed stellar outflow measurements).  However, much higher resolution simulations that include a fair treatment of conduction and other micro-physics will be needed to test for robustness. 

The broadest contribution of this work is in positing the possibility that outflow stars may populate the stellar halos of Milky Way-mass galaxies.  The degree to which this happens will depend in detail on star formation and galaxy-formation physics; this means that local searches for such a population may offer a new way to test global galaxy formation models.  The quantitative results presented here provide a starting point for such an endeavor. With these caveats in mind, we can move on to conclude and summarize.

\section{Conclusions} \label{s:conclusion}

Using six zoom simulations of Milky Way-mass galaxies, we have found that stellar outflows associated with starburst activity contribute a significant fraction to the stellar halo of each galaxy.  These stars form with large positive radial velocities ($\gtrsim 150 ~\kms$, Figure \ref{fig:Vbirth}) inside powerful galactic outflows (Figure \ref{fig:gas_visual_step}).   As shown in Figure \ref{fig:SFH}, stellar outflows are more common during the early, bursty star formation phases of our galaxies ($z \gtrsim 0.3$).  While the overall fraction of stars formed in outflows is small ($\sim 1 \%$), during starbursts the outflow fraction ranges from $\sim 1-50 \%$ of the overall star formation rate.

The majority of outflow stars fall back in on radial orbits that are similar to the bulk of the stellar halo at $z=0$.  Though the fraction varies from halo to halo, we find that $5-40\%$ of outer stellar halos ($r>50$ kpc) of our galaxies formed in radial outflows (Figure \ref{fig:all_frac}). The rest of the outer halo is dominated by stars from accreted, disrupted satellites, which tend to be metal poor.  If we restrict to the most metal rich ([Fe/H] $>-1$) outer-halo stars, an even higher fractions were born in outflows ($\sim 30-80\%$) owing to the fact that these outflow stars originated from the more metal rich gas in the massive central galaxy.

In the inner halo ($r \sim 8$ kpc), we find that outflows contribute substantially only the the most metal-rich portion, making up $\sim 10-40\%$ of [Fe/H] $> -0.5$ stellar halo material in the solar neighborhood (Figure \ref{fig:localhalofraction}).  The lowest metallicity component of the local stellar halo in our simulations tends to be dominated by accretions, while the metal-rich portion is made of a combination of accreted stars,  kicked-out (heated disk) stars, and \insitu stars born in outflows (Figure \ref{fig:haloRbirthVbirth}). 

The outer stellar halo of the Milky Way and M31 hold some promise in for finding evidence for outflow stars.   Not only are the outflow stars more metal rich than the dominant, accreted stellar halo, they tend to be more smoothly distributed (Figure \ref{fig:rvr_spatialiw}) and $\alpha$-enhanced.  They stand out clearly in [Mg/Fe]-[Fe/H] diagrams (Figure \ref{fig:all_mgfe_feh}). Efforts such as the H3 Spectroscopic Survey for the Milky Way \citep{Conroy19} and ongoing work using resoled stars around M31 from Keck \citep[][Kirby et al. \textit{in prep}]{Escala19} may enable direct comparisons to these predictions.
 
Distinguishing between outflow stars and other halo components in the {\em local} stellar halo ($r \sim 8$ kpc) may be more difficult. The distributions of their current tangential velocities are almost identical. We find only small differences in the $z=0$ velocity distributions or the chemical properties of outflow stars (see Appendix \ref{app:2}). Indeed, outflow stars that have remained bound to the inner stellar halo appear to be quite similar to heated disk stars in everything other than their birth velocities.  We find that outflow stars dominate the local halo population only among stars that are both high in radial velocity {\em and} high in metallicity (see Figure \ref{fig:joint}).  This may be the space where outflow stars are most easily identified in the stellar halo.
 
Could an outflow event be directly observed?  At least one of our six galaxies has a prominent visible plume of young stars born in an outflow that is visible at $z=0$ (halo \texttt{m12w}). A stream of this kind could be observable in imaging of the nearby universe as a blue stream of young stars (see Figure \ref{fig:m12w_image}, for example). A potential avenue of future work is to make predictions on the detection probability of such events and the expectations for the surface brightness of young plumes of outflow stars. 

Stellar outflows may also provide a potential source for extreme-velocity or hypervelocity stars \cite[][]{Brown05,Hawkis2018}.  
The radius-velocity diagram of \texttt{m12w} shows that some outflow stars could have radial velocities  as high as $\sim 300\, \kms$ (Figure	\ref{fig:rvr_spatialiw}).  We also find that a small fraction ($\sim 5 \%)$ of outflow stars generated in outflow events are unbound. 
While most HVS discoveries are are metal-poor (and therefore are not good matches for the stellar outflows we predict in this scenario), there are HVS candidates that have metallicity higher than expected in the stellar halo \citep[e.g.,][]{Du18}. These high-metallicity HVSs could potentially be sourced by outflows.

A related observational constraint will be to consider the direct rates of outflows predicted in our simulations compared to ongoing efforts to detect stellar outflows in spectroscopic surveys.  As discussed in the introduction, \citet{Gallagher2018} have used integral field MaNGA data to show that star formation inside outflows accounts for $5 - 30 \% $ of the total star formation in their galaxies when detected. This is consistent with the rough estimate presented in Figure \ref{fig:SFH}, but direct mock observations of the simulations would be required for accurate comparisons. 

In work related to ours, \cite{Ma2019} have used FIRE-2 simulations of high-redshift galaxies to study star formation in clusters.  They find star formation at the edges of shells in high density gas clouds, compressed by feedback-driven winds, which is quite similar to the physical conditions we find fundamental to launching outflow stars in this paper.  It is encouraging that \cite{Ma2019} find similar results using a simulation suite that includes a run with $\sim 70$ times better mass resolution than our own.  

Another implication for bubble-driven star formation outflow is that it may enable higher ionizing photon escape fractions than might otherwise be possible.  Ma et al. \textit{in prep} have studied this effect in detail using FIRE-2 simulations to estimate the escape fraction of  stars formed in this manner and find a large time-average escape fraction ($\gtrsim 10\%$), for several different star formation models.  Specifically, stars formed at the edges of evacuated shells have much larger escape fractions.  They also find age gradients for stars born at the shell front, which is again consistent with the idea of an accelerating shell of star formation, consistent with the outflow picture presented here.

In summary, we have introduced the possibility that stellar outflows contribute a non-negligible fraction of stars to the hot, extended stellar halos of Milky Way-mass galaxies.  Such a population is plausible given the evidence that dense molecular outflows and stellar outflows exist in nature \citep{Maiolino17,Gallagher2018}.  The characteristics of such a halo population are likely sensitive to the nature of feedback and star-formation physics (Section \ref{s:discuss}). Given this, we believe that local observational searches for stellar-halo outflow stars may provide yet another place where near-field cosmological probes can begin to test global models of galaxy formation.

\section{Acknowledgments} \label{s:Acknowledgements}
This project was developed in part at the 2019 Santa Barbara Gaia Sprint, hosted by the Kavli Institute for Theoretical Physics (KITP) at the University of California, Santa Barbara.
This research was supported in part at KITP by the Heising-Simons Foundation and the National Science Foundation (NSF) under grant No. PHY-1748958.

SY and JSB were  supported by NSF AST-1910346, AST-1518291, HST-AR-14282, and HST-AR-13888. AW received support from NASA, through ATP grant 80NSSC18K1097 and HST grants GO-14734 and AR-15057 from STScI, the Heising-Simons Foundation, and a Hellman Fellowship. ASG is supported by the McDonald Observatory at the University of Texas at Austin, through the Harlan J. Smith fellowship. MBK acknowledges support from NSF grants AST-1517226, AST-1910346, and CAREER grant AST-1752913 and from NASA grants NNX17AG29G and HST-AR-14282, HST-AR-14554, HST-AR-15006, HST-GO-14191, and HST-GO-15658 from the Space Telescope Science Institute, which is operated by AURA, Inc., under NASA contract NAS5-26555.
Support for PFH was provided by NSF Collaborative Research Grants 1715847 \&\ 1911233, NSF CAREER grant 1455342, NASA grants 80NSSC18K0562, JPL 1589742. Numerical calculations were run on the Caltech compute cluster ``Wheeler,'' allocations from XSEDE TG-AST130039 and PRAC NSF.1455342 supported by the NSF, and NASA HEC SMD-16-7592.
DK was supported by NSF grant AST-1715101 and the Cottrell Scholar Award from the Research Corporation for Science Advancement.
CAFG was supported by NSF through grants AST-1517491, AST-1715216, and CAREER award AST-1652522; by NASA through grant 17-ATP17-0067; and by a Cottrell Scholar Award from the Research Corporation for Science Advancement.

This work also made use of Astropy \footnote{\url{https://www.astropy.org}}, a community-developed core Python package for Astronomy \citep{Astropy, Astropy18}, matplotlib \citep{Matplotlib}, numpy \citep{Numpy}, 
and the NASA Astrophysics Data System.




\bibliographystyle{mnras}
\bibliography{Galactic_halo_refs} 




\appendix

\section{Outflow stars with  more conservative star formation criteria}
\label{app:1}
\begin{figure}
\includegraphics[width=0.48\textwidth,trim = 0.0 80.0 0.0 0.0]{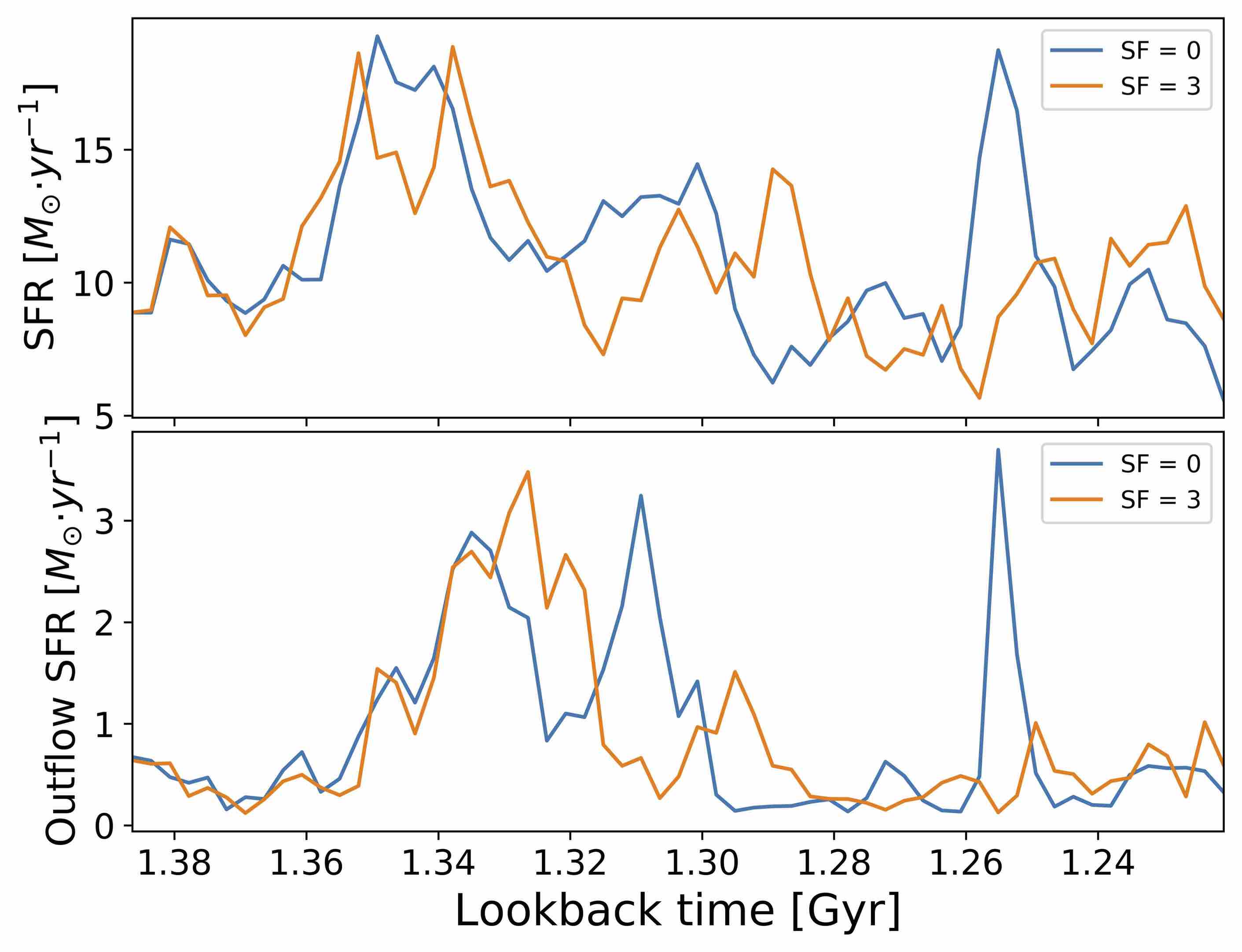}
\caption[]{Instantaneous star formation rates for total stars (top panels) compared to stellar outflow star formation rate (bottom panels) for two reruns of \texttt{m12w} with different star formation criterion. Star formation rates are averaged over 2.8 Myr.  Outflows are defined as stars that were formed with $V_{\rm rad}^{\rm birth} > 150 \, \kms$.  Lines labeled "SF=0'' use the standard FIRE-2 star formation criteria.  Lines labeled "SF=3'' are more conservative, demanding convergent flow, completely zero star formation in any unbound regions and that the thermal Jeans mass be below 1000\Msun \citep[see][]{Grudic18}.}
\label{fig:rerun_SFR}
\end{figure}

In order to test the robustness of our stellar outflow predictions to star formation recipe, we have rerun simulation \texttt{m12w} over a period spanning a lookback time from 1.4 to 1.25 Gyr using stricter star formation criteria.   The results are shown in Figure \ref{fig:rerun_SFR}, which shows the star formation histories as in Figure \ref{fig:SFH} with global star formation rate vs. time on top and outflow star formation rate on the bottom.

The standard FIRE-2 star formation criterion (labeled ``SF = 0'') demands that star formation occurs within gas clouds that are  locally self-gravitating.  In our re-runs, we  also require that the thermal Jeans mass be below 1000\Msun,  that the gas be part of a convergent flow, and that the gas be gravitationally bound as evaluated from the usual virial parameter, but smoothed over 0.125 local freefall times \citep[][labeled ``SF = 3'']{Grudic18}.   The top panel shows the total star formation rate averaged over 2.8 Myr for the two runs and the bottom panels show the stellar outflow star formation rate. Here outflows are defined as stars that were formed with $V_{\rm rad}^{\rm birth} > 150 \, \kms$. 

The number of total new stars and outflow stars formed during this period is approximately the same for the two criterion.  The difference between global and outflow star formation is smaller than $2\%$.  Although the detailed star formation histories for the two runs are different, the general trend seems very similar. The outflow events correlate with starburst activity in the main galaxy.  This is consistent with the idea that these events are triggered by clustered supernova feedback.

\begin{figure*}
\includegraphics[width=0.48\textwidth,trim = 30.0 0.0 120.0 0.0]{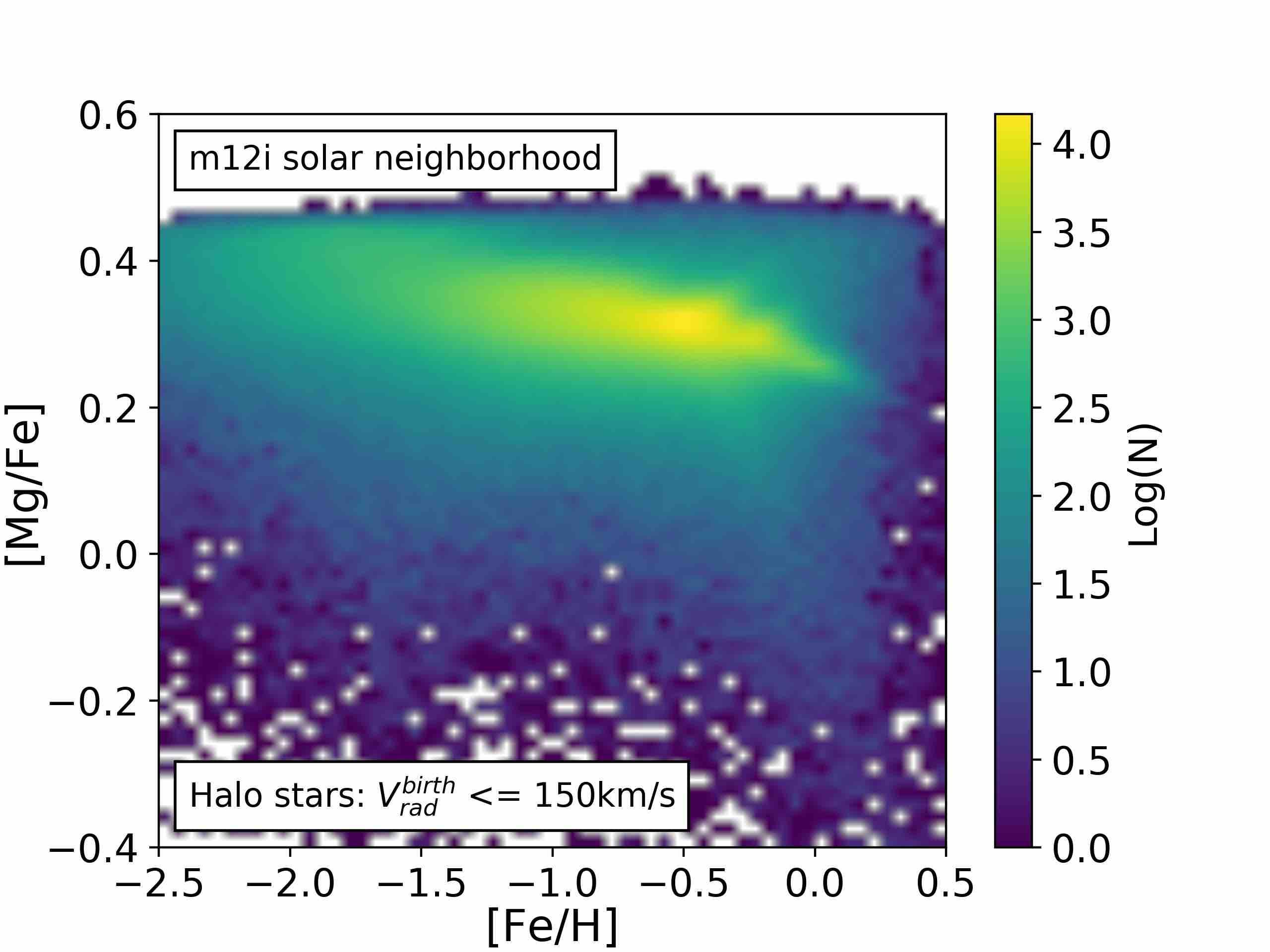}
\includegraphics[width=0.48\textwidth,trim = 0.0 0.0 150.0 0.0]{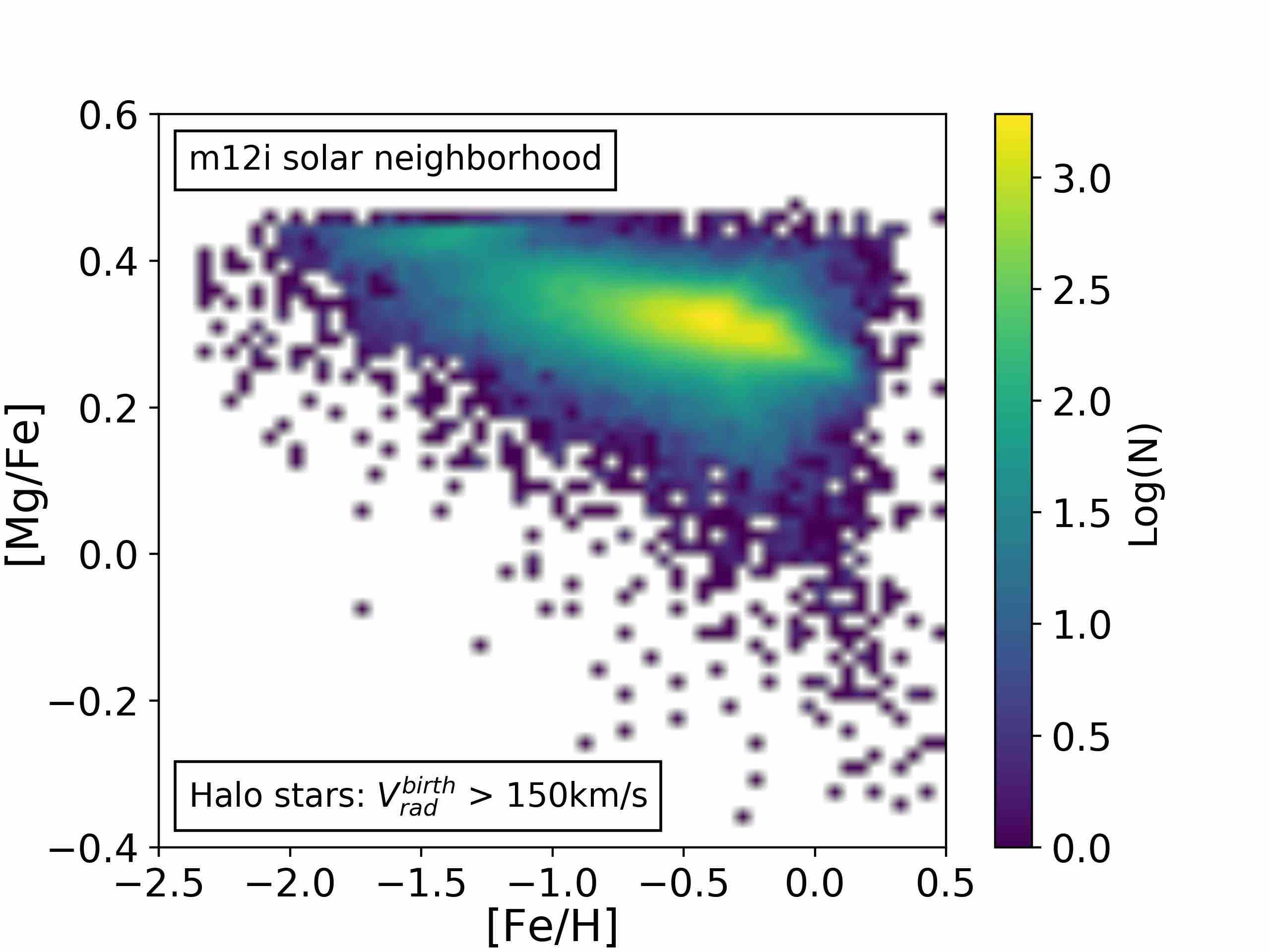}
\caption[]{Chemical abundances for the local stellar halo in \texttt{m12i} split by birth radial velocity, with stars identified as being born in outflows shown on the right and other stars shown on the left. At fixed [Fe/H], differences are not substantial.}
\label{fig:localalpha}
\end{figure*}

\begin{figure*}
\includegraphics[width=0.48\textwidth,trim = 60.0 0.0 0.0 0.0]{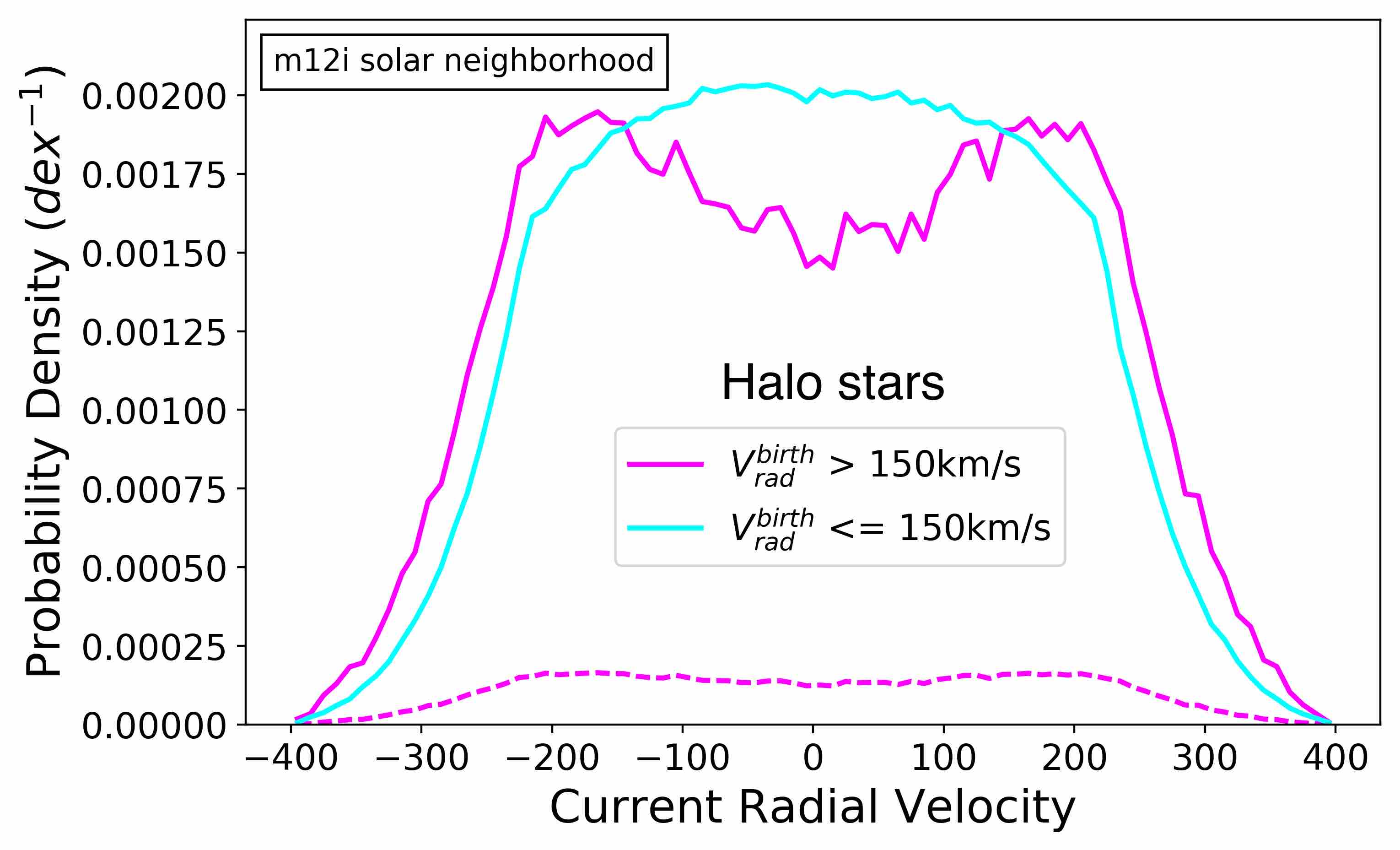}
\includegraphics[width=0.4685\textwidth,trim = 0.0 0.0 30.0 0.0]{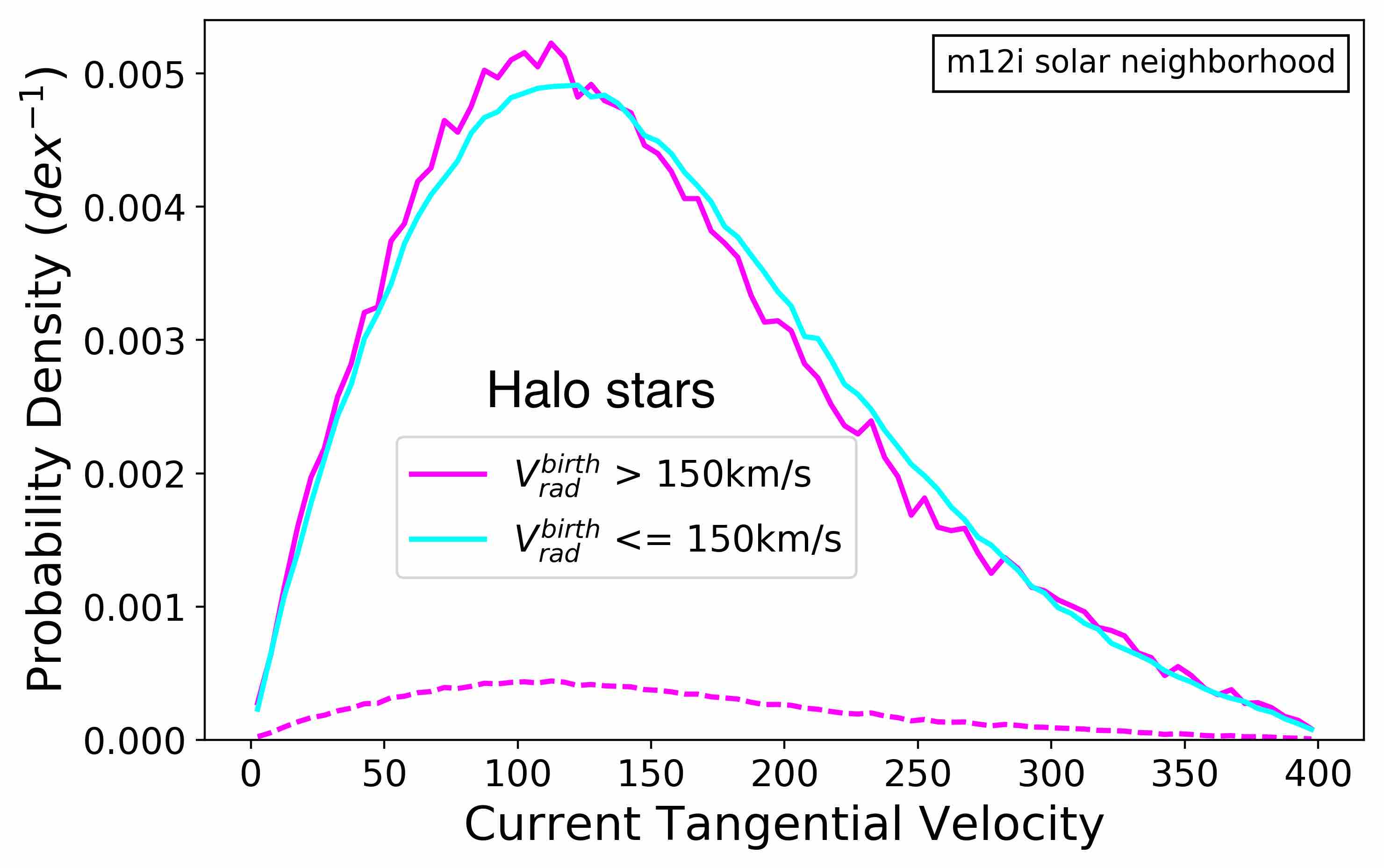}
\caption[]{Current radial and tangential velocity distribution for the local stellar halo in \texttt{m12i} split by birth radial velocity. Dotted lines in both plots show the distribution of the outflow stars set to the same scale as the other stars. No substantial differences spotted in both distribution.}
\label{fig:localkinematic}
\end{figure*}

\section{Observable properties of inner stellar halo stars: outflow stars vs. bulk}
\label{app:2}
In Section \ref{sec:local}, we discussed the contribution of stars made in outflows to the kinematically-identified local stellar halos of our simulated galaxies.  We found that the bulk of the local halo was comprised of heated stars and accreted stars, with a minority coming from stars born in outflows.  Here we show that it may be difficult to distinguish between outflow stars and other local halo stars using either chemical properties (Figure \ref{fig:localalpha}) or kinematic properties (Figure \ref{fig:localkinematic}) separately.  In Section \ref{sec:local}, we showed that the joint combination of high metallicty and high radial velocity helps to single out stars that were born in outflows. 

Figure \ref{fig:localalpha} shows the densities of  stars in \texttt{m12i} identified as being born in outflows (right, defined as $V_{\rm rad}^{\rm birth} > 150 \, \kms$) and other stars (left) in the space of [Mg/Fe] versus [Fe/H]. Note that the number of stars made in outflows is relatively small compared to the other stars. There are no obvious differences between the two population. 

In Figure \ref{fig:localkinematic}, we make similar comparison between the same population for their current radial (left) and tangential (right) velocities. The two populations are virtually identical in tangential velocity space.  In radial velocity, the distributions show some differences (with outflow material peaked towards both high and low velocities).  The dotted lines show the distribution of stars born in outflows scaled relative to the other stars based on the numbers of the two population. Given the relatively small number of outflow stars, it would be difficult to distinguish any individual stars based on kinematics alone.


\bsp	
\label{lastpage}
\end{document}